\documentclass{aa}  

\usepackage{graphicx}
\usepackage{txfonts}

\usepackage{multirow}
\usepackage{natbib,twoopt}
\usepackage[breaklinks=true]{hyperref} 
\usepackage{lscape}
\usepackage{booktabs}
\usepackage{threeparttablex}

\hypersetup{
  colorlinks=true,   
  urlcolor=blue,     
  linkcolor=blue     
}

\bibpunct{(}{)}{;}{a}{}{,} 
\makeatletter
\newcommandtwoopt{\citeads}[3][][]{\href{http://adsabs.harvard.edu/abs/#3}%
{\def\hyper@linkstart##1##2{}%
\let\hyper@linkend\@empty\citealp[#1][#2]{#3}}}
\newcommandtwoopt{\citepads}[3][][]{\href{http://adsabs.harvard.edu/abs/#3}%
{\def\hyper@linkstart##1##2{}%
\let\hyper@linkend\@empty\citep[#1][#2]{#3}}}
\newcommandtwoopt{\citetads}[3][][]{\href{http://adsabs.harvard.edu/abs/#3}%
{\def\hyper@linkstart##1##2{}%
\let\hyper@linkend\@empty\citet[#1][#2]{#3}}}
\newcommandtwoopt{\citeyearads}[3][][]%
{\href{http://adsabs.harvard.edu/abs/#3}
{\def\hyper@linkstart##1##2{}%
\let\hyper@linkend\@empty\citeyear[#1][#2]{#3}}}
\makeatother

\begin{document} 
  \title{
    Comparison of dynamical and kinematic reference frames 
    via pulsar positions from timing, \textit{Gaia}, and interferometric
        astrometry}
  
  \author{N. Liu\inst{1,2}
    \and
        Z. Zhu\inst{1}
    \and
    J. Antoniadis\inst{3,4}
    \and
        J.-C. Liu\inst{1}
    \and
        H. Zhang\inst{1}
    \and
        N. Jiang\inst{1}
          }
  \titlerunning{Comparison between dynamical and kinematic reference frames}
  \authorrunning{N. Liu et al.}

  \institute{School of Astronomy and Space Science,
   		     Key Laboratory of Modern Astronomy and Astrophysics (Ministry of Education), 
   		     Nanjing University, Nanjing 210023, P. R. China\\
              \email{[niu.liu;zhuzi]@nju.edu.cn}
         \and
             School of Earth Sciences and Engineering, 
             Nanjing University, Nanjing 210023, P. R. China
        \and
            Institute of Astrophysics, Foundation for Research and Technology-Hellas, Voutes, 71110 Heraklion, Greece \\
            \email{john@ia.forth.gr}
        \and
            Max-Planck-Institut für Radioastronomie, Auf dem Hügel 69, D-53121 Bonn, Germany
             }

  \date{Received; accepted}

  \abstract
    {
     Pulsars are special objects whose positions can be determined independently from timing, radio interferometric, and \textit{Gaia} astrometry at sub- milliarcsecond (mas) precision; 
     thus, they provide a unique way to monitor the link between dynamical and kinematic reference frames.
    }
   {
    We aimed to assess the orientation consistency between the dynamical reference frame represented by the planetary ephemeris and the kinematic reference frames constructed by \textit{Gaia} and VLBI through pulsar positions. 
   }
   {
   We identified 49 pulsars in \textit{Gaia} Data Release 3 and 62 pulsars with very long baseline interferometry (VLBI) positions from the PSR$\pi$ and MSPSR$\pi$ projects and searched for the published timing solutions of these pulsars.
    We then compared pulsar positions measured by timing, VLBI, and \textit{Gaia} to estimate the orientation offsets of the ephemeris frames with respect to the \textit{Gaia} and VLBI reference frames by iterative fitting.
   }
   {
   We found orientation offsets of $\sim$10\,mas in the DE200 frame with respect to the \textit{Gaia} and VLBI frame.
   Our results depend strongly on the subset used in the comparison and could be biased by underestimated errors in the archival timing data, reflecting the limitation of using the literature timing solutions to determine the frame rotation. 
   }
   {}


\maketitle


\section{Introduction}

    The celestial reference system provides the basic positioning standard widely used in astrometry, geodesy, and navigation for spacecraft.
    The International Celestial Reference System (ICRS) was adopted by the International Astronomical Union (IAU) in 1998 as the new fundamental reference system to replace the Fundamental Katalog No 5 (FK5) system  \citepads{1998A&A...331L..33F}.
    According to the latest B3 resolution adopted by the IAU in 2021\footnote{\url{https://www.iau.org/static/archives/announcements/pdf/ann21040c.pdf}.}, the fundamental realization of the ICRS consists of the third realization of the International Celestial Reference Frame \citepads[ICRF3;][]{2020A&A...644A.159C} constructed using very long baseline interferometry (VLBI) observations and the \textit{Gaia} celestial reference frame (\textit{Gaia}-CRF3) based on the \textit{Gaia} (Early) Data Release 3 \citepads[\textit{Gaia} DR3;][]{2016A&A...595A...1G,2022A&A...667A.148G}.
    The ICRS concept is built on the assumption that the Universe does not show a global rotation; such a definition of a nonrotating celestial reference system is purely kinematic.
    Therefore, ICRF3 and \textit{Gaia}-CRF3 fall into the category of kinematic celestial reference frames.
    
    On the other hand, observations of Solar System objects can be used to construct an inertial reference system, in which the motions of these objects do not present any acceleration reflected in the rotation of the celestial reference system \citepads{2012fuas.book.....K}.
    In this case, the nonrotation of the celestial reference system is defined dynamically.
    The dynamical celestial reference frame is mainly materialized by numerical planetary ephemerides, for example, the Jet Propulsion Laboratory's (JPL) planetary and lunar Development Ephemeris (DE) series \citepads{2021AJ....161..105P}, the Ephemerides of Planets and the Moon (EPM) at the Institute of Applied Astronomy, Russian Academy of Sciences \citepads{2014CeMDA.119..237P}, and the INPOP (Int\'egration Num\'erique Plan\'etaire de l’Observatoire de Paris) ephemerides from the Paris Observatory \citepads{2019NSTIM.109.....F}.
    
    Accurate alignments among celestial reference frames at different wavelengths are important for identifying multiwavelength counterparts \citepads[e.g.,][]{2021ApJ...914...48T}, aligning images at different bands or made by various instruments \citepads[e.g.,][]{2021PASA...38...50D}, and studying the frequency-dependent relation of the centroid location \citepads[e.g.,][]{2008A&A...483..759K,2017A&A...598L...1K}.
    An accurate alignment between the dynamical and kinematic celestial frames is also required in other applications, for instance, VLBI tracking of the spacecraft in deep space missions \citepads{2022AdSpR..69.1060Y}.
    
    The alignment of the dynamical frames based on the numerical ephemerides for the inner planets onto the ICRS is mainly achieved by VLBI measurements of the planetary spacecraft relative to the nearby extragalactic sources and range measurements \citepads{2015HiA....16..219F}, for example, the DE series since DE403 \citepads{standish1995jpl}.
    For the ephemerides of the outer planets, the alignment is carried out by optical observations with referred to various stellar catalogs, which is less accurate than those of the inner planets.
    The first version of DE ephemerides linked to the ICRF1 \citepads{1998AJ....116..516M} is DE405 \citepads{standish1998jpl}, whose alignment of the inner planet ephemeris system uses the single-baseline VLBI observations of the Magellan Spacecraft orbiting Venus and the Phobos Spacecraft approaching Mars \citepads{1994TDAPR.119...46H,standish1995jpl,folkner2007planetary}.
    These VLBI observations mostly come from only two baselines, which are the Goldstone-Madrid baseline (nearly in the right ascension direction) and Goldstone-Australia baseline (toward approximately the midway between the right ascension and declination).
    The newest DE440 and DE441 are tied to ICRF3 with an average accuracy of 0.2\,mas \citepads{2021AJ....161..105P} for the inner planets through the very long baseline array (VLBA) measurements of Mars-orbiting satellites \citepads{2015AJ....150..121P} combined with the traditional single-baseline VLBI observations and ranging data.
    Other indirect alignments are also used, for instance, by comparison of the Earth orientation parameters between the VLBI and lunar laser ranging (LLR) observations \citepads{1994A&A...287..279F}.
    
    Pulsars are special pointlike objects whose position can be determined independently from pulse timing, VLBI, and \textit{Gaia} astrometry at sub- milliarcsecond (mas) precision.
    There are two broad classes of pulsars.
    The recycled millisecond pulsars are old neutron stars with very short and stable spin periods ($\lesssim\,30\,{\rm ms}$) and can be timed precisely, resulting in the timing position measurement with a precision now reaching a few microarcseconds for the best cases \citepads[e.g.,][]{2019MNRAS.490.4666P}.
    Another class is the young, nonrecycled pulsars with spin periods of approximately 1\,s, whose timing position precision is several orders of magnitude worse than that of the millisecond pulsars \citepads{2017MNRAS.469..425W}.
  
    The timing observations of the pulsars have various applications, for example, searching nanohertz (nHz) gravitational waves \citepads{2021MNRAS.508.4970C,2021ApJ...917L..19G,2022MNRAS.510.4873A}, testing General Relativity \citepads{2021ApJ...921L..19D}, constructing a pulse-base time standard \citepads{2020MNRAS.491.5951H}, measuring the mass of the Solar System planets \citepads{2010ApJ...720L.201C}, constraining the acceleration of the Solar System \citepads{2018MNRAS.481.5501C}, and evaluating the errors in Solar System ephemerides \citepads{2020ApJ...893..112V}.
    The introduction of VLBI astrometric data in the timing solution can improve the precision of measurements for parameters such as timing irregularities \citepads{2020PASJ...72...70L} and increase the detection sensitivity of nHz gravitational waves \citepads{2013ApJ...777..104M}.
    Recently, \citetads{2021MNRAS.502..915C} used the astrometric data in \textit{Gaia} Data Release 2 to boost the efficiency of their gamma-ray pulsation search.
    In these cases, the systematics in the ephemeris reference frames and the potential frame tie issue among these frames may influence the interpretation of the timing solution \citepads{2013ApJ...777..104M,2020ApJ...893..112V}.
    An assessment of the tie precision among the timing, VLBI, and \textit{Gaia} frames can therefore help to clarify the error contribution from the misalignment of the celestial reference frames.  
    
    Since the \textit{Gaia} and VLBI positions refer to the kinematic reference frames and the timing positions refer to the dynamic frames, the comparison of these positions of pulsars thus provides a unique way to evaluate the frame tie accuracy between these two kinds of reference frames.
    For the purpose of the tie between dynamical and kinematic frames, the usage of the pulsar positions has several advantages:\\
    (i) using pulsar positions can achieve a direct tie between the ephemeris and the extragalactic reference frames \citepads{1996AJ....112.1690B,2009ApJ...698..250C};\\
    (ii) pulsar positions are fully independent of the production of the ephemerides; thus, it can serve as an external check on the frame-tie \citepads{2009A&A...507.1675F,2011CeMDA.111..363F};\\
    (iii) the relatively uniform sky distribution of the pulsars and the strong geometry of the observed VLBI network may be less sensitive to the systematics along the specific direction when using single baselines and a few targets;\\
    and (iv) the ongoing and future pulsar timing array (PTA) and VLBI observing campaigns permit regular monitoring of the frame-tie status.
    
    Comparisons between timing and radio interferometric positions date back to the 1980s, for example, in \citetads{1984MNRAS.210..113F} and \citetads{1985AJ.....90..318B}.
    In a recent comparison of timing and VLBI reference frames, \citetads{2017MNRAS.469..425W} considered not only millisecond pulsars but also young pulsars.
    They reported that the alignment accuracy between the VLBI and timing frames was mainly limited by uncertainties in the VLBI position (of a few mas).
    On the other hand, although no pulsar is intrinsically brighter than the limiting magnitude of \textit{Gaia} ($G\,\simeq\,21\,{\rm mag}$), the bright companions in the binary pulsar systems can be observed with \textit{Gaia} astrometry. 
    It would be interesting to check whether the arrival of the \textit{Gaia} data can, to some extent, contribute to improving the alignment precision between the dynamic and kinematic frames. 
    \citetads{2018ApJ...864...26J} found optical counterparts for 22 binary pulsars in \textit{Gaia} Data Release 2 (DR2).
    \citetads{2020RNAAS...4..223A} also searched for pulsars in the \textit{Gaia} catalog and reported a list of 41 close astrometric pairs in \textit{Gaia} Early Data Release 3, although eight pairs were only candidate associations.
    These samples are sufficiently large to permit an evaluation of the alignment agreement between the \textit{Gaia} and timing celestial reference frames, which is the main motivation of this work.
    
    We aimed to compare the dynamical celestial reference frames constructed by the planetary ephemerides and the kinematic celestial reference frames constructed by VLBI and \textit{Gaia} through the pulsar positions.
    Throughout the paper, we used the pulsar name based on their J2000 coordinate.
    All necessary data and codes to reproduce all the results and figures in this paper are publicly available online\footnote{\url{https://git.nju.edu.cn/neo/TimingCRF-vs-GaiaVLBICRF}}.


\section{Materials and methods}

\subsection{Search for \textit{Gaia} and VLBI pulsars} \label{subsec:sample}

    Our first step was to search pulsars in the \textit{Gaia} and VLBI catalogs.
    The main guideline was to find as many pulsars as possible with astrometric parameters estimated in a uniform and consistent manner.
    We did not limit our search to millisecond pulsars; instead, we planned to include all the available pulsars in the comparison as done in \citetads{1984MNRAS.210..113F}. 
    \citetads{2017MNRAS.469..425W} found that the inclusion of young pulsars in addition to millisecond pulsars in the comparison of VLBI and timing positions did not improve their results.
    Including young pulsars in our analyses can allow us to check if this assertion holds valid for the comparison between \textit{Gaia} and timing positions. 
    
    To look for the \textit{Gaia} pulsars,
    we used the same method as described in \citetads{2020RNAAS...4..223A}
    but updated their work by using a newer version (version 1.68) of the Australia Telescope National Facility (ATNF) Pulsar Catalogue \citepads{2005AJ....129.1993M}\footnote{\url{http://www.atnf.csiro.au/research/pulsar/psrcat}. Accessed on 2022 November 28.}.
    We found 49 astrometric close pairs, including all identified binary pulsars except PSR J1628--3205 in \citetads{2018ApJ...864...26J} and three of five millisecond radio pulsars in \citetads{2019MNRAS.486.4098I}.
    Astrometric pairs for two pulsars -- PSR J1435--6100 and PSR J1955+2908 -- were considered the most unlikely true associations \citepads{2021MNRAS.501.1116A,2018ApJ...864...26J}.
    We kept these two pulsars in our \textit{Gaia} pulsar sample, leaving them to be verified in future studies, but removed them from the list in the comparison between \textit{Gaia} and timing positions.
    We also noticed that PSR J1024-0719 was found to belong to a wide binary \citepads{2016ApJ...826...86K,2016MNRAS.460.2207B}, for which the orbital period is not known but is likely to be a few kilo-years.
    The position of the optical companion as observed by \textit{Gaia} will differ significantly from the position of the pulsar determined by timing, which is validated in our comparison (Sect.~\ref{subsec:timing-vs-gaia}).
    Including this source would strongly bias the determination of the orientation offset between ephemeris reference frames and \textit{Gaia}-CRF.
    This pulsar was excluded from the sample used for comparison of timing versus \textit{Gaia}. 
    
    Table~\ref{tab:gaia-psr-info} tabulates the information for all 49 astrometric pairs.
    The full five astrometric parameters (i.e., position, proper motion, and parallax) are given in \textit{Gaia} DR3 for all pulsars except for PSR J1546--5302.
    Since the proper motion was required by the position correction for the reference epoch difference in the next step, this pulsar was excluded in the comparison of celestial reference frames.
    We noted that all the sources used in this work were treated as singular objects in the astrometric solution of \textit{Gaia} DR3 \citepads{2021A&A...649A...2L}, while most of the \textit{Gaia} pulsars are not isolated.
    As seen in Table~\ref{tab:gaia-psr-info} (Col.~9), the orbital periods for the \textit{Gaia} pulsars, if known, are generally too small compared to the length of the \textit{Gaia} DR3 data collection window ($\sim$1000~d).
    As a result, we could safely assume that the effect from the orbital motion on the \textit{Gaia} pulsar position was largely averaged out, as already discussed in \citetads{2018ApJ...864...26J}, and hence would not alter our results much.
    
    For the VLBI pulsar sample, we used the latest data release from the PSR$\pi$ project \citepads{2019ApJ...875..100D} and the published solutions for PSR J1012+5307 and PSR J1537+1155 (based on the bootstrap method) in the MSPSR$\pi$ project \citepads{2020ApJ...896...85D,2021ApJ...921L..19D}, which contains 62 pulsars in total.
    The five astrometric parameters for these pulsars together with the orbital parameters for two binary pulsars (PSR J1022+1001 and PSR J2145--0750) were derived from the relative astrometric observations (images) made by the VLBA \footnote{\url{https://safe.nrao.edu/vlba/psrpi/release.html}.}.
    Asymmetric uncertainties were assigned to these measurements; we took the largest uncertainty value as the formal uncertainty.    
    Since the positional uncertainty for PSR J1537+1155 given in \citetads{2021ApJ...921L..19D} did not include the positional errors of the phase calibrators, we inflated the positional uncertainty by considering systematic uncertainties due to both the core shift and the phase referencing error from the out-of-beam calibrator to the in-beam calibrator.
    We empirically adopted a systematics of 0.8\,mas from \citetads{2019ApJ...875..100D} and \citetads{2020ApJ...896...85D} to account for the core shift in each coordinate.
    To estimate the phase referencing error, we subtracted from the absolute position uncertainty of all pulsars in the PSR$\pi$ sample the uncertainty due to the core shift (i.e., 0.8~mas) and absolute position uncertainty of their out-of-beam calibrators in the RFC2019a solution (the reference catalog used for the absolute astrometric solution in the PSR$\pi$ data).
    The median value was 1.26\,mas in the right ascension and 0.59\,mas in declination; these two values were then used to inflate the positional uncertainty in the corresponding coordinate.
    
    There were only two common pulsars (PSR J0614+2229 and PSR J1012+5307) between the VLBI and \textit{Gaia} pulsar samples.
    It is also possible to increase the VLBI pulsar sample size by including solutions published by other authors \citepads[e.g.,][]{2002ApJ...571..906B,2003AJ....126.3090B,2009ApJ...701.1243D,2009ApJ...698..250C}.
    We finally decided to only consider the PSR$\pi$ and MSPSR$\pi$ data for two reasons:
    (i) a mixture of VLBI solutions from various authors might introduce some additional systematics to the resulting VLBI celestial reference frame, considering that the observation setup and scheduling, treatments of the data, and method for parameter estimation were usually different;
    (ii) the PSR$\pi$ solution surpassed the previous solutions in terms of accuracy, and the size of the PSR$\pi$ and MSPSR$\pi$ sample is sufficiently large for a meaningful comparison of the celestial reference frame as required by this work.

\subsection{Search for timing solutions in the literature}

    \begin{figure}
      \resizebox{\hsize}{!}{\includegraphics{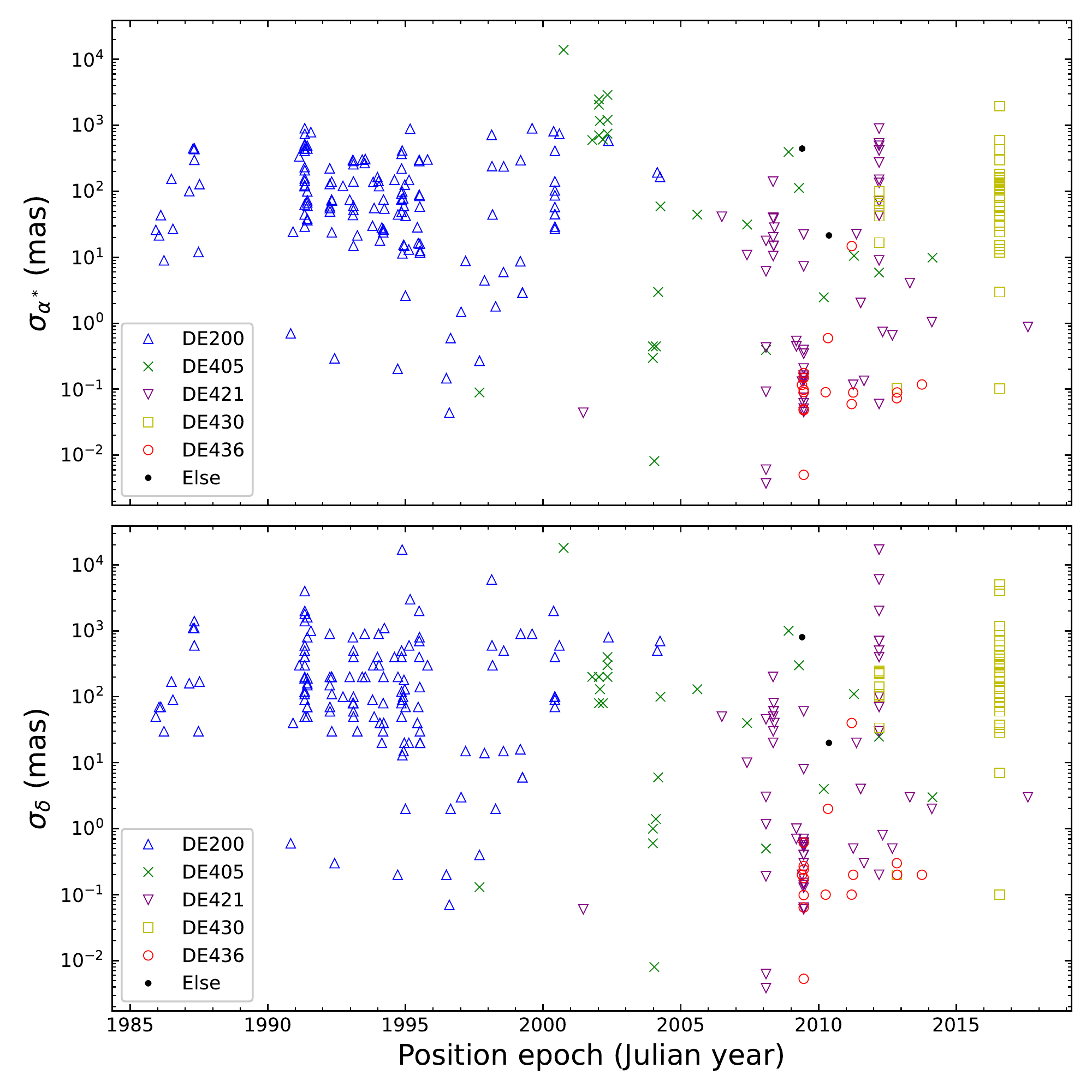}}
      \caption{
            Formal uncertainty of the timing positions in right ascension (top) and declination (bottom) quoted from the literature as a function of their position epochs.
            The measurements are distinguished by the reference ephemerides, that is, blue open triangles for DE200, green crosses for DE405, purple open inverted-triangles for DE421, yellow open squares for DE430, red open circles for DE436, and black filled circles for other ephemerides.      
      }
      \label{fig:timing-err}
    \end{figure}

    \begin{figure}
        \resizebox{\hsize}{!}{\includegraphics{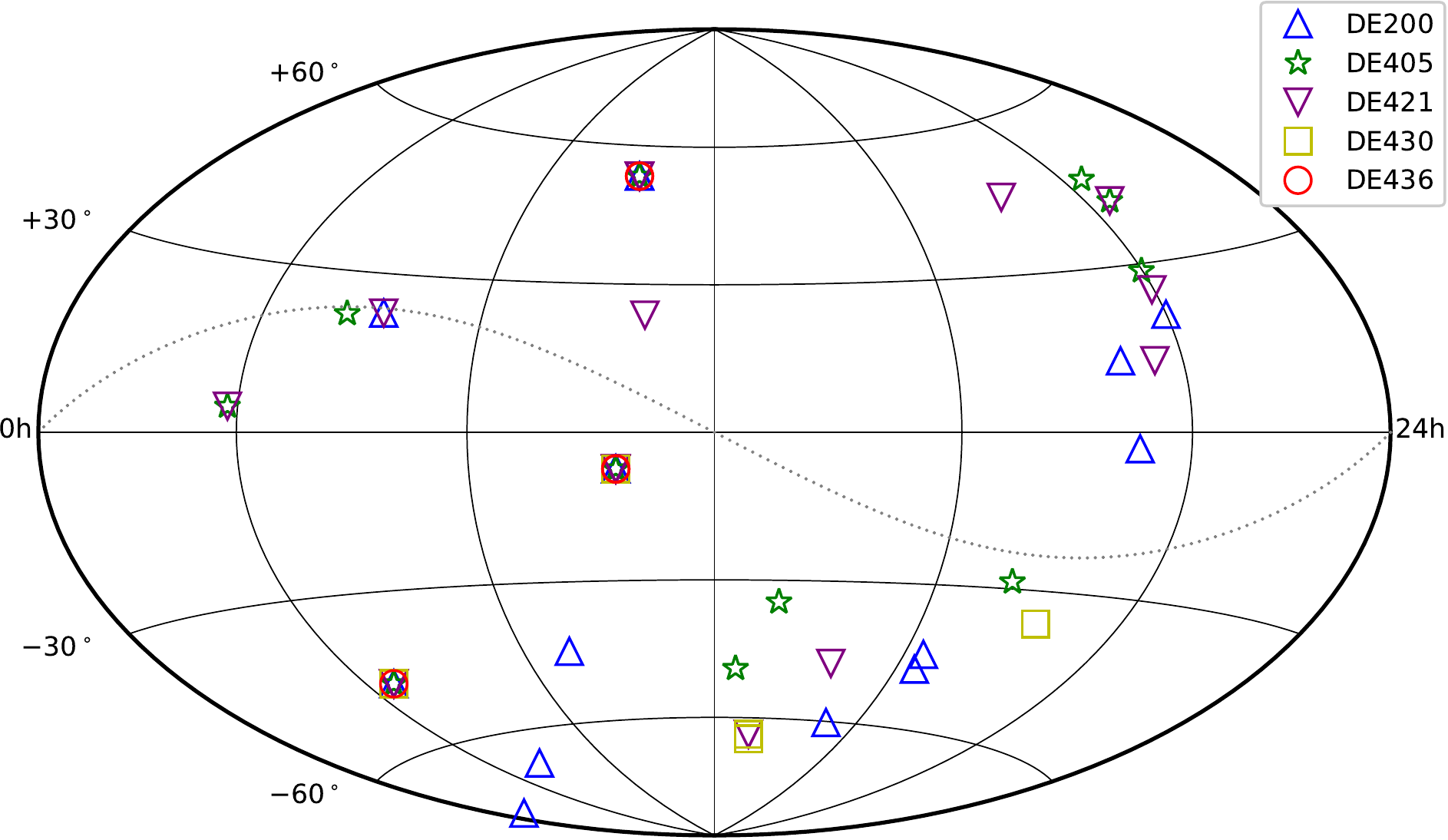}}
        \resizebox{\hsize}{!}{\includegraphics{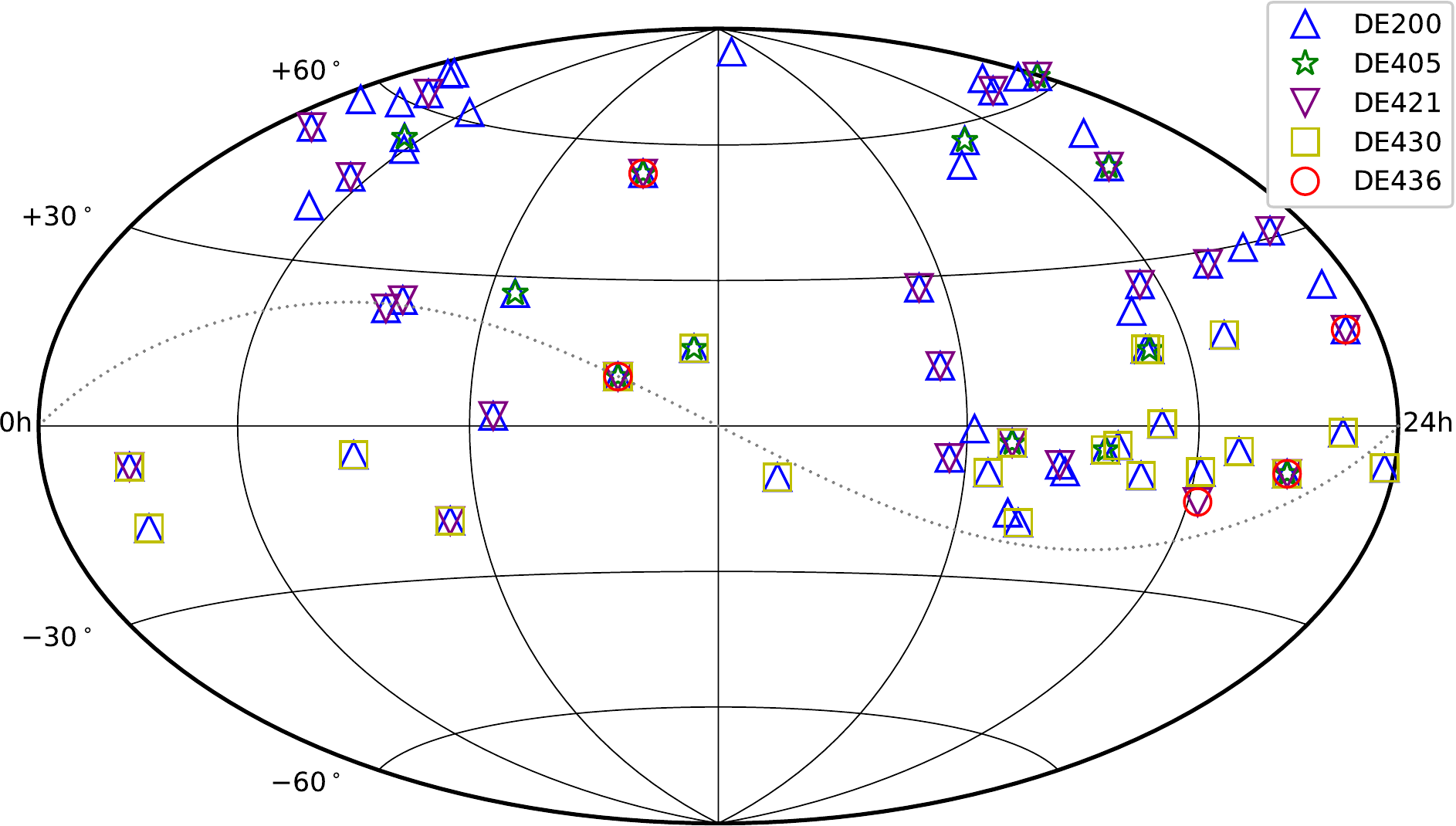}}
        \caption{%
            All-sky distribution of pulsars used in this work in the equatorial coordinate system.
            Top: Pulsars in \textit{Gaia} DR3;
            Bottom: Pulsars in the PSR$\pi$ and MSPSR$\pi$ projects.
            Different marks are used to distinguish pulsars whose timing solutions refer to different ephermerides, that is, blue open triangles for DE200, green open stars for DE405, purple open inverted-triangles for DE421, yellow open squares for DE430, and red open circles for DE436. 
            The dotted curves indicate the location of the ecliptic plane.
                  }
       \label{fig:psr-sky-dist}
    \end{figure}

    \begin{table*}
     \centering
     \caption[]{\label{tab:psr-pos-err}
     Overview of pulsar positions measured by timing, VLBI, and \textit{Gaia}.
     
     }
    \begin{tabular}{lccrrccccccl}
        \hline \hline
        CRF & Subset & Epoch & $N_{\rm PSR}$  & $N_{\rm obs}$
        & $\sigma_{\alpha^*}$  & $\sigma_{\delta}$ & $\sigma_{\rm pos,max}$ & $\sigma_{\varpi}$
        & $\sigma_{\mu,\alpha^*}$  & $\sigma_{\mu,\delta}$
        & References \\
        & & (yr) & & & (mas) & (mas) & (mas) & (mas) & ($\mathrm{mas~yr^{-1}}$) & ($\mathrm{mas~yr^{-1}}$) & \\ 
        \hline
        DE200
                    &All     &1994.1  &75    &142  &75   &120   &173 &5                     & 8  &13 & 1--23 \\
                    &MSP     &1996.9  &10    & 20  &2.8  &4.5   &5.0 &5.0                     &0.7 &1.4 &  \\
                    &Non-MSP &1993.5  &65    &122  & 99  &165   &194 &\dots \tablefootmark{a}   &15  &22 &  \\
                    \cline{2-12}
        DE405
                    &All  &2004.0    &23    & 26  & 52 & 90   &127  &0.50  &0.15    &0.30 & 24--34 \\
                    &MSP  &2004.3    &20    & 13  &2.5 &3.0   &4.7  &0.50  &0.15    &0.30 &  \\
                    &Non-MSP &2002.3 &13    & 13  &749 &200   &1075 &\dots \tablefootmark{a}  &\dots \tablefootmark{a}    &\dots \tablefootmark{a} &  \\
                    \cline{2-12}
        DE421
                    &All     &2009.5 &35    & 55  &1.0    &3.0    &3.1  &0.20                 &0.19 & 0.44 & 26,35--51 \\
                    &MSP     &2009.5 &15    & 33  &0.16   &0.50   &0.59 &0.20                 &0.08 & 0.17 &  \\
                    &Non-MSP &2010.4 &20    & 22  &41     &65     &76    &\dots \tablefootmark{a} &9.0  & 18.5 &  \\
                    \cline{2-12}
        DE430     
                    &All  &2016.6    &27    & 42  &77   &186   &225   &0.23     &2.0     &3.0 & 26,52--53 \\
                    &MSP  &2016.6    & 4    &  5  &3.0  &7.0   &7.6   &0.23     &0.06    &0.13 &   \\
                    &Non-MSP &2016.6 &23    & 37  &86   &200   &237   &\dots \tablefootmark{a}    &3.0     &4.5 &   \\
                     \cline{2-12}
        DE436\tablefootmark{b}
                    &All (MSP)  &2009.9        &8    & 16  &0.09   &0.20   &0.23 &0.14    &0.03     &0.07 & 54--56 \\
        \hline
        \textit{Gaia}
                    &All     &2016.0 &49    & 49  &0.16   &0.18   &0.22  &0.23    &0.22     &0.23 & 57 \\
                    &MSP     &2016.0 &27    & 27  &0.18   &0.19   &0.27  &0.25    &0.24     &0.24 &  \\
                    &Non-MSP &2016.0 &22    & 22  &0.14   &0.14   &0.20  &0.15    &0.20     &0.17 &  \\
                    \cline{2-12}
        VLBI      
                    &All     &2012.2 &62    & 62  &1.4    &1.0    &1.8   &0.06    &0.09     &0.12 & 58--61 \\
                    &MSP     &2012.2 & 6    &  6  &1.5    &1.5    &2.1   &0.08    &0.07     &0.12 &  \\
                    &Non-MSP &2012.2 &56    & 56  &1.4    &1.0    &1.8   &0.05    &0.09     &0.12 &  \\
        \hline
    \end{tabular}
    \tablefoot{
    The first two columns give the reference frames to which the astrometric parameters of pulsars are referred and the classicification of pulsars.
     The next three columns tabulate the median position epoch, the number of pulsars, and the number of astrometric measurements for these pulsars (some pulsars have more than one timing measurement).
     Columns 6--11 display the median formal uncertainty of the astrometric parameter measurements.
     The last column provides the references for the data.\\
    \tablefoottext{a}{The parallax or proper motion was not measured for this sample.}\\
    \tablefoottext{b}{All pulsars in this sample are millisecond pulsars.}
    }
    \tablebib{
(1)~\citetads{2004MNRAS.353.1311H};
(2)~\citetads{1994ApJ...422..671A};
(3)~\citetads{1993MNRAS.262..449S};
(4)~\citetads{1996ApJS..106..611D};
(5)~\citetads{2001MNRAS.328..855W};
(6)~\citetads{1998ApJ...505..352S};
(7)~\citetads{2002ApJ...581..501S};
(8)~\citetads{2003ApJ...589..495K};
(9)~\citetads{2000ApJ...528..907W};
(10)~\citetads{1994ApJ...426L..85A};
(11)~\citetads{1996Natur.381..584K};
(12)~\citetads{1997MNRAS.286..463B};
(13)~\citetads{2006ApJ...649..235M};
(14)~\citetads{1998MNRAS.297...28D};
(15)~\citetads{2001MNRAS.326..274L};
(16)~\citetads{1996ApJ...469..819C};
(17)~\citetads{1999MNRAS.307..925T};
(18)~\citetads{2001ApJ...548L.187C};
(19)~\citetads{2001MNRAS.328...17M};
(20)~\citetads{2003MNRAS.342.1299K};
(21)~\citetads{2006MNRAS.372..777L};
(22)~\citetads{1998MNRAS.295..397I};
(23)~\citetads{1994ApJ...425L..41B};
(24)~\citetads{2005MNRAS.362.1189Z};
(25)~\citetads{2011ApJ...743..102G};
(26)~\citetads{2020MNRAS.494..228L};
(27)~\citetads{2006MNRAS.369.1502H};
(28)~\citetads{2009MNRAS.400..805L};
(29)~\citetads{2009MNRAS.400..951V};
(30)~\citetads{2015ApJ...800L..12R};
(31)~\citetads{2012Sci...338.1314P};
(32)~\citetads{2013ApJ...776...20C};
(33)~\citetads{2011ApJS..194...17R};
(34)~\citetads{2009A&A...498..223J};
(35)~\citetads{2020ApJ...896..140D};
(36)~\citetads{2020PASJ...72...70L};
(37)~\citetads{2014MNRAS.437.3255S};
(38)~\citetads{2014ApJ...787...82F};
(39)~\citetads{2016MNRAS.458.1267V};
(40)~\citetads{2016MNRAS.458.3341D};
(41)~\citetads{2016ApJ...818...92M};
(42)~\citetads{2016MNRAS.460.4011L};
(43)~\citetads{2013Sci...340..448A};
(44)~\citetads{2016MNRAS.455.1751R};
(45)~\citetads{2016MNRAS.460.2207B};
(46)~\citetads{2021ApJ...909....6D};
(47)~\citetads{2015MNRAS.446.4019B};
(48)~\citetads{2014ApJ...791...67S};
(49)~\citetads{2019ApJ...886..148P};
(50)~\citetads{2016ApJ...833..192S};
(51)~\citetads{2017MNRAS.464.1211H};
(52)~\citetads{2019MNRAS.484.3691J};
(53)~\citetads{2016ApJ...826...86K};
(54)~\citetads{2019MNRAS.490.4666P};
(55)~\citetads{2018ApJS..235...37A};
(56)~\citetads{2021MNRAS.507.2137R};
(57)~\citetads{2021A&A...649A...1G};
(58)~\citetads{2016ApJ...828....8D};
(59)~\citetads{2019ApJ...875..100D};
(60)~\citetads{2020ApJ...896...85D};
(61)~\citetads{2021ApJ...921L..19D}.
    }
    \end{table*}
    
    \begin{table}
    \centering
    \caption[]{\label{tab:psr-pos-err-best}
    Typical (``Med'' for median) and best (``Min'' for minumin) level of the timing positional precision for pulsars used in this work.
     }
    \begin{tabular}{ccccccc}
        \hline \hline
        Subset  &\multicolumn{2}{c}{All}  &\multicolumn{2}{c}{MSPs}  &\multicolumn{2}{c}{Non-MSPs} \\
        \cmidrule(r){2-3} \cmidrule(r){4-5} \cmidrule(r){6-7}
        & Med & Min & Med & Min & Med & Min \\
        & (mas) & (mas) & (mas) & (mas) & (mas) & (mas) \\ 
        \hline
        \textit{Gaia} & 31 &0.005 &2.26 &0.005 &628 &11\\
        VLBI          &134 &0.074 &0.68 &0.074 &196 &15\\
        \hline
    \end{tabular}
    \end{table}

    We noticed that the pulsar positions given in the ATNF pulsar catalog were not always quoted from the timing solutions, and the ephemerides used in the timing solutions were not uniform.
    Therefore, we collected the published timing solutions for our sample by using the SIMBAD query service \footnote{\url{http://simbad.u-strasbg.fr/simbad/sim-fid}.}
    \citepads{2000A&AS..143....9W}.
    The ephemerides used in the timing solutions were not explicitly pointed out in some publications; the timing positions therein were not used in this work.
    For some pulsars, various authors published the timing positions at different epochs referring to the same ephemeris, based on independent or partly shared observations and utilizing identical analysis software packages \citepads[e.g., TEMPO2;][]{2006MNRAS.369..655H}.
    It was difficult (and even impossible) for us to determine exactly the correlations among these timing positions.
    Therefore, we assumed that these measurements were independent and used them all in the following analysis.
    Finally, we found 283 astrometric timing solutions for 93 pulsars, including 72 for 33 \textit{Gaia} pulsars and 221 for 62 VLBI pulsars.
    For most pulsars, only the celestial coordinates were estimated in their timing solutions.
    
    Figure~\ref{fig:timing-err} depicts the formal uncertainty of the timing positions quoted directly from the original papers.
    A declining trend from several hundreds milliarcseconds down to a few microarcseconds can be seen.
    The most popular ephemerides used in the timing solutions were DE200, D405, DE421, DE430, and DE436.
    We considered these five ephemeris frames as representatives of the dynamical celestial frames for our comparison.
    
    Using a similar method to \citetads{2021MNRAS.501.1116A}, we divided the pulsar samples into two subsets, that is, millisecond pulsars (MSPs) and the others (non-MSPs).
    The MSP sample consisted of both fully and mildly recycled millisecond pulsars distinguished in \citetads{2021MNRAS.501.1116A}.
    
    Table~\ref{tab:psr-pos-err} displays an overview of the pulsar catalogs, including the sample size and median formal uncertainty for the astrometric parameters.
    We computed the overall positional precision for each timing solution as
    \begin{equation} \label{eq:sig-max}
        \sigma_{\rm pos,max} = \sqrt{\left(\sigma_{\alpha^*} \right)^2 + \sigma^2_{\delta}}.
    \end{equation}
    where $\sigma_{\alpha^*}\,=\,\sigma_{\alpha}\cos\delta$.
    We noted that the precision of pulsar positions measured by the latest timing observations is close to or even better than those from \textit{Gaia}.
    The precision of VLBI position for pulsars is approximately 1\,mas, which is mostly due to the conjunction of the errors in the absolute positions of calibrators, core-shift and radio structure effect of calibrators, and bias in the extrapolation of calibrations \citepads{2019ApJ...875..100D}.
    The timing position errors of the MSPs are at least one order of magnitude smaller than those of the non-MSPs, while there is no such discrepancy for these two subsets in the \textit{Gaia} and VLBI catalogs.
    
    We also wanted to know the best precision of the timing position achieved for individual pulsars.
    For this purpose, we picked the lowest value of the overall positional precision, denoted $\sigma^{\rm best}_{\rm pos,max}$, from all available timing solutions for each pulsar.
    We used the median and minimum of $\sigma^{\rm best}_{\rm pos,max}$ as the estimators for the typical and best timing precision for the \textit{Gaia} and VLBI pulsars, as presented in Table~\ref{tab:psr-pos-err-best}.
    The best timing positional precision is achieved for PSR J0437$-$4715, which now approaches $5\,{\rm \mu as}$.
    This pulsar was observed by both \textit{Gaia} and VLBI \citepads[by the Australian Long Baseline Array;][]{2009ApJ...701.1243D}, but was not included in the PSR$\pi$ archive because it was at a declination of $-47^{\circ}$, which is not visible to the VLBA.
    
    Figure~\ref{fig:psr-sky-dist} depicts the distribution of pulsar samples on the celestial sphere.
    For some pulsars that are observed frequently, their positions are referred to different ephemerides (usually published by different authors).
    A noticeable feature in the lower panel is that the pulsars common to the VLBI sample are all above $\delta=-30~^{\circ}$, which is inherited from the PSR$\pi$ sample \citepads{2019ApJ...875..100D}.


\subsection{Computation and modeling of pulsar positional offsets} \label{sect:pos-oft}

    We used the \textit{Gaia} and VLBI positions as a reference and calculated the offsets of the timing position referring to them.
    The reference epoch of the timing position is always different from those of the \textit{Gaia} and VLBI positions, although the coordinate epochs are all J2000.0.
    The position propagation was thus necessary to enable a meaningful comparison of the timing positions with the \textit{Gaia} and VLBI positions.
    We used the \textit{Gaia} and VLBI proper motions to propagate the respective positions from their own reference epochs to those of the timing position, and then we calculated the offsets of the timing position with respect to the new \textit{Gaia} and VLBI positions.
    There were two reasons for using the \textit{Gaia} and VLBI proper motions for propagation in addition to the fact that the proper motion measurement was absent in most timing solutions (see Sect.~\ref{subsect:pm-error}).
    One was that the global spin (changing rate of orientation) of the \textit{Gaia} and VLBI celestial reference frames is known to a certain extent from previous studies \citepads{2018A&A...616A..14G,2021arXiv211210079L}.
    The other reason was that the uncertainties in the \textit{Gaia} and VLBI proper motions are generally smaller than those from timing solutions, leading to smaller formal uncertainty of the resulting positional differences.
    The formulae for computing the position offset and the associated uncertainty are given as follows:
    \begin{align} 
        \Delta \alpha^{*} &= \left(\alpha_{\rm T} - \alpha_{\rm R} \right)\,\cos\delta_{\rm R} 
        - \mu_{\alpha^*, {\rm R}} \left(t_{\rm T} - t_{\rm R}\right), \label{eq:pos-oft-ra} \\
        \Delta \delta &= \delta_{\rm T} - \delta_{\rm R} - \mu_{\delta} \left(t_{\rm T} - t_{\rm R}\right), \label{eq:pos-oft-dec} \\
        \sigma_{\Delta\alpha^{*}} &= \sqrt{\sigma^2_{\alpha,{\rm T}}\cos\delta_{\rm T}^2 + \sigma^2_{\alpha,{\rm R}}\cos\delta_{\rm R}^2
        + \sigma_{\mu_{\alpha^*},{\rm R}}^2\left(t_{\rm T} - t_{\rm R}\right)^2}, \label{eq:sigma-pos-oft-ra} \\
        \sigma_{\Delta\delta} &= \sqrt{\sigma^2_{\delta,{\rm T}} + \sigma^2_{\delta,{\rm R}}
        + \sigma_{\mu_\delta,{\rm R}}^2\left(t_{\rm T} - t_{\rm R}\right)^2}, \label{eq:sigma-pos-oft-dec} 
    \end{align}
    where $\Delta\alpha^{*}=\Delta\alpha\cos\delta$ and $\mu_{\alpha^*}=\mu_{\alpha}\cos\delta$.
    The subscripts ${\rm T}$ and ${\rm R}$ indicate data from the timing and reference (\textit{Gaia} or VLBI) solutions, respectively.
    
    We modeled the positional offsets as a rigid rotation $\boldsymbol{R}=(R_1, R_2, R_3)^{\rm T}$ in a global sense by
    \begin{equation} \label{eq:vsh01}
        \begin{array}{ll}
            \Delta\alpha^*  &= -R_1\cos\alpha_{\rm R}\sin\delta_{\rm R}  - R_2\sin\alpha_{\rm R}\sin\delta_{\rm R} + R_3\cos\delta_{\rm R}, \\
            \Delta\delta    &= +R_1\sin\alpha_{\rm R}            - R_2\cos\alpha_{\rm R}, \\
        \end{array}
    \end{equation}
    despite that the cause for differences in pulsar positions measured by timing, \textit{Gaia}, and VLBI may vary for individual cases \citepads{1984MNRAS.210..113F}.
    The rotation parameters were estimated by least-squares fitting to all pulsars.
    For pulsars with multiple-epoch positions, we considered each one as an independent measurement and used them all in the fitting.
    The position offsets were weighted by the inverse of the square of the combined formal uncertainties.
    Although the correlations between the right ascension and declination for the \textit{Gaia} and VLBI positions are available, they were not published with the timing positions in most cases.
    Therefore, the covariances between right ascension and declination were assumed to be zero for all pulsar positions.
    This assumption would be dangerous for pulsars near the ecliptic plane because the timing error ellipse, whose major axis is approximately aligned with ecliptic longitude, becomes extremely elongated for pulsars located near the ecliptic plane, making the timing measurements of the right ascension and declination highly correlated for these sources \citepads{2006MNRAS.369..655H}.
    Therefore, we removed severely affected pulsars, that is, those with ecliptic latitudes less than $5\,^{\circ}$ in an absolute sense.
    
    To cross-check our results with those provided by previous studies, we divided our samples into three subsets: all pulsars, MSPs only, and non-MSPs only.
    Another reason for such a division is that all candidate associations for \textit{Gaia} pulsars are non-MSPs \citepads{2021MNRAS.501.1116A}. 
    A separate analysis may allow our results to be less affected by the contamination of foreground or background stars in the \textit{Gaia} pulsar sample.
    Noting that our sample size is rather small and the estimation of the rotation parameters could be biased by individual measurements, as reported in \citetads{2017MNRAS.469..425W}, we used an iterative fitting.
    We computed the normalized position offsets for each measurement before and after the fitting as 
    \begin{align} 
        X_{{\rm pre},~i} &= \sqrt{\left( \frac{\Delta\alpha_i^*}{\sigma_{\Delta\alpha_i^{*}}} \right)^2 + \left( \frac{\Delta \delta_i}{\sigma_{\Delta\delta_i}} \right)^2}, \label{eq:pre_chi2} \\
        X_{{\rm post},~i} &= \sqrt{ \left( \frac{\Delta\alpha_i^*-\Delta\alpha_i^{*,c}}{\sigma_{\Delta\alpha_i^{*}}} \right)^2 + \left( \frac{\Delta \delta_i-\Delta \delta^{c}_i}{\sigma_{\Delta\delta_i}} \right)^2}, \label{eq:post_chi2}
    \end{align}
    where $\Delta\alpha_i^{*,c}$ and $\Delta \delta^{c}_i$ were computed according to Eq.~(\ref{eq:vsh01}) using the estimates of the rotation parameters.
    We ruled out the measurement with the largest $X_{{\rm pre},~i}$ once and then re-estimated the rotation parameters in each iteration.
    The fitting was repeated until there were two pulsars left in the sample.
    We considered the median value and the interquartile range divided by a factor of 1.35\footnote{The interquartile range (IQR) is the distance between the 25\% and 75\% quartiles. 
    For a normal distribution, the standard deviation equals 1.35 times the IQR.} of parameters from the iteration solutions as the final estimation and the corresponding uncertainty.
    The overall reduced chi-squared for all measurements before and after the fitting was calculated as 
    \begin{align} \label{eq:reduce_chi2}
        \chi^2_{\rm pre} &= \sum_i X_{{\rm pre},~i}^2 / \left( 2N_{\rm obs} - 1\right), \\
        \chi^2_{\rm post} &= \sum_i X_{{\rm post},~i}^2 / \left( 2N_{\rm obs} - 4\right),
    \end{align}
    where $N_{\rm obs}$ stands for the number of measurements.
    However, when the number of pulsars was less than five, a single least-squares fitting was performed instead.
    In the case of the sample with fewer than three pulsars, we did not estimate the rotation parameters.
    
\section{Results}

    \begin{table*}[htbp!]
     \centering
     \caption[]{\label{tab:rot-ang}
     Rotation parameters of the ephemeris frames with respect to the \textit{Gaia} and VLBI celestial reference frames.
     
     }
    \begin{tabular}{clrrrrrrrr}
        \hline \hline
        Ephemeris &Subset & $N_{\rm PSR}$  & $N_{\rm obs}$
        & $R_1$  & $\pm$ & $R_2$ & $\pm$ & $R_3$ & $\pm$ \\
        & & & & mas & mas & mas & mas & mas & mas \\ 
        \hline
        \multicolumn{10}{l}{wrt. \textit{Gaia}-CRF} \\
        DE200       &All    &10    &11  &$9$ &11  &$-25$ &21  &$-27$  &49  \\
                    &MSP\tablefootmark{a}    & 3    & 4  &8   &10  &$-24.9$  &5.9  &$-26$ &14  \\
                    &Non-MSP   & 7    & 7  &$-158$ &63 &$-402$ & 29  &$-292$ &128  \\
                    \cline{2-10}
        DE405       &All    & 7    & 7  &$0.9$ &9.5    &$-5.4$   &9.9    &$0.6$   &3.1    \\
                    &MSP    & 5    & 5  &$0.9$ &4.9    &$-5.4$   &4.9    &$0.7$   &3.3    \\
                    &Non-MSP\tablefootmark{b}  & 2     & 2  &\dots     &\dots     &\dots     &\dots   &\dots   &\dots    \\
                    \cline{2-10}
        DE421       &All    & 9    &13  &$0.7$ &0.5    &$-3.0$   &0.9    &$0.1$    &1.4    \\
                    &MSP    & 7    &11  &$0.7$ &0.7    &$-2.8$   &0.6    &$0.1$    &0.7    \\
                    &Non-MSP\tablefootmark{b}   & 2    & 2  &\dots     &\dots     &\dots     &\dots   &\dots   &\dots   \\
                    \cline{2-10}
        DE430       &All\tablefootmark{a}    & 4    & 5  &29  &30    &79    &78    &$-93$   &90    \\
                    &MSP\tablefootmark{b}    & 1    & 1  &\dots     &\dots     &\dots     &\dots   &\dots   &\dots     \\
                    &Non-MSP\tablefootmark{a}   & 3    & 4  &$-148$  &186    &$-329$    &333    &$-647$   &404    \\
                    \cline{2-10}
        DE436       &All (MSP)\tablefootmark{b} & 2    & 3  &\dots     &\dots     &\dots     &\dots   &\dots   &\dots    \\
        \hline
        \multicolumn{10}{l}{wrt. VLBI-CRF} \\
        DE200       &All                     &55    &106 &$-0.1$  &5.8    &$-6$     &15    &$-5.9$   &5.6    \\ 
                    &MSP\tablefootmark{a}    & 4    &  9 &$-0.9$  &1.1    &$-13.0$  &1.2   &$-12.1$  &0.6    \\ 
                    &Non-MSP                 &51    &97  &$0.9$   &6.0    &$-3$     &17    &$-5.0$   &4.3    \\ 
                    \cline{2-10}
        DE405       &All                     &11    &12  &$-2.7$  &6.0   &1.9  &2.1  &2.5    &1.5    \\
                    &MSP\tablefootmark{b}    & 2    & 3  &\dots     &\dots     &\dots     &\dots   &\dots   &\dots    \\
                    &Non-MSP                 & 9    & 9  &$-30$   &15    &9.5  &8.0  &$2$    &13   \\
                    \cline{2-10}
        DE421       &All                     &21    &31  &$-1.1$ &0.9    &$-0.6$   &0.7    &$-0.1$   &0.4    \\
                    &MSP    & 5    &14  &$-0.3$ &0.9    &$-1.2$   &0.6    &$-0.2$   &0.2    \\
                    &Non-MSP                 &16    &17  &$-9.0$ &9.5    &$-2.3$   &1.6    &$-0.4$   &5.5    \\
                    \cline{2-10}
        DE430       &All                    &18    &28  &$4.8$   &9.0    &17      &15     &$-0.3$     &3.2    \\
                    &MSP\tablefootmark{b}   & 1    & 2  &\dots     &\dots     &\dots     &\dots   &\dots   &\dots    \\
                    &Non-MSP                &17    &26  &$6$     &11     &9       &11     &$1.8$      &1.0    \\
                    \cline{2-10}
        DE436       &All (MSP)\tablefootmark{a}    & 4    & 9  &$-2.1$   &0.5    &$0.3$   &0.5    &$0.7$    &0.4    \\
        \hline
        \multicolumn{10}{l}{wrt. combined \textit{Gaia}/VLBI-CRF} \\
        DE200       &All    &64    &117 &$-0.4$   &5.4     &$-6$     &16    &$-6.3$   &5.4    \\ 
                    &MSP    & 6    & 13 &$1.8$    &9.2     &$-16.4$  &7.2   &$-14.5$  &3.7    \\ 
                    &Non-MSP   &58 &104 &$0.6$    &6.1     &$-5$     &17    &$-5.3$   &4.8    \\ 
                    \cline{2-10}
        DE405       &All    &17    &19  &$-0.8$ &1.0   &0.2     &1.1    &0.1      &1.4    \\
                    &MSP    & 6    &8  &$-0.7$  &0.8   &$-0.3$  &2.0    &$0.9$    &1.1    \\
                    &Non-MSP   &11    &11  &$-36$ &13  &5.0     &5.1    &$-40$    &132   \\
                    \cline{2-10}
        DE421       &All        &29     &44  &$-0.5$    &0.6    &$-0.9$   &0.6    &$-0.1$      &0.4    \\
                    &MSP        &11     &25  &$-0.4$    &0.6    &$-1.2$   &0.5    &$-0.1$      &0.4    \\
                    &Non-MSP    &18     &19  &$-4$      &12     &$-4$     &12     &$-0.4$   &5.4    \\
                    \cline{2-10}
        DE430       &All                    &22    &33  &$1.4$  &4.6    &4.5    &7.0  &$-2.9$   &3.8    \\
                    &MSP\tablefootmark{b}   & 2    & 3  &\dots     &\dots     &\dots     &\dots   &\dots   &\dots    \\
                    &Non-MSP                &20    &30  &$7$    &11     &12     &21   &$1.2$   &3.1    \\
                    \cline{2-10}
        DE436       &All (MSP)    & 5    &12  &$-1.3$   &1.4    &$-0.5$   &0.9    &$0.4$    &0.2    \\
        \hline
    \end{tabular}
    \tablefoot{
    The first two columns display the ephemeris name and the subset, followed by  the number of pulsars and the number of astrometric timing measurements used in the least-squares fitting.
     Columns 5--10 tabulate the estimates of rotation parameters and the corresponding uncertainties (see Sect.~\ref{sect:pos-oft} for details).\\
    \tablefoottext{a}{The rotation parameters were obtained from a single least-squares fitting for this sample.}\\
    \tablefoottext{b}{The number of pulsars in this sample is so small (less than three) that the corresponding rotation parameters were not estimated.}
    }
    \end{table*}

\subsection{Ephemeris frames versus \textit{Gaia}-CRF} \label{subsec:timing-vs-gaia}

    \begin{figure*}
        \includegraphics[width=\columnwidth]{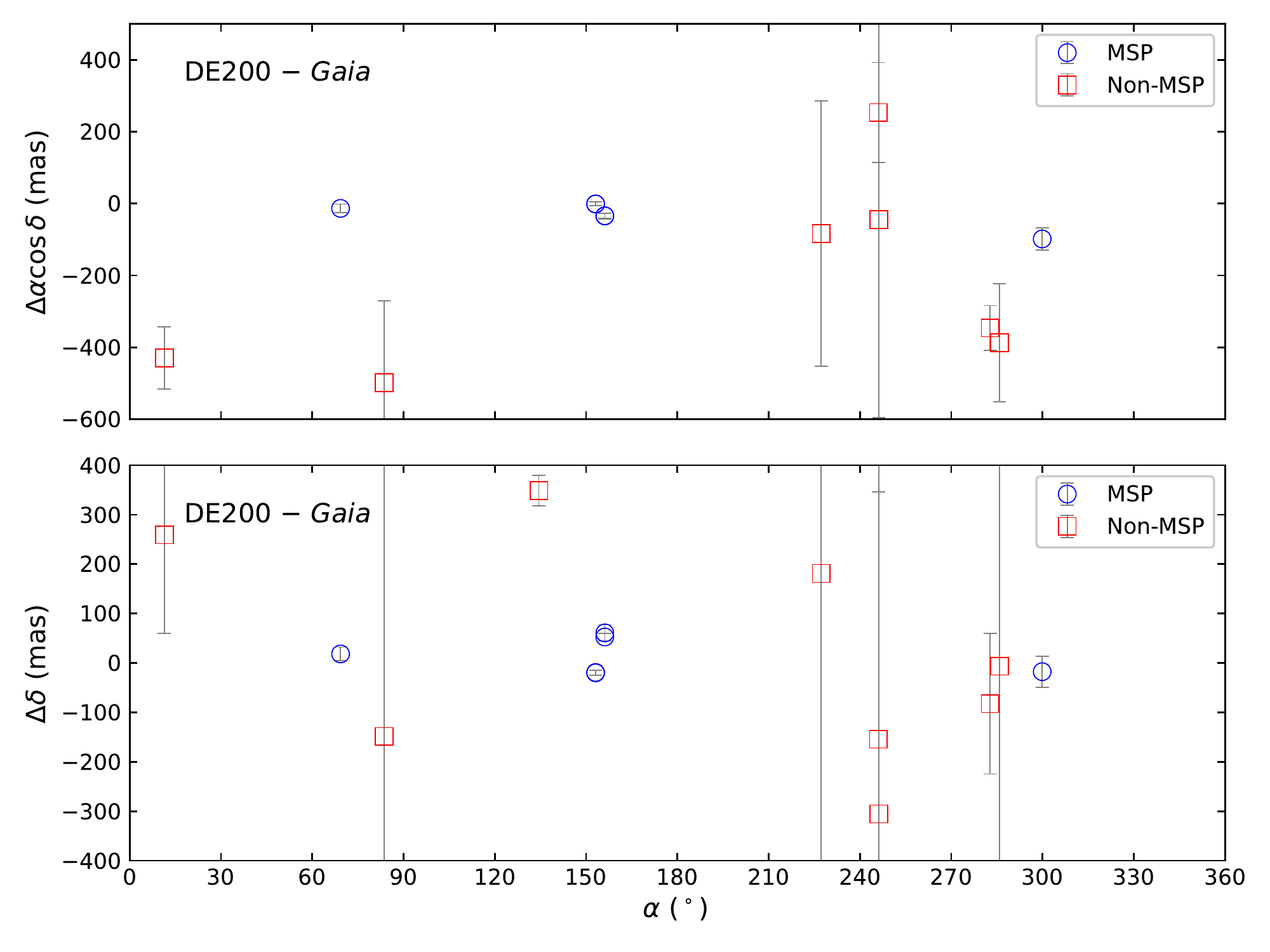}
        \includegraphics[width=\columnwidth]{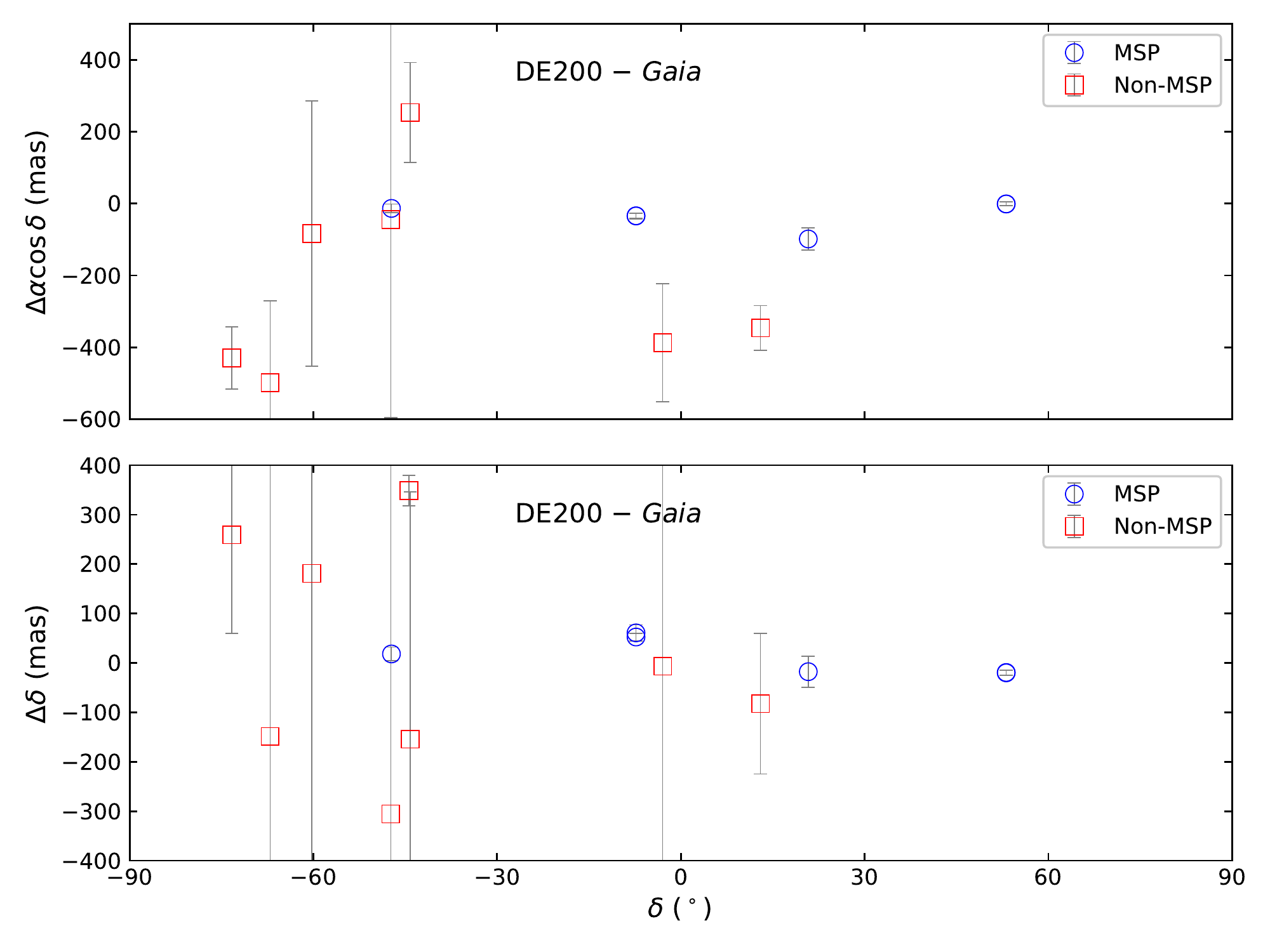}
        \caption{
            Positional differences between timing positions in the DE200 frame and \textit{Gaia} DR3 positions as a function of right ascension (left) and declination (right).
            Data points for MSPs and non-MSPs are indicated by blue circles and red squares, respectively.
            The error bars show the associated formal uncertainties calculated from Eqs.~(\ref{eq:sigma-pos-oft-ra})--(\ref{eq:sigma-pos-oft-dec}), corresponding to a confidence level of 68\%.
                  }
        \label{fig:timing-vs-gaia-pos-oft-de200}%
    \end{figure*}

    \begin{figure*}
        \centering
        \includegraphics[width=\columnwidth]{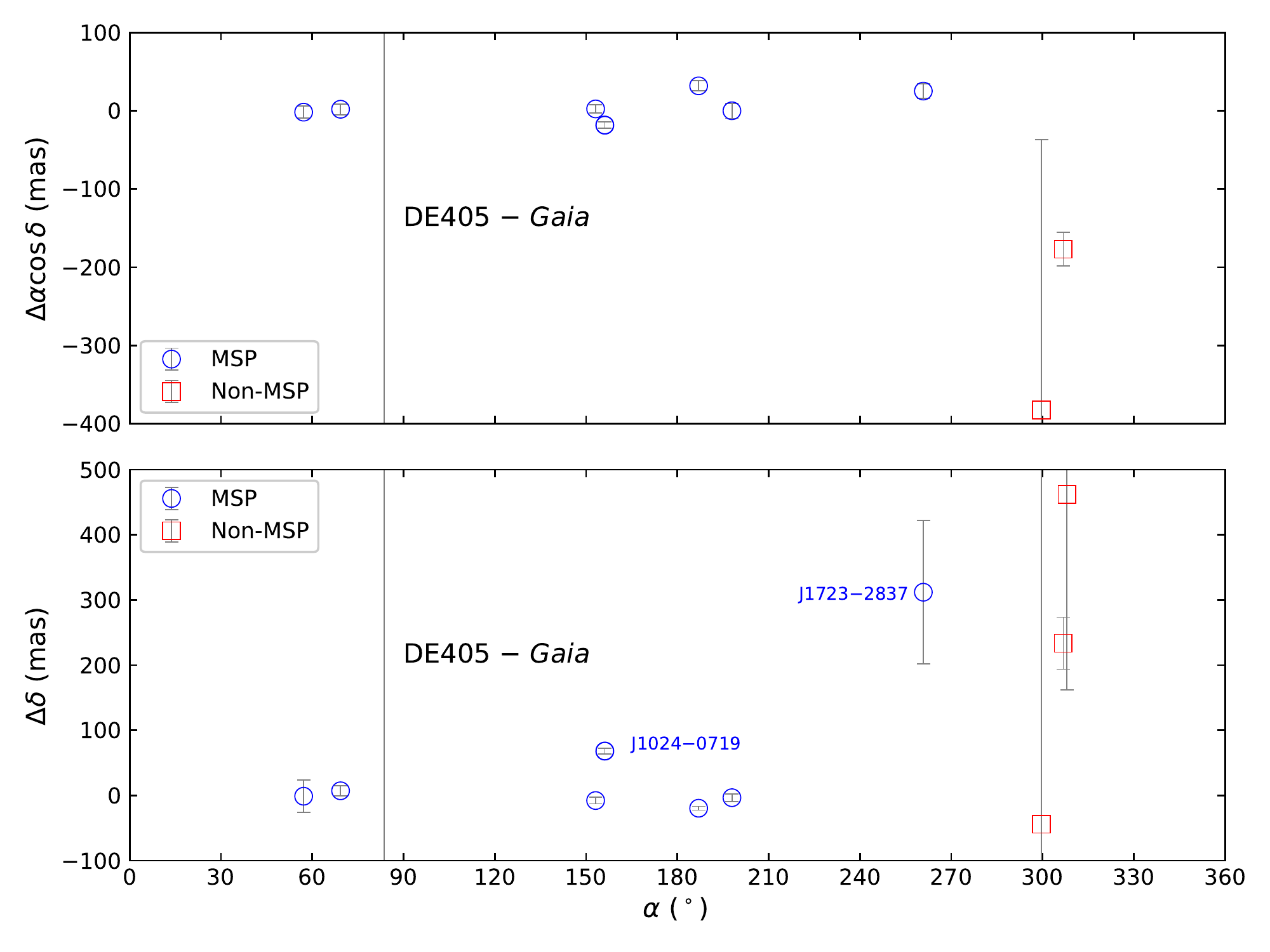}
        \includegraphics[width=\columnwidth]{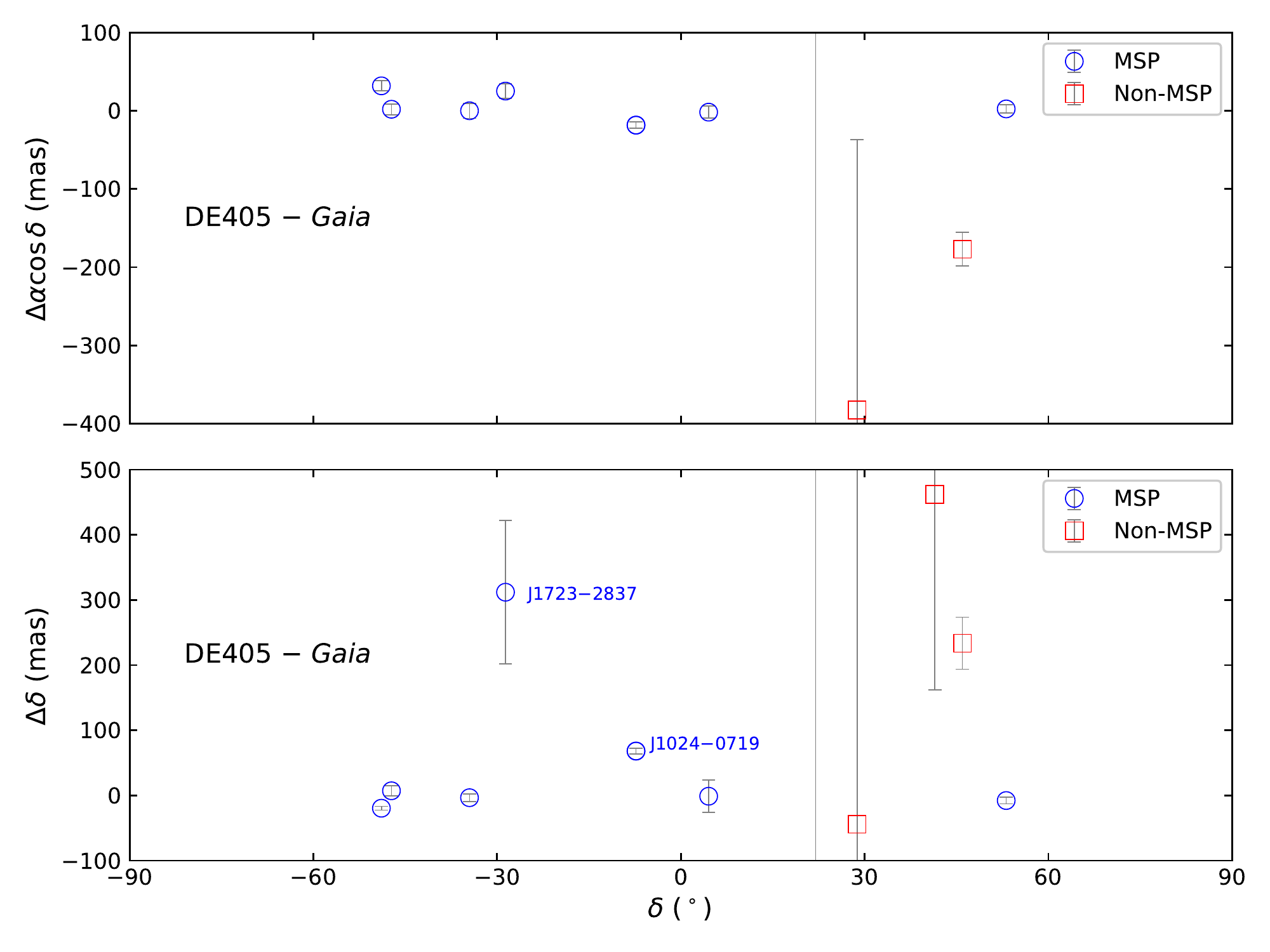}
        \caption{
            Positional differences between timing positions in the DE405 frame and \textit{Gaia} DR3 positions as a function of right ascension (left) and declination (right).
            Data points for MSPs and non-MSPs are indicated by blue circles and red squares, respectively.
            The error bars show the associated formal uncertainties calculated from Eqs.~(\ref{eq:sigma-pos-oft-ra})--(\ref{eq:sigma-pos-oft-dec}), corresponding to a confidence level of 68\%.
                  }
       \label{fig:timing-vs-gaia-pos-oft-de405}%
    \end{figure*}
    
    \begin{figure*}
        \centering
        \includegraphics[width=\columnwidth]{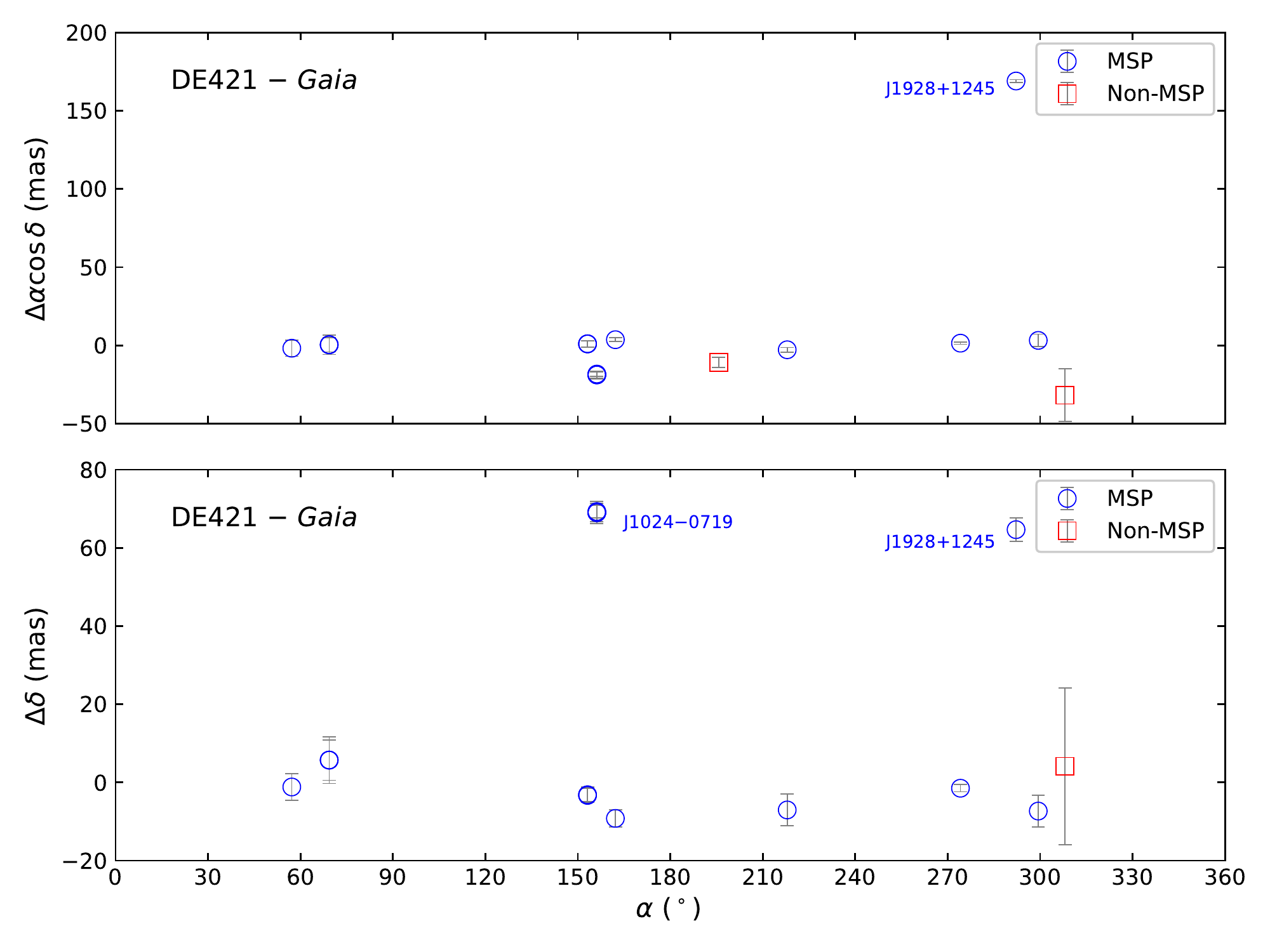}
        \includegraphics[width=\columnwidth]{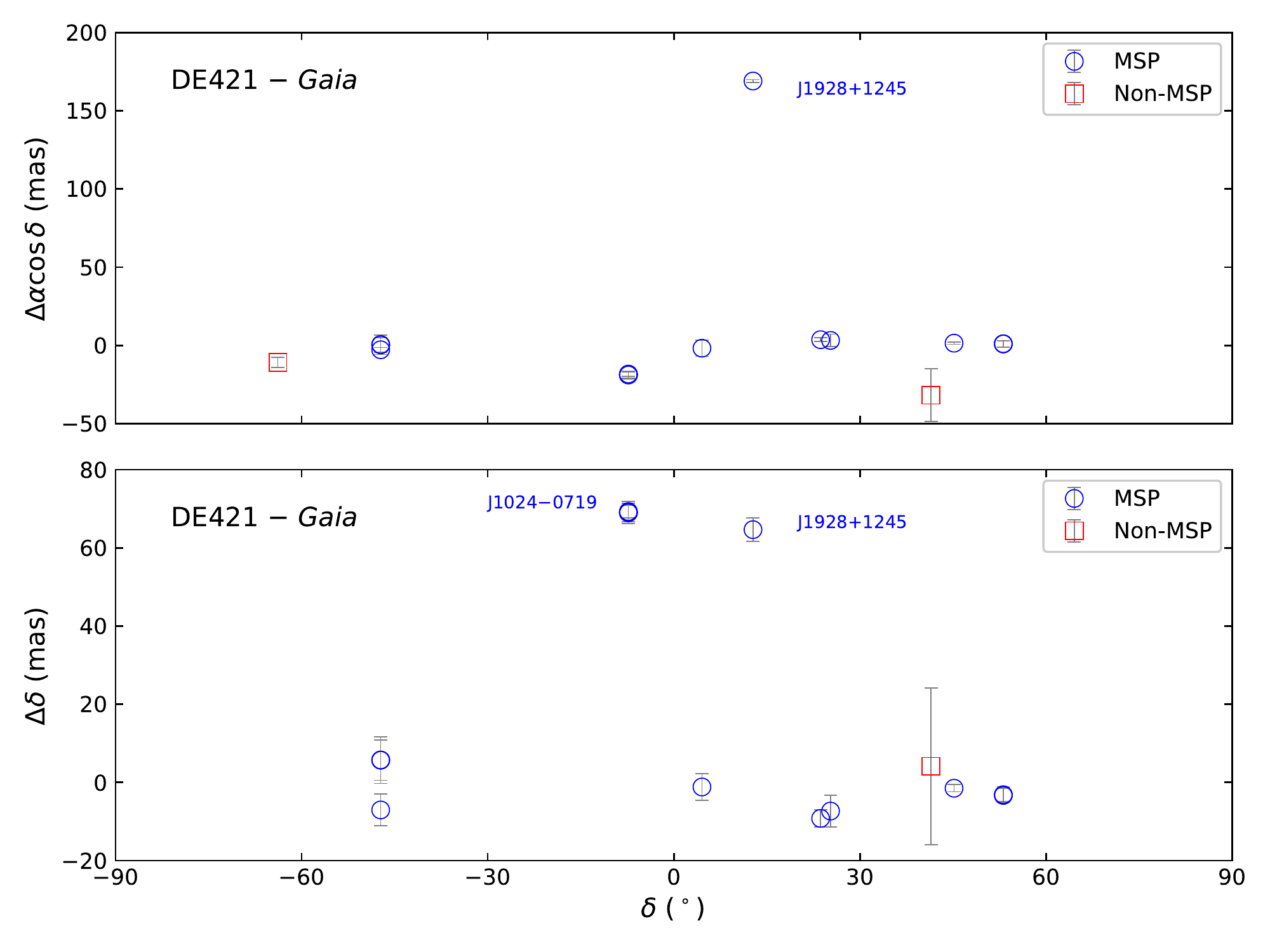}
        \caption{
            Positional differences between timing positions in the DE421 frame and \textit{Gaia} DR3 positions as a function of right ascension (left) and declination (right). 
            Data points for MSPs and non-MSPs are indicated by blue circles and red squares, respectively.
            The error bars show the associated formal uncertainties calculated from Eqs.~(\ref{eq:sigma-pos-oft-ra})--(\ref{eq:sigma-pos-oft-dec}), corresponding to a confidence level of 68\%.
                  }
        \label{fig:timing-vs-gaia-pos-oft-de421}%
    \end{figure*}
    
    \begin{figure*}
        \centering
        \includegraphics[width=\columnwidth]{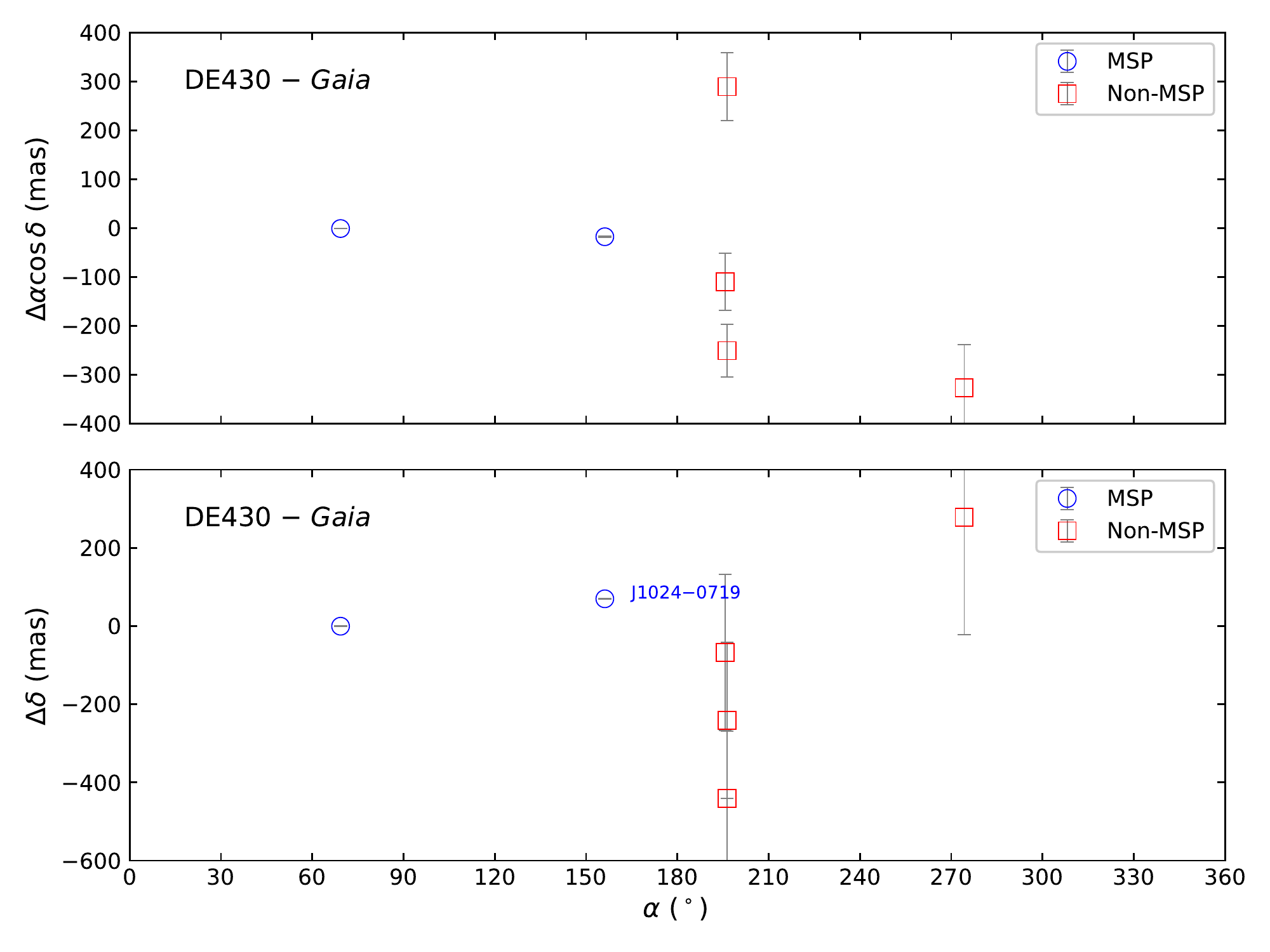}
        \includegraphics[width=\columnwidth]{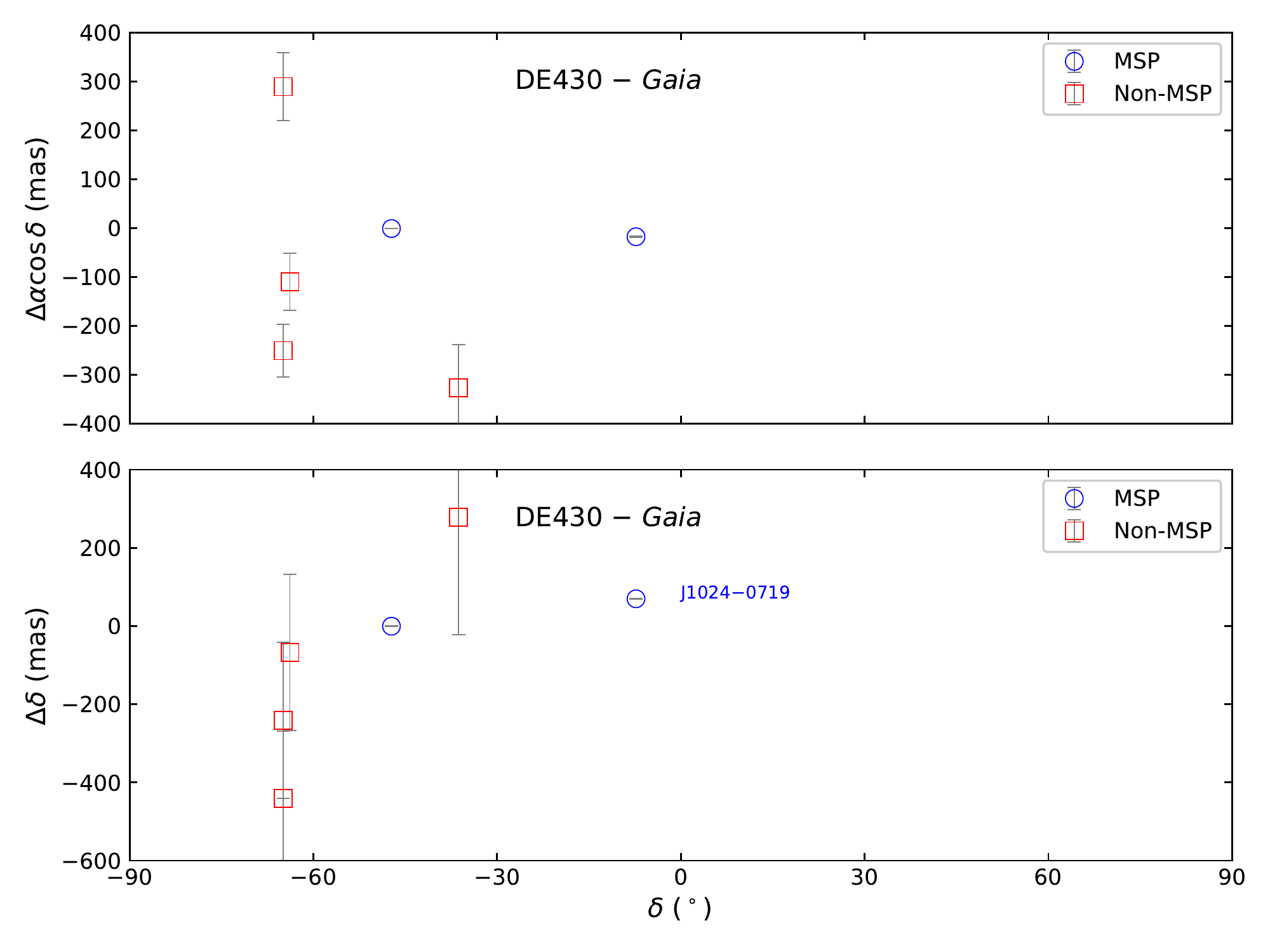}
        \caption{
            Positional differences between timing positions in the DE430 frame and \textit{Gaia} DR3 positions as a function of right ascension (left) and declination (right). 
            Data points for MSPs and non-MSPs are indicated by blue circles and red squares, respectively.
            The error bars show the associated formal uncertainties calculated from Eqs.~(\ref{eq:sigma-pos-oft-ra})--(\ref{eq:sigma-pos-oft-dec}), corresponding to a confidence level of 68\%.
                  }
        \label{fig:timing-vs-gaia-pos-oft-de430}%
    \end{figure*}
    
    \begin{figure*}
        \centering
        \includegraphics[width=\columnwidth]{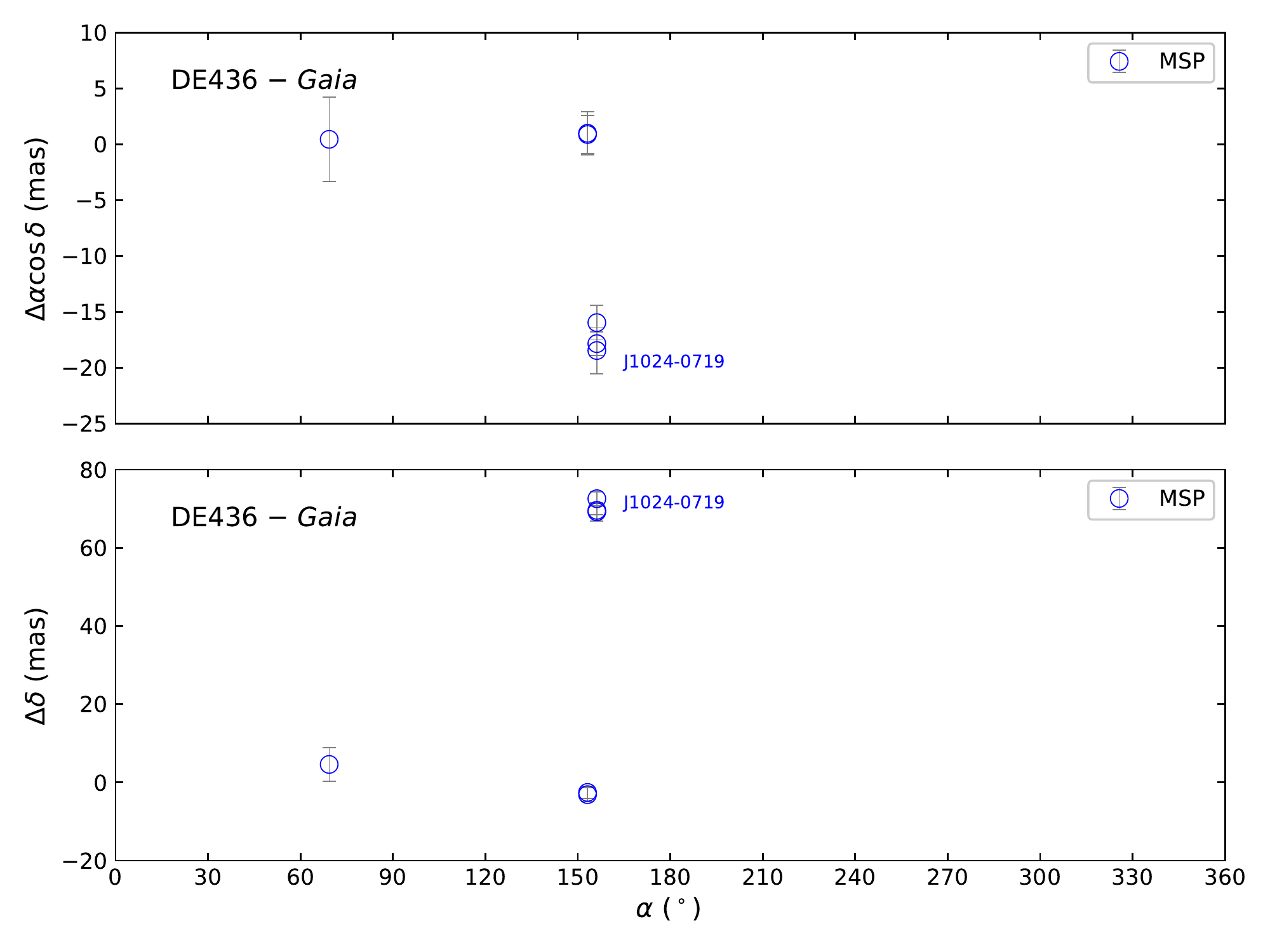}
        \includegraphics[width=\columnwidth]{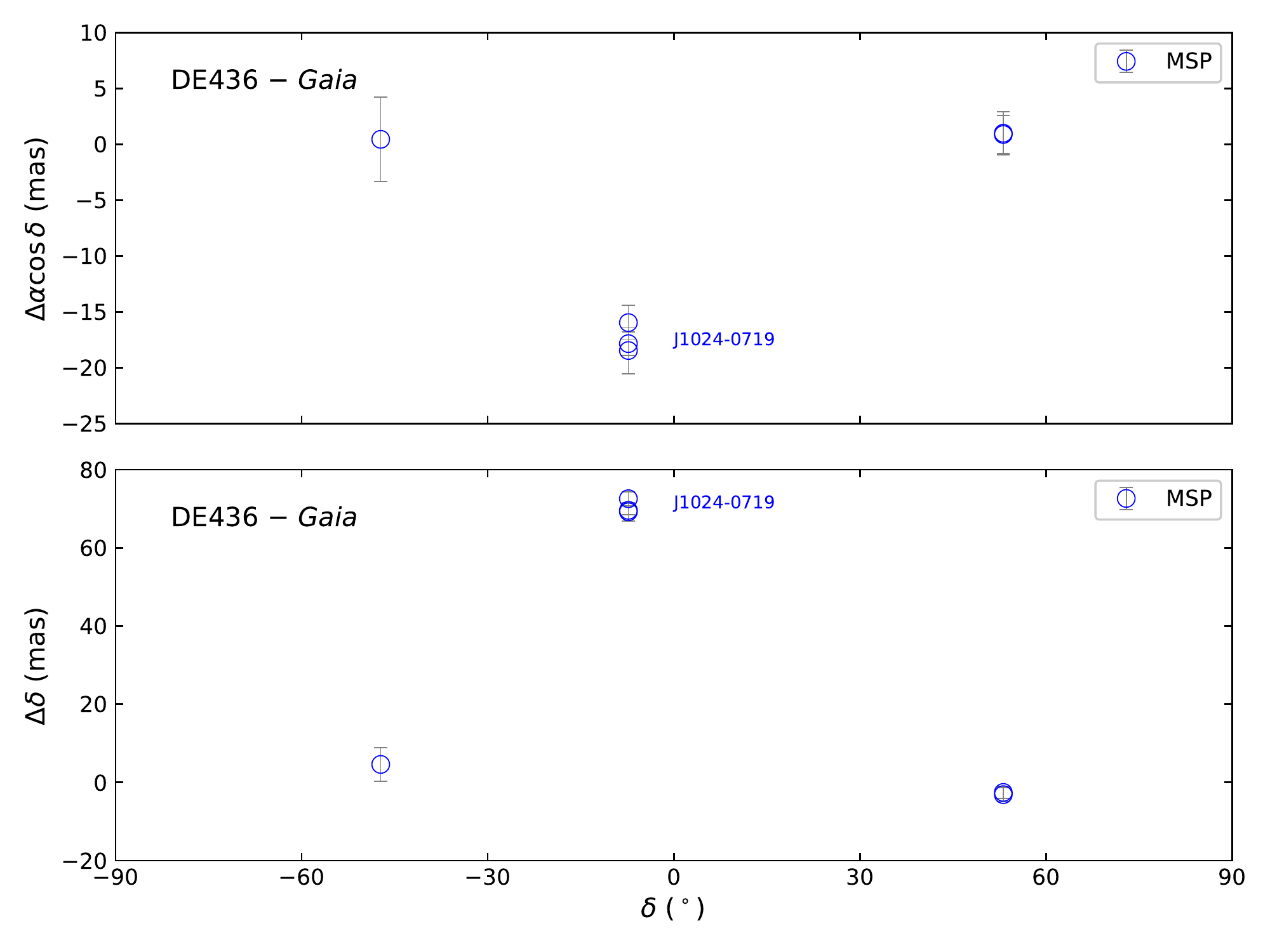}
        \caption{\label{fig:timing-vs-gaia-pos-oft-de436}%
            Positional differences between timing positions in the DE436 frame and \textit{Gaia} DR3 positions as a function of right ascension (left) and declination (right). 
            Data points for MSPs and non-MSPs are indicated by blue circles and red squares, respectively.
            The error bars show the associated formal uncertainties calculated from Eqs.~(\ref{eq:sigma-pos-oft-ra})--(\ref{eq:sigma-pos-oft-dec}), corresponding to a confidence level of 68\%.
                  }
    \end{figure*}
    
    \begin{figure}[htbp!]
        \centering
        \includegraphics[width=\columnwidth]{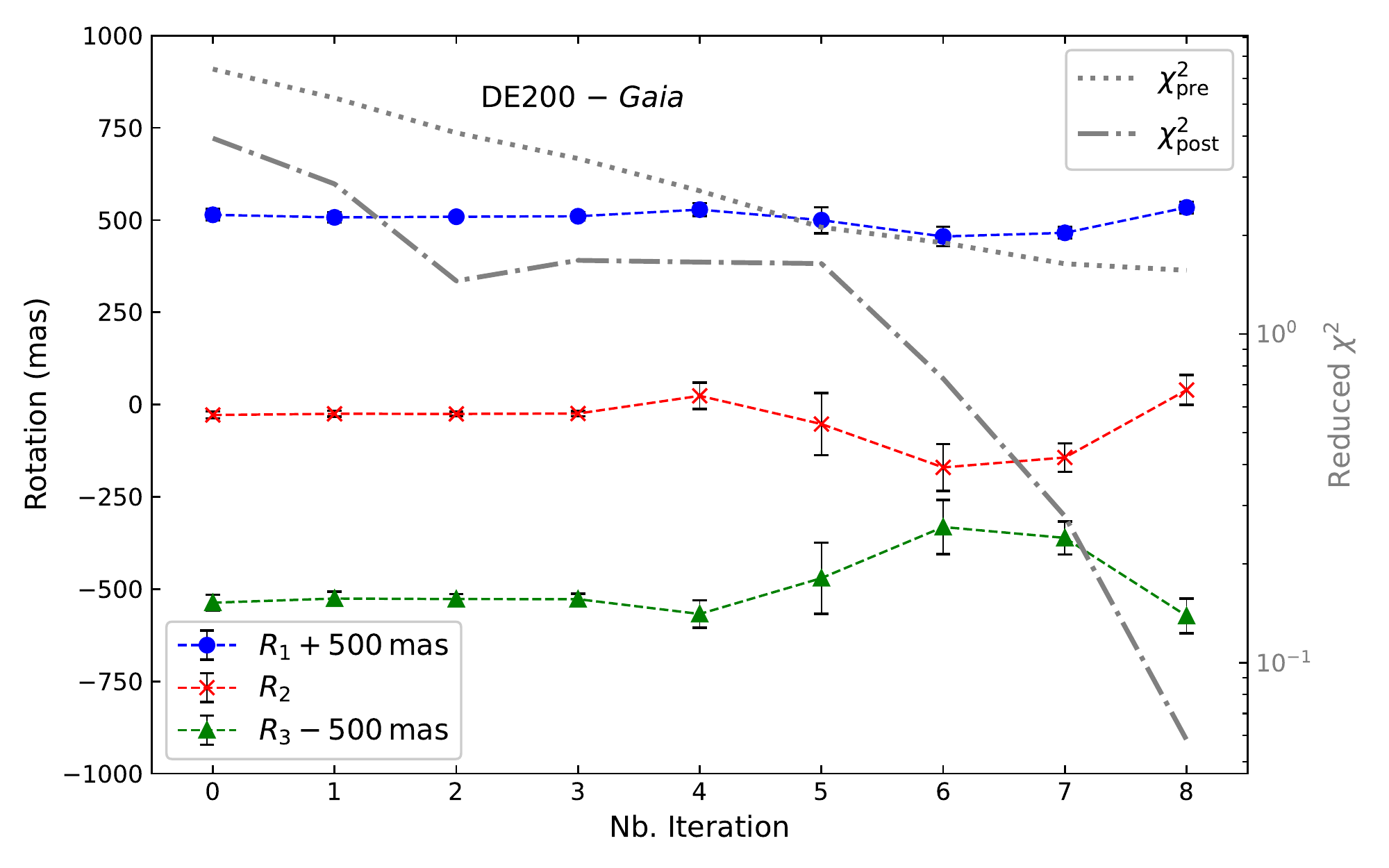}
        \includegraphics[width=\columnwidth]{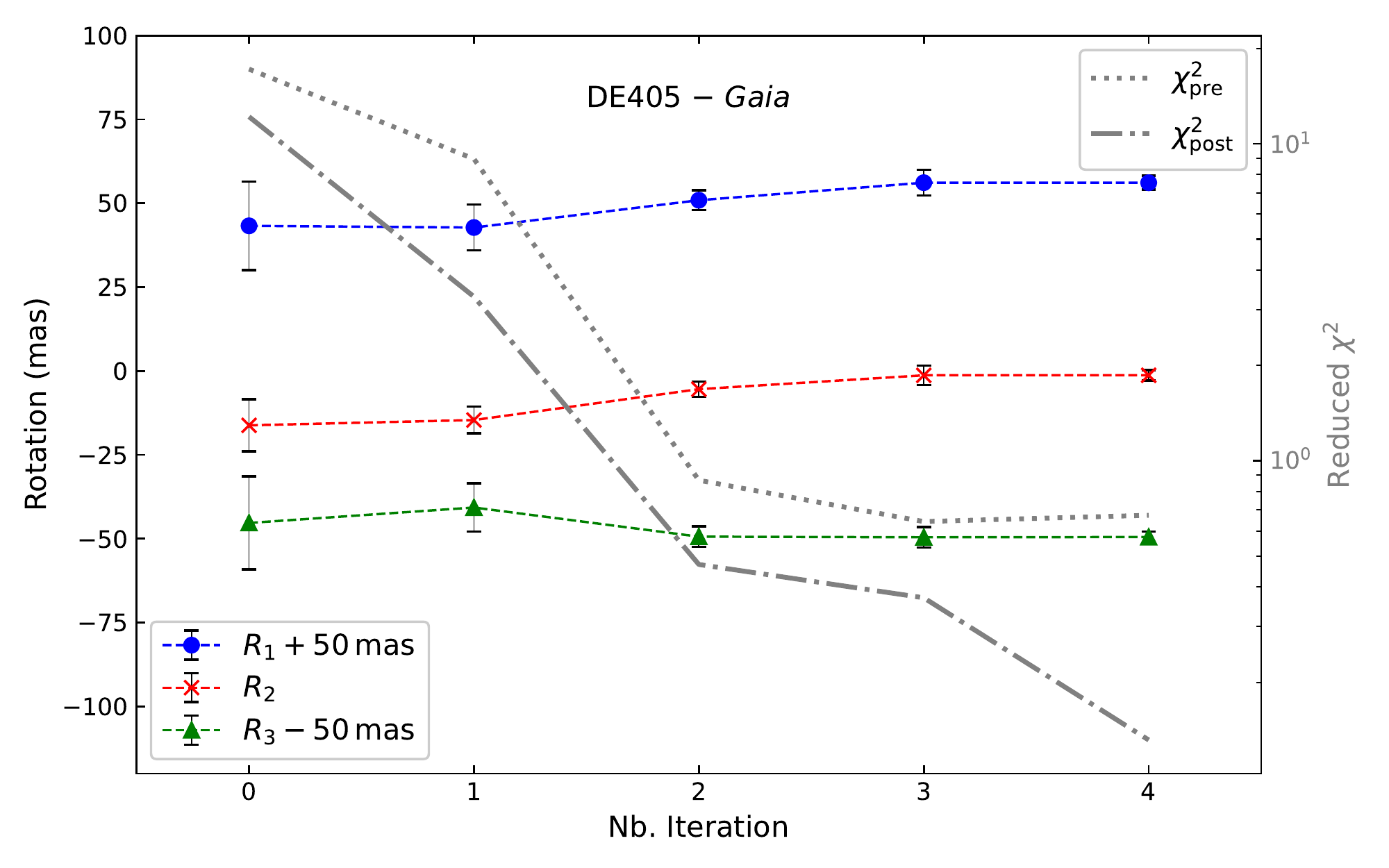}
        \includegraphics[width=\columnwidth]{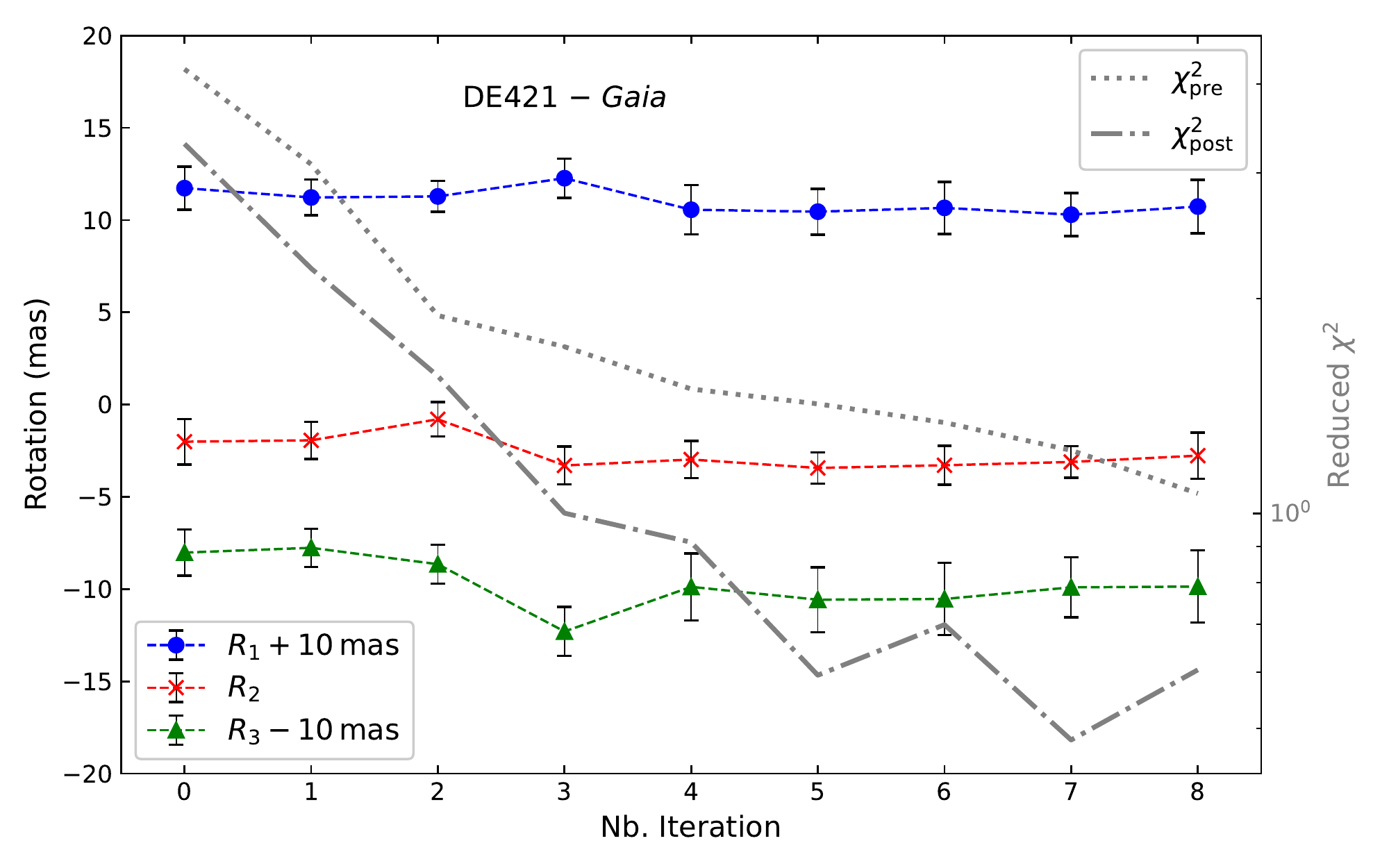}
        \caption{
            Orientation offsets of DE200 (top), DE405 (middle), and DE421 (bottom) frames referred to the \textit{Gaia}-CRF3 as a function of the number of iterations.
            The prefit and postfit reduced chi-squared for the whole sample are also plotted with reference to the vertical axis on the right.
                  }
       \label{fig:timing-vs-gaia-rot}%
    \end{figure}

    Figures~\ref{fig:timing-vs-gaia-pos-oft-de200}--\ref{fig:timing-vs-gaia-pos-oft-de436} display the differences between the timing and \textit{Gaia} positions computed using Eqs.~(\ref{eq:pos-oft-ra})-(\ref{eq:sigma-pos-oft-dec}). 
    For approximately half of these timing measurements, the positional offsets are less than 100\,mas in either right ascension or declination.
    The MSPs show obviously better agreements between the timing and \textit{Gaia} positions than the non-MSPs.
    Offsets greater than 1 arcsec are not shown in these plots; they all come from the non-MSP sample and can be summarized into two cases.
    The first case is the extremely large offsets accompanied by large timing positional uncertainties.
    For example, the largest difference between the timing and the \textit{Gaia} positions is as large as 39 arcsec in declination for PSR J0614+2229 based on the timing solutions from  \citetads{2004MNRAS.353.1311H} and \citetads{2020PASJ...72...70L}.
    We noticed a large formal uncertainty of 17 arcsec in the declination of these timing positions.
    Therefore, it was not surprising to see large differences between the timing and \textit{Gaia} positions of these pulsars.
    
    The second case is that these large offsets of some pulsars only appear in one or a few timing solutions, which are usually based on few early observations.
    For instance, the timing position of PSR J0857--4424 in the DE200 frame given in \citetads{1998MNRAS.297...28D} yields an offset of $-7.7$ arcsec in right ascension, while the right ascension offset is less than 0.5 arcsec for the latest timing solution \citepads{2019MNRAS.489.3810P}.
    Therefore, the large right ascension difference between \textit{Gaia} and the DE200 timing position is most likely due to the errors in the timing solution from \citetads{1998MNRAS.297...28D}.
    These offsets exceeding 1 arcsec are unlikely to be related to frame alignment issues; the corresponding measurements were thus not used to determine the rotation parameters between the \textit{Gaia} and ephermeris frames.
    We noted that these large offsets usually appeared in the early timing solutions, for which DE200 was used as the position reference.
    This implied an issue that comes with the underlying correlation between improvements in data and analysis quality with time and the ephemerides used, which will be discussed in Sect.~\ref{subsect:limits}.
    
    Some MSPs were also found to show statistically significant differences between timing and \textit{Gaia} positions.
    The timing positions for PSR J1024--0719 differ from the \textit{Gaia} position by approximately $-69$\,mas in declination, which is significant at 10$\sigma$ or more.
    These significant offsets can be expected from the fact that PSR J1024--0719 was in a wide binary system as explained in Sect.~\ref{subsec:sample}.
    There is only a timing solution for PSR J1723--2837 \citepads{2013ApJ...776...20C}, for which a declination offset of more than 300\,mas was found (Fig.~\ref{fig:timing-vs-gaia-pos-oft-de405}), several times greater than the combination of the timing and \textit{Gaia} uncertainties in declination (110\,mas and 0.03\,mas, respectively).
    In addition, the DE421 timing solution of PSR J1928+1245 given in \citetads{2019ApJ...886..148P} yields positional offsets of $169\,\pm\,1$\,mas in right ascension and $65\,\pm\,3$\,mas in declination (Figs.~\ref{fig:timing-vs-gaia-pos-oft-de421}).
    We did not have clear explanations for these significant positional offsets yet, leaving them to be revisited in future studies.
    However, these offsets were most unlikely to be caused by frame misalignment.
    Therefore, we excluded these timing solutions from the comparison of the ephemeris reference frames against \textit{Gaia}-CRF.
    
    We removed four pulsars near the ecliptic plane, which are PSR J0337+1715, PSR J0534+2200, PSR J0614+2229, and PSR J2339--0533, and then estimated the rotation parameters.
    Figure~\ref{fig:timing-vs-gaia-rot} plots the evolution of estimates of the rotation parameters for DE200, DE405, and DE421 frames against the \textit{Gaia}-CRF in the iteration process for all pulsars.
    Since the number of pulsars in the DE430 sample common to \textit{Gaia} DR3 was relatively small, the rotation parameters of the DE430 frame were estimated via a single least-squares fitting.
    We did not estimate the orientation offset between the DE436 frame and \textit{Gaia}-CRF because there were only two pulsars in common.
    We performed similar analyses to the MSP and non-MSP samples.
    
    Table~\ref{tab:rot-ang} reports the estimate of rotation parameters between the ephemeris frames and \textit{Gaia}-CRF.
    The DE200 frame shows orientation offsets of 10\,mas--20\,mas referred to the \textit{Gaia}-CRF using samples of all common pulsars and MSPs only when taking the formal uncertainty into consideration.
    No statistically significant rotation is found for the DE405, DE421, and DE430 frames with respect to \textit{Gaia}-CRF in almost all subsets, except for $R_2\simeq-3\,{\rm mas}$ ($>3\,\sigma$) of DE421.
    The solution based on the sample of non-MSPs differs significantly from those of the other two subsets.
    Comparing the results of using all pulsars and MSPs only, we found that MSPs are dominant in determining the rotation parameters thanks to the small uncertainty of their timing positions.
    We also found that inclusion of the non-MSPs into the sample increased the uncertainty of the rotation parameters in several cases.

\subsection{Ephemeris frames versus VLBI-CRF} \label{subsec:timing-vs-vlbi}

    \begin{figure*}
        \centering
        \includegraphics[width=\columnwidth]{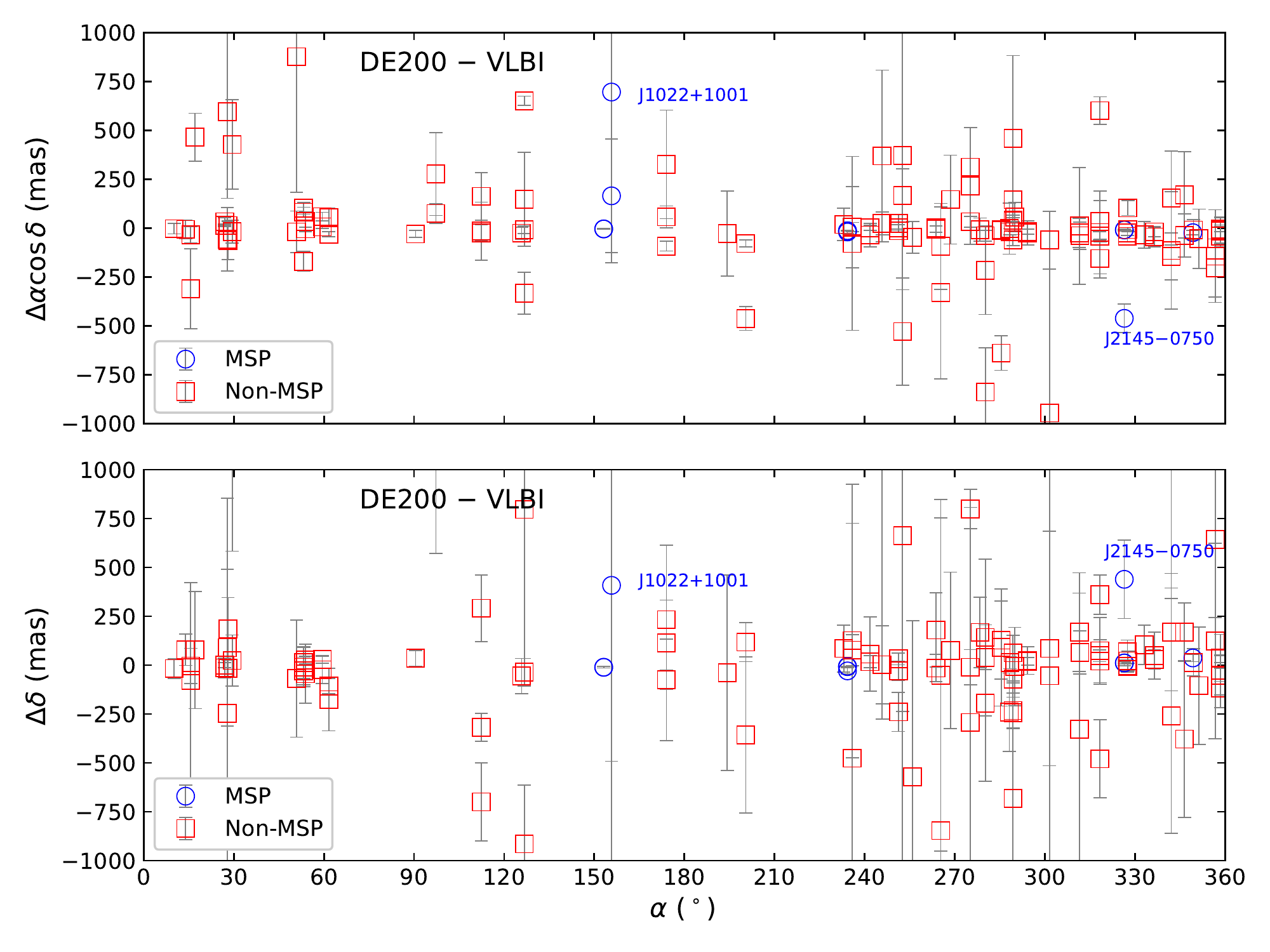}
        \includegraphics[width=\columnwidth]{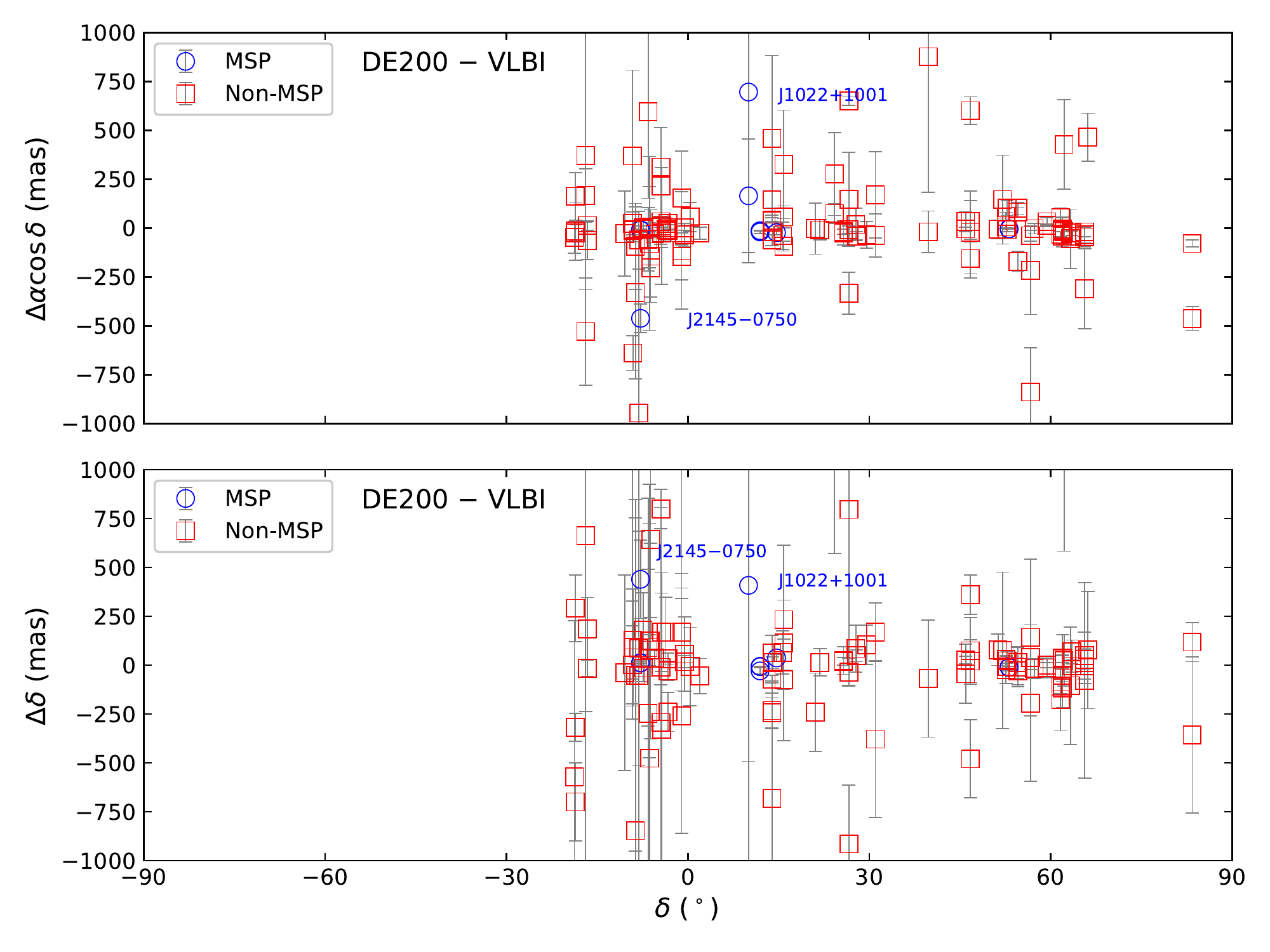}
        \caption{
            Positional differences between timing positions in the DE200 frame and the VLBI positions taken from the PSR$\pi$ data archive as a function of right ascension (left) and declination (right). 
            Data points for MSPs and non-MSPs are indicated by blue circles and red squares, respectively.
            The error bars show the associated formal uncertainties calculated from Eqs.~(\ref{eq:sigma-pos-oft-ra})--(\ref{eq:sigma-pos-oft-dec}), corresponding to a confidence level of 68\%.
                  }
       \label{fig:timing-vs-vlbi-pos-oft-de200}%
    \end{figure*}
    
    \begin{figure*}
        \centering
        \includegraphics[width=\columnwidth]{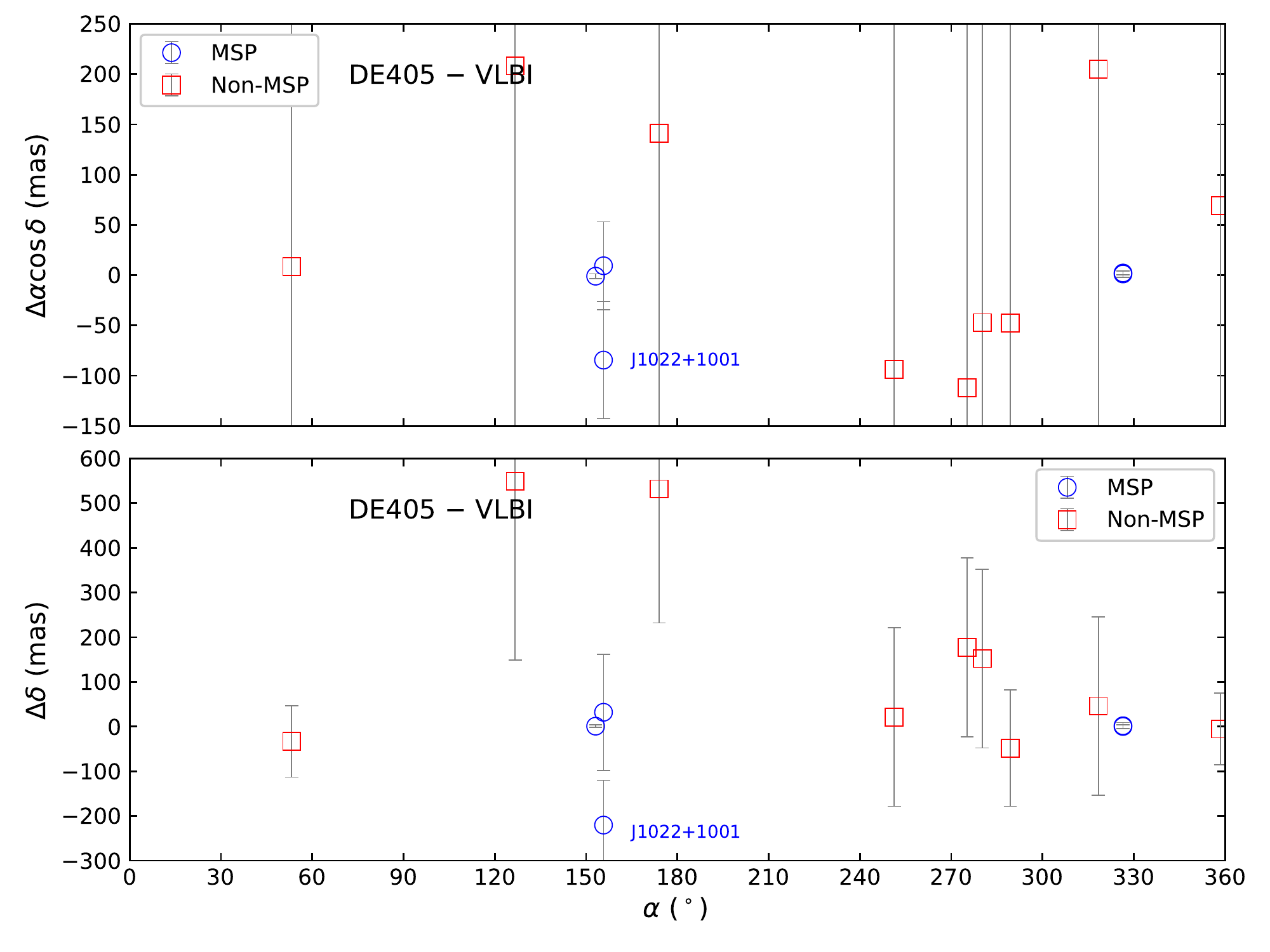}
        \includegraphics[width=\columnwidth]{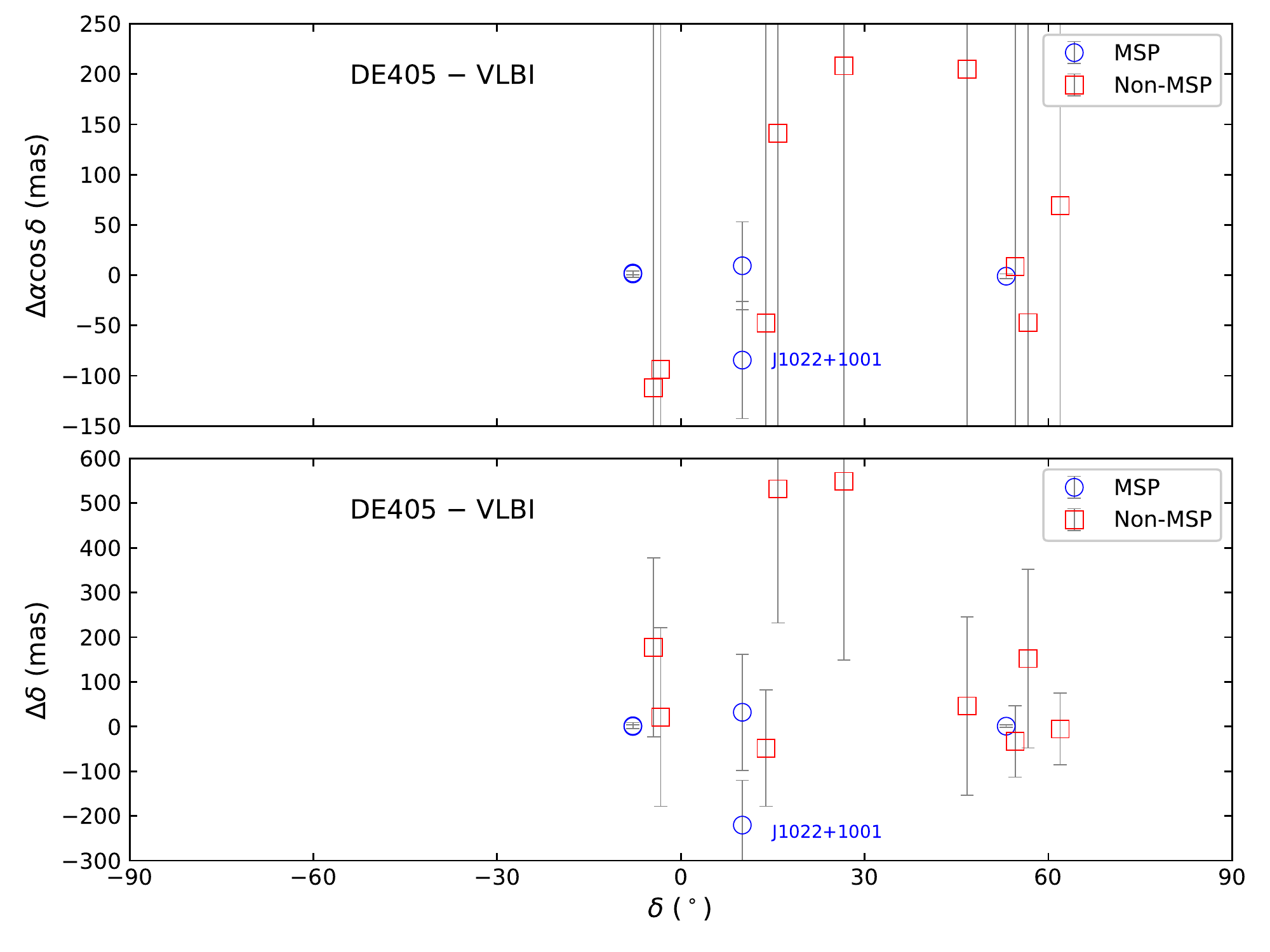}
        \caption{
            Positional differences between timing positions in the DE405 frame and the VLBI positions taken from the PSR$\pi$ data archive as a function of right ascension (left) and declination (right). 
            Data points for MSPs and non-MSPs are indicated by blue circles and red squares, respectively.
            The error bars show the associated formal uncertainties calculated from Eqs.~(\ref{eq:sigma-pos-oft-ra})--(\ref{eq:sigma-pos-oft-dec}), corresponding to a confidence level of 68\%.
                  }
       \label{fig:timing-vs-vlbi-pos-oft-de405}%
    \end{figure*}

    \begin{figure*}
        \includegraphics[width=\columnwidth]{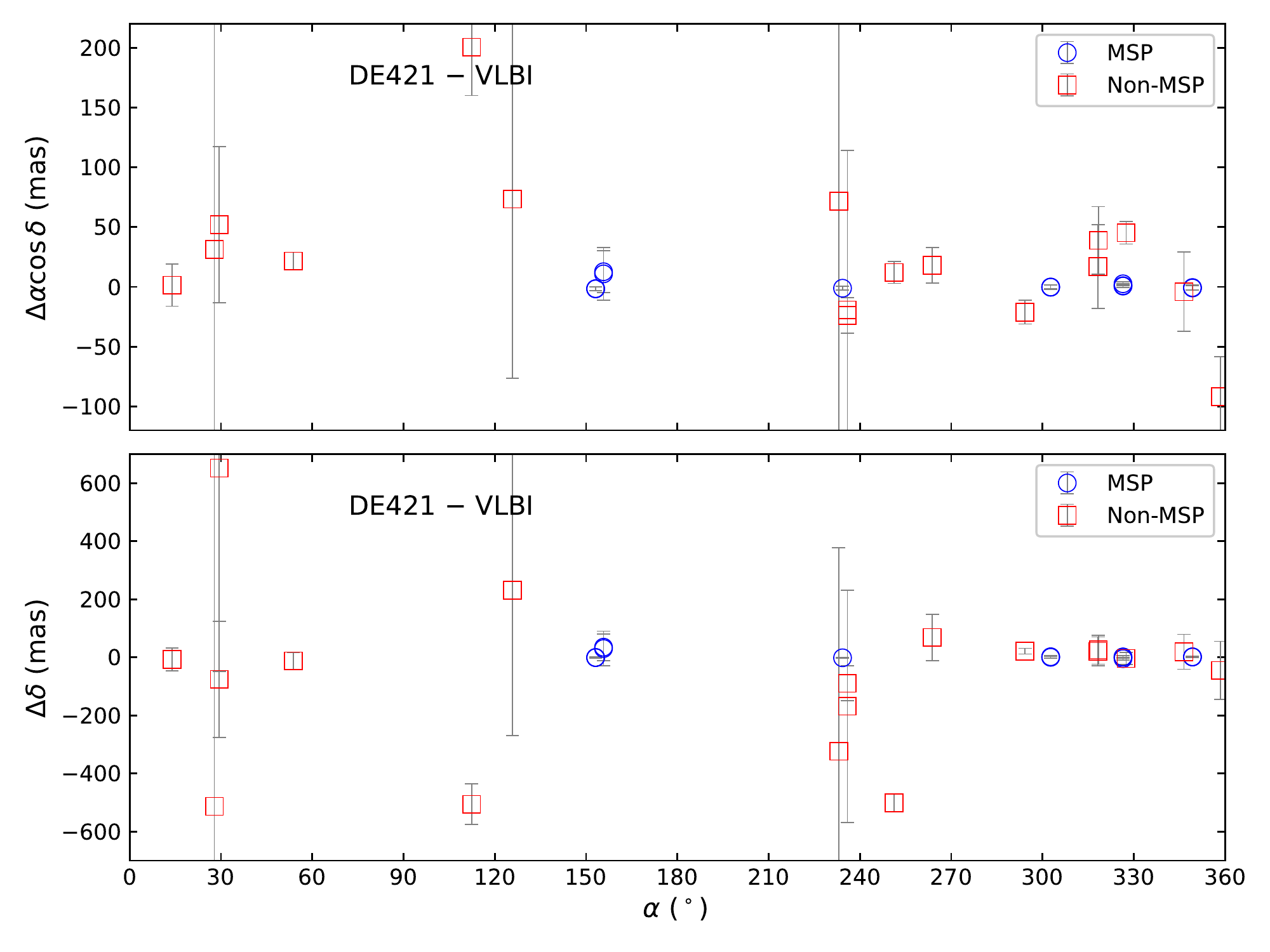}
        \includegraphics[width=\columnwidth]{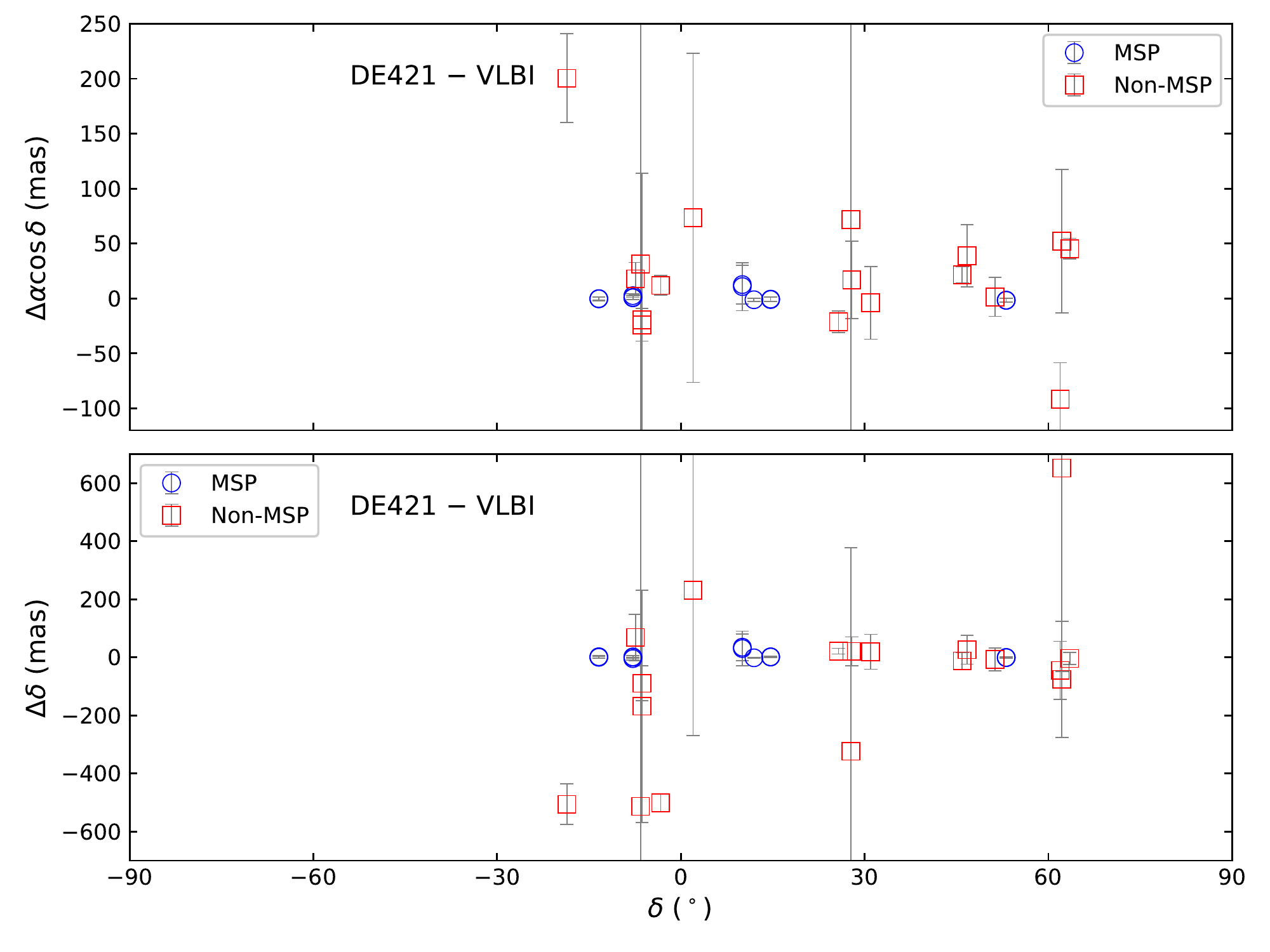}
        \caption{
            Positional differences between timing positions in the DE421 frame and the VLBI positions taken from the PSR$\pi$ data archive as a function of right ascension (left) and declination (right). 
            Data points for MSPs and non-MSPs are indicated by blue circles and red squares, respectively.
            The error bars show the associated formal uncertainties calculated from Eqs.~(\ref{eq:sigma-pos-oft-ra})--(\ref{eq:sigma-pos-oft-dec}), corresponding to a confidence level of 68\%.
                  }
       \label{fig:timing-vs-vlbi-pos-oft-de421}%
    \end{figure*}

    \begin{figure*}
        \includegraphics[width=\columnwidth]{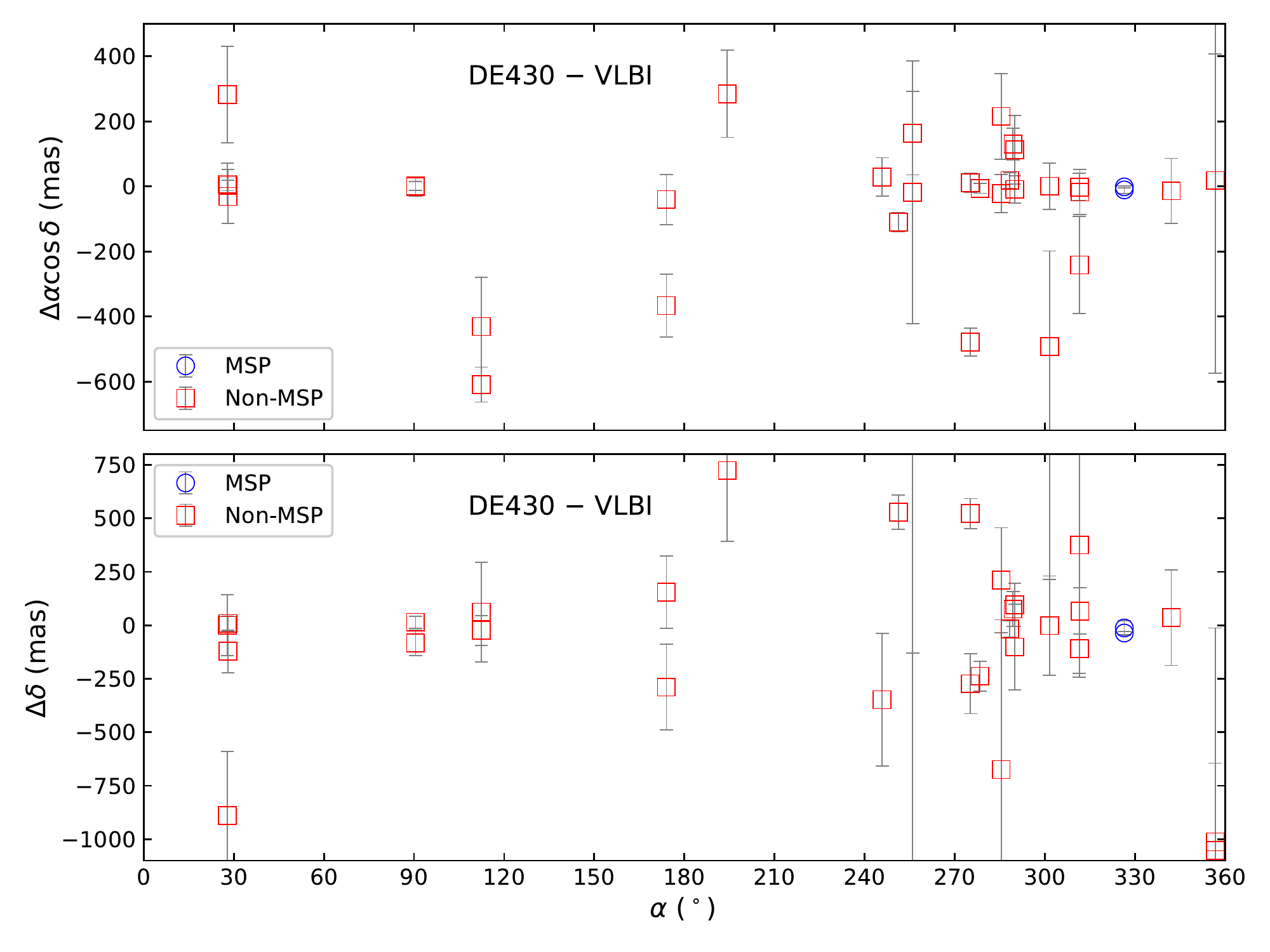}
        \includegraphics[width=\columnwidth]{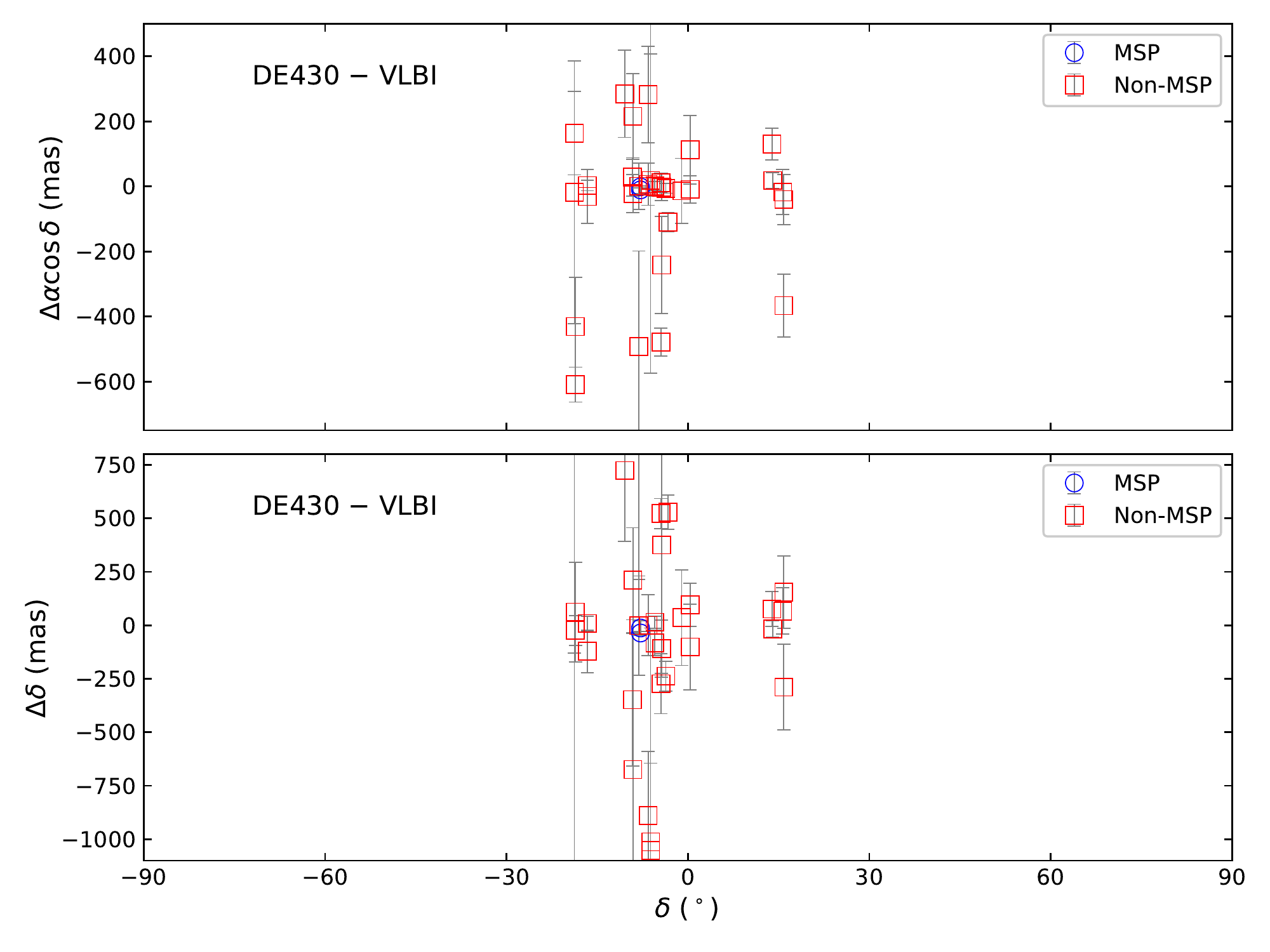}
        \caption{
            Positional differences between timing positions in the DE430 frame and the VLBI positions taken from the PSR$\pi$ data archive as a function of right ascension (left) and declination (right). 
            Data points for MSPs and non-MSPs are indicated by blue circles and red squares, respectively.
            The error bars show the associated formal uncertainties calculated from Eqs.~(\ref{eq:sigma-pos-oft-ra})--(\ref{eq:sigma-pos-oft-dec}), corresponding to a confidence level of 68\%.
                  }
       \label{fig:timing-vs-vlbi-pos-oft-de430}%
    \end{figure*}

    \begin{figure*}
        \includegraphics[width=\columnwidth]{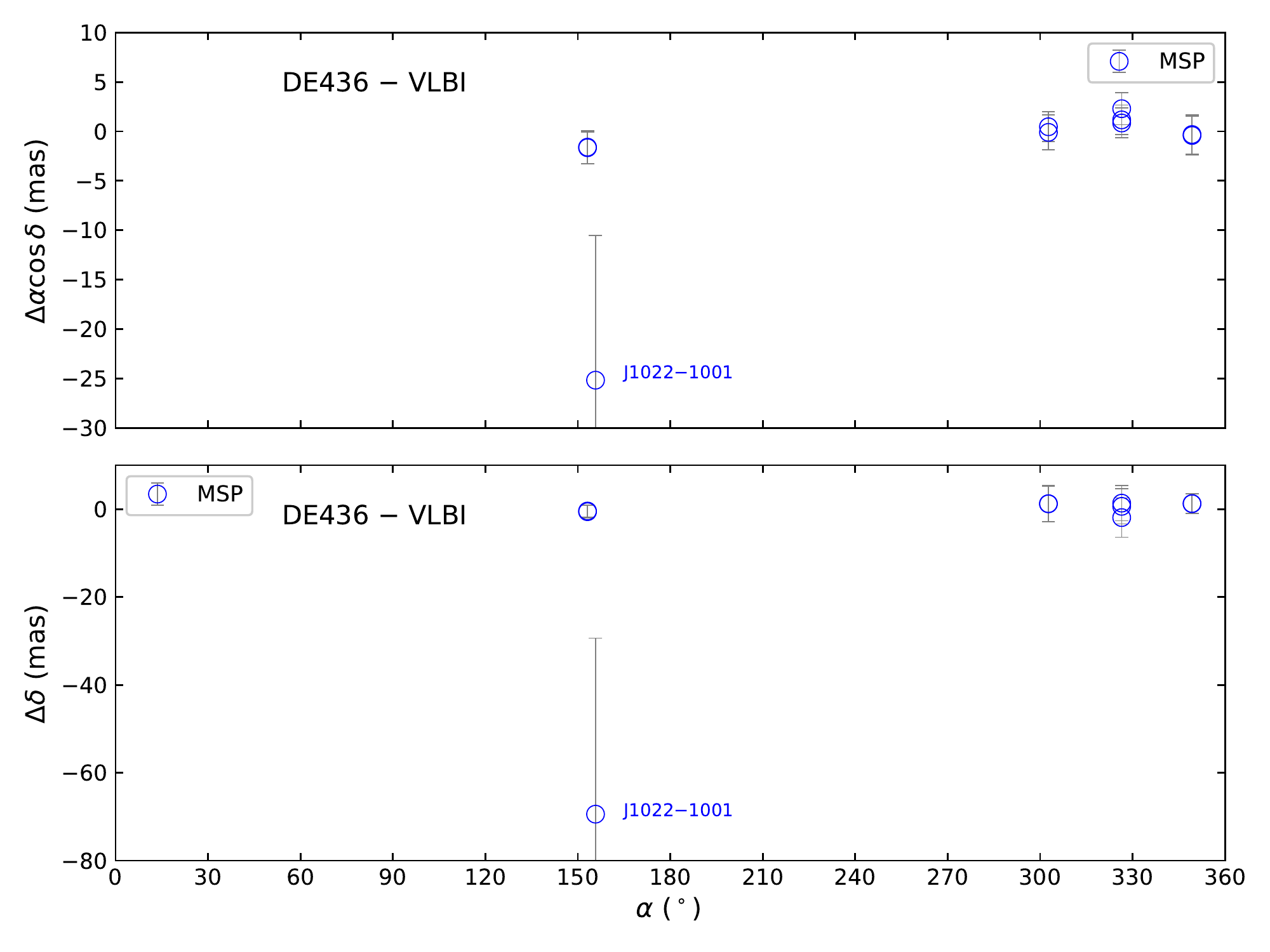}
        \includegraphics[width=\columnwidth]{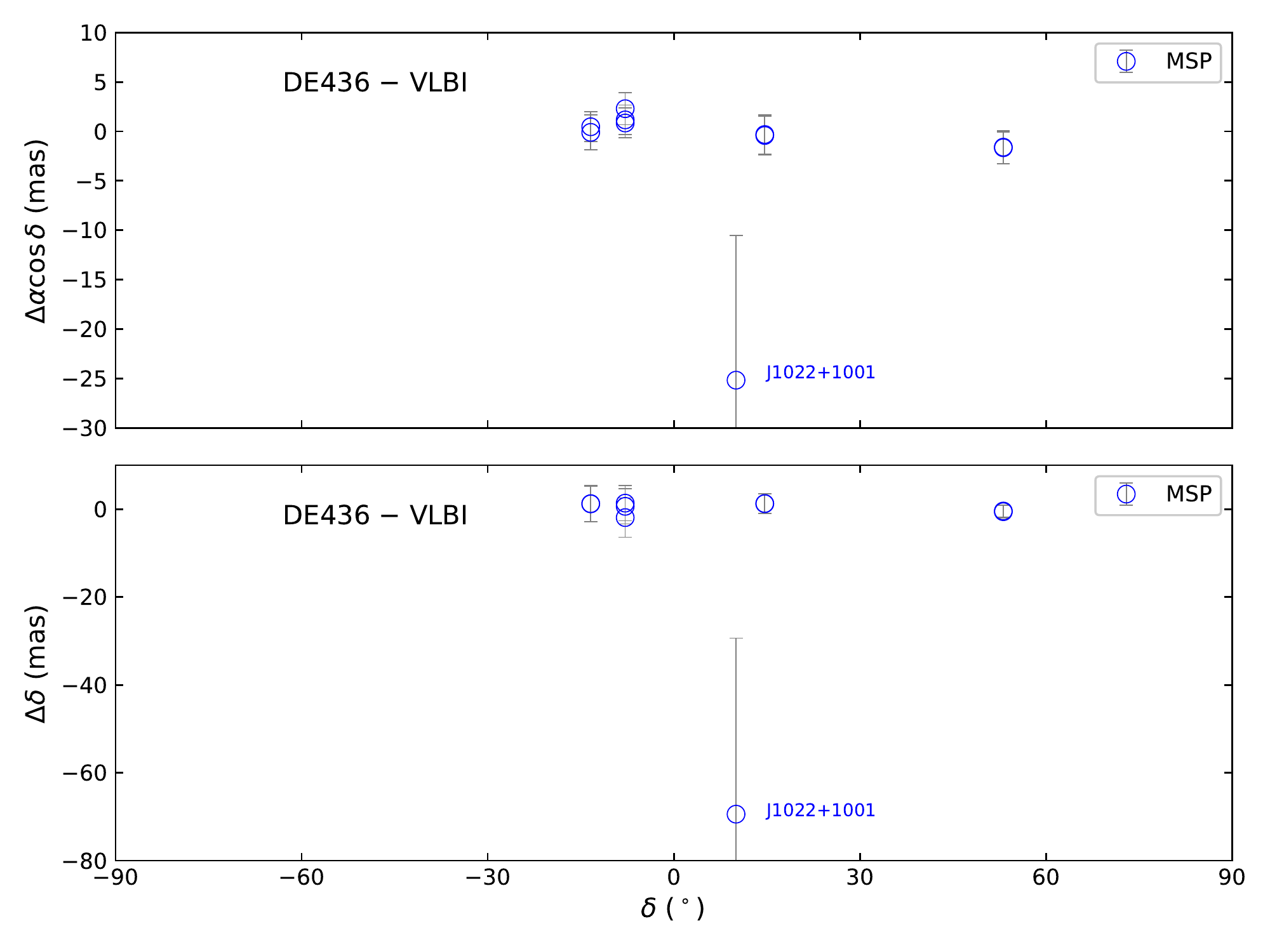}
        \caption{
            Positional differences between timing positions in the DE436 frame and the VLBI positions taken from the PSR$\pi$ data archive as a function of right ascension (left) and declination (right). 
            Data points for MSPs and non-MSPs are indicated by blue circles and red squares, respectively.
            The error bars show the associated formal uncertainties calculated from Eqs.~(\ref{eq:sigma-pos-oft-ra})--(\ref{eq:sigma-pos-oft-dec}), corresponding to a confidence level of 68\%.
                  }
       \label{fig:timing-vs-vlbi-pos-oft-de436}%
    \end{figure*}

    \begin{figure*}
        \includegraphics[width=\columnwidth]{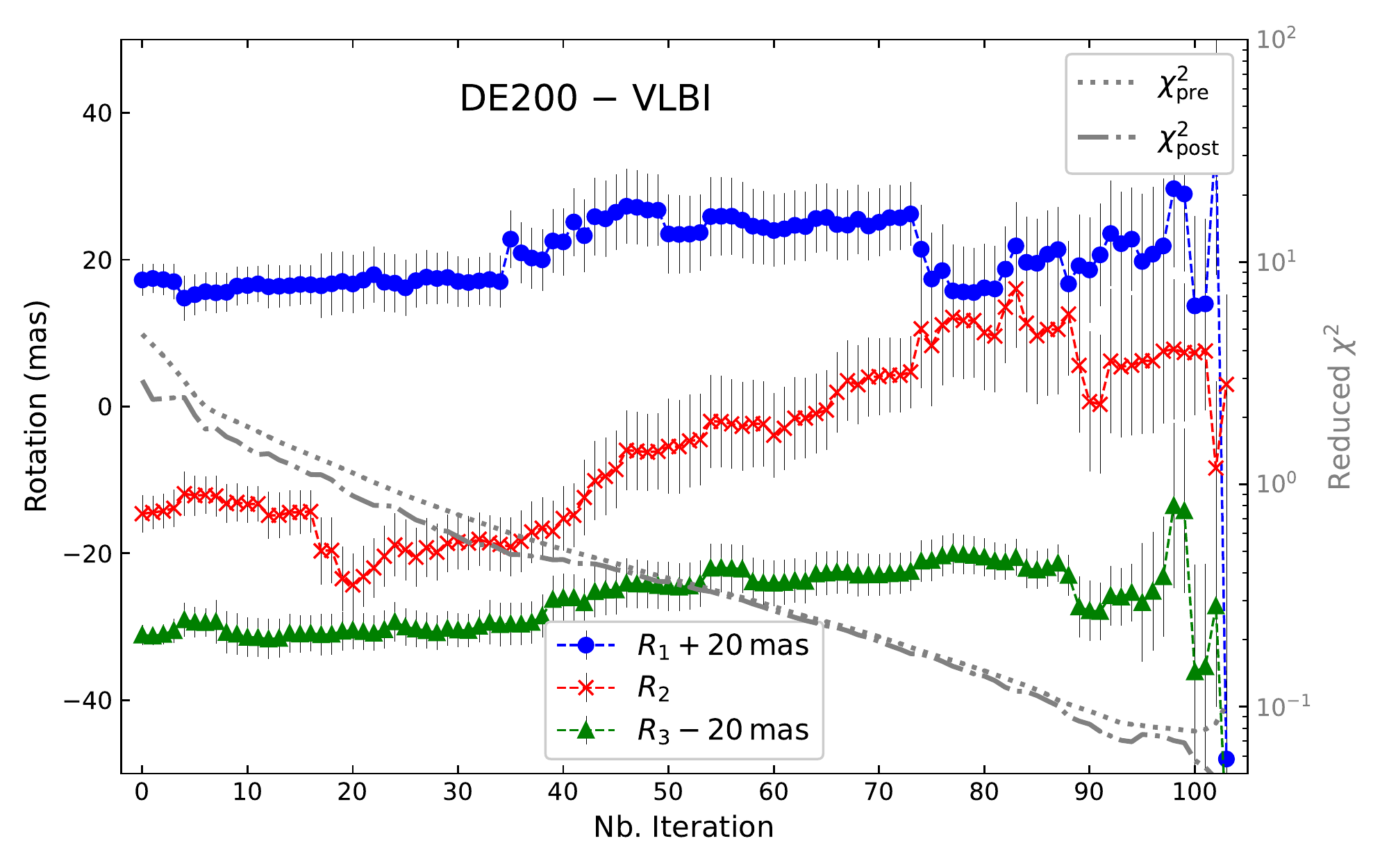}
        \includegraphics[width=\columnwidth]{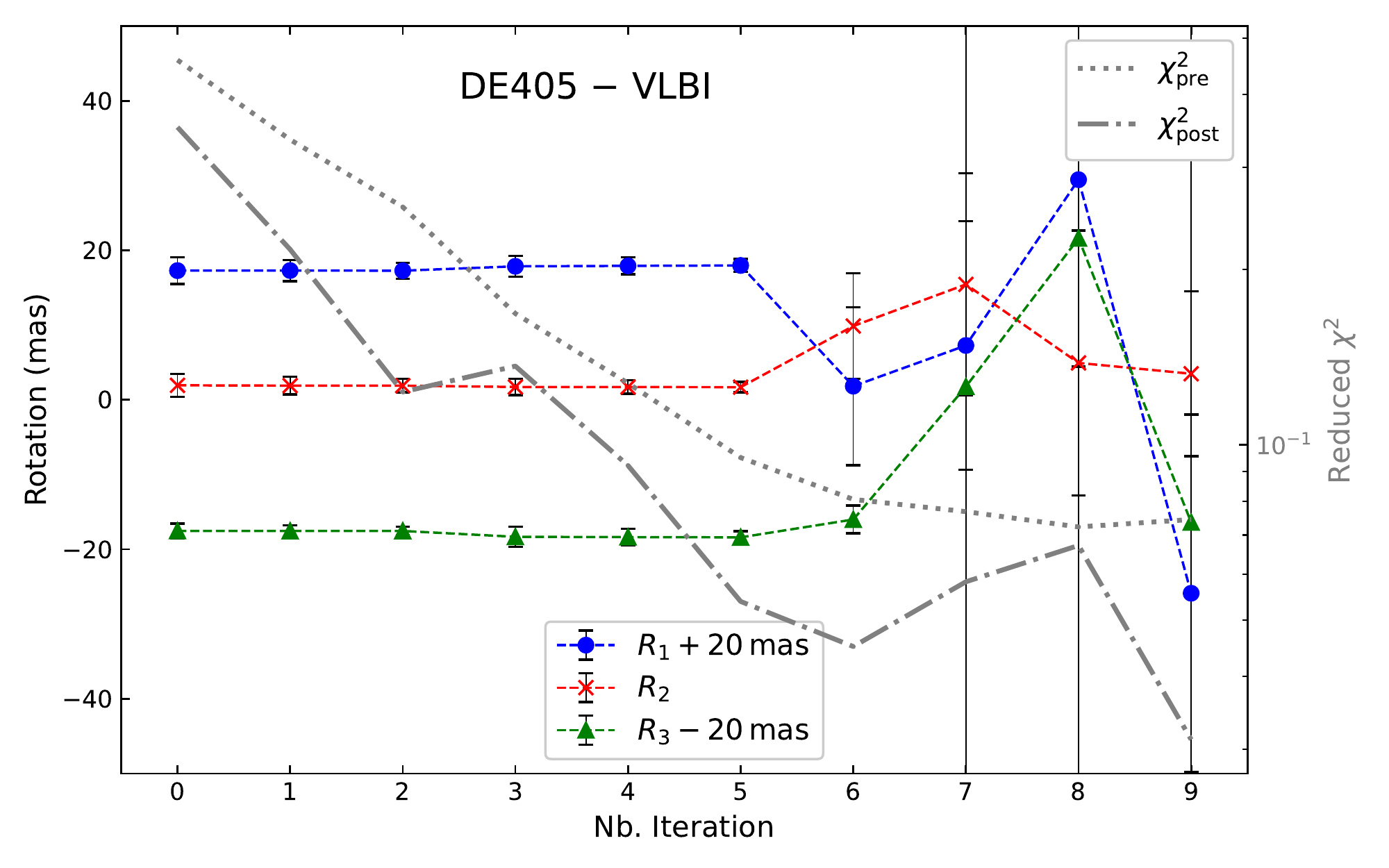}
        \includegraphics[width=\columnwidth]{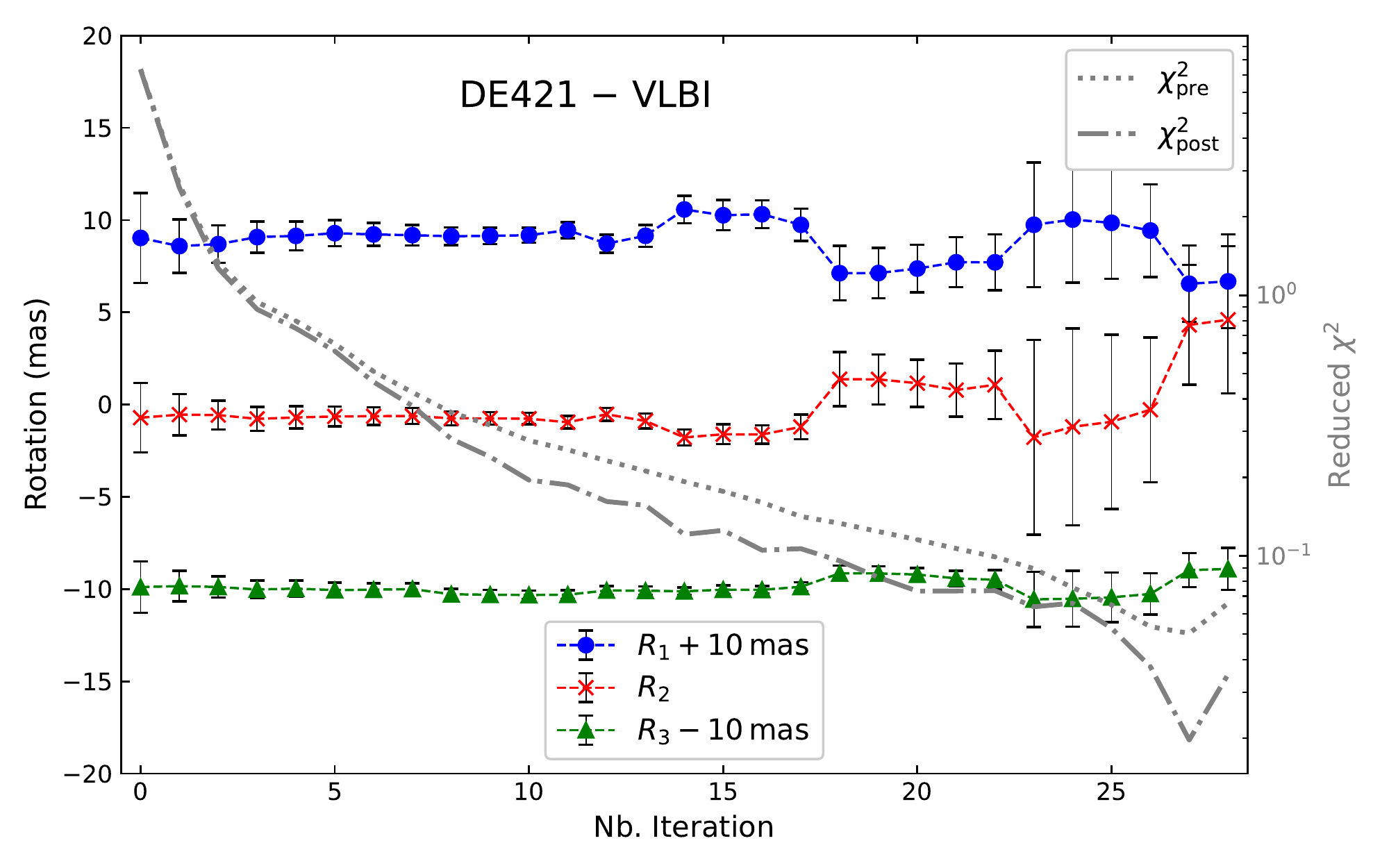}
        \includegraphics[width=\columnwidth]{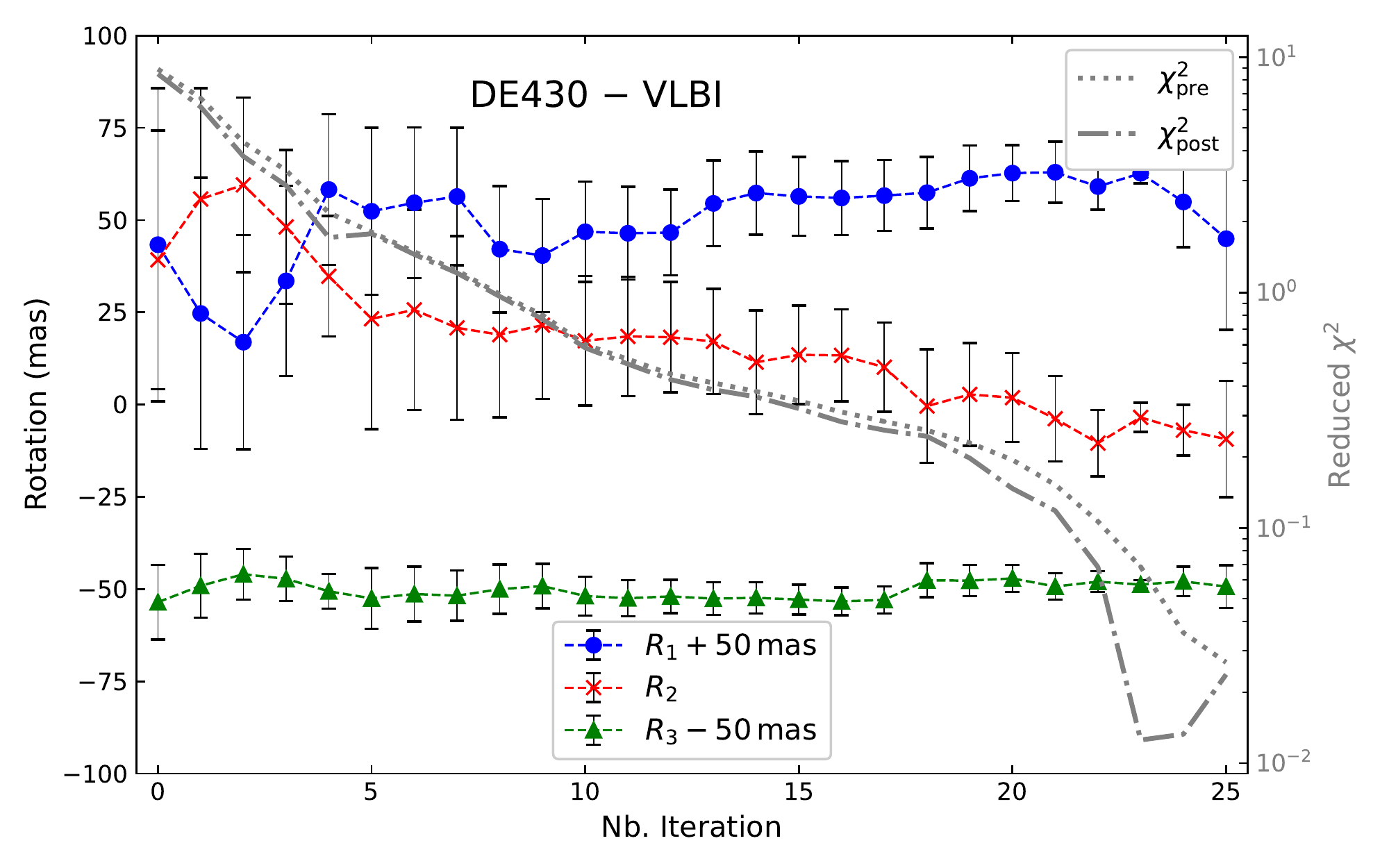}
        \caption{
            Orientation offsets of DE200 (top left), DE405 (top right), DE421 (bottom left), and DE430 (bottom right) frames referred to the VLBI celestial frame as a function of the number of iterations.
            The prefit and postfit reduced chi-squared for the whole sample are also plotted with reference to the vertical axis on the right.
                  }
       \label{fig:timing-vs-vlbi-rot}%
    \end{figure*}

    The differences between timing and VLBI positions are presented in Figs.~\ref{fig:timing-vs-vlbi-pos-oft-de200}--\ref{fig:timing-vs-vlbi-pos-oft-de436}, which mostly fell into the range of $-$50\,mas to 50\,mas for MSPs and on the order of a few 100\,mas for non-MSPs.
    These results are consistent with those in \citetads{1984MNRAS.210..113F,1992MNRAS.258..497F} based on the comparison between the Very Large Array (VLA) and timing positions.
    Similar to the comparison of timing and \textit{Gaia} positions, we found that for the same non-MSPs, larger positional offsets greater than 1~arcsec seen in some of the timing solutions did not appear in the other solutions. 
    This kind of discrepancy among timing solutions was also observed in one MSP -- PSR J1022+1001.
    For instance, the timing solutions of this pulsar given in \citetads{1996ApJ...469..819C} and \citetads{2020MNRAS.494..228L} yield declination offsets of 1.3~arcsec and 13~arcsec, respectively, with respect to the VLBI position.
    We noted that this pulsar was right on the ecliptic plane, whose timing positions were not used in the analysis of the frame tie.
    In addition, the positional difference between VLBI and timing positions of PSR J2145--0750 computed based on the timing solution in \citetads{1994ApJ...425L..41B} reaches beyond 400\,mas, much greater than those computed from the timing solutions of other authors.
    This timing solution seemed to be problematic and thus was not used for the following analysis.
    
    Before the fit, we removed six pulsars that were close to the ecliptic plane within $\pm~5^{\circ}$, namely, PSR J1022+1001, PSR J0614+2229, PSR J0629+2415, PSR J1257--1027, PSR J1703--1846, and PSR J2346--0609.
    Figure~\ref{fig:timing-vs-vlbi-rot} presents the rotation parameters of the ephemeris frames referred to the VLBI celestial frame, except for the DE436 frame, for which only five pulsars are common to the VLBI sample.
    The estimates of the rotation parameters generally converge as the fitting iterates.
    Larger scatters together with postfit reduced $\chi^2_{\rm post}$ less than unity appear at the end of the iteration processes for DE200, DE405, and DE421.
    
    Table~\ref{tab:rot-ang} reports the estimate of the rotation parameters in the middle panel.
    The MSP sample suggested orientation offsets of $\sim$10\,mas around the $Y$- and $Z$-axes of the DE200 frame compared to the VLBI frame, which are confident at $>10\sigma$.
    The orientation of the DE405 frame differs from that of the VLBI frame by approximately 2\,mas in all three axes based on all pulsars, but none was significant at $2\sigma$.
    In addition, we detected a nonzero rotation of approximately $-2$\,mas around the $X$-axis in the DE436 frame with an uncertainty of 0.7\,mas.
    Similar to the comparison between \textit{Gaia} and ephemeris frames, the non-MSP sample generally yields inconsistent results with those based on the MSP sample.
    
    \citetads{2017MNRAS.469..425W} reported that the orientation offsets of the DE405, DE421 and DE436 frames in the $X$-axis are mainly due to PSR J1012+5307.
    We tested to remove this MSP from our sample and reran the fitting.
    The orientation offset of $-2$\,mas on the $X$-axis for the DE436 frames was reduced to below 1\,mas as expected.
    However, the new fittings yielded an orientation offset of $-1.5$\,mas on the $Y$-axis.
    This experiment suggested that in the case of a few pulsars in the sample, the determination of the orientation offset would be affected significantly by adding or removing individual sources.

\subsection{Comparison of parallaxes}
    
    There are four pulsars in the \textit{Gaia} and six in the VLBI samples that have timing parallax measurements published along with the timing positions.
    Figure~\ref{fig:plx-comparison} compares the parallax measurement from the timing solutions to those from the \textit{Gaia} and VLBI catalogs.
    
    When comparing the timing and \textit{Gaia} parallaxes, we found that more than half of the measurements were consistent within the combined formal uncertainties.
    For only one measurement, the parallax difference exceeded three times the corresponding uncertainty:
    the parallax for PSR J1012+5307 was estimated to be 0.71~$\pm$~0.17\,mas in \citetads{2016MNRAS.458.3341D} using DE421, while \textit{Gaia} DR3 gave an estimate of 1.745~$\pm$~0.291\,mas.
    We noted that other parallax measurements for this pulsar agreed with the \textit{Gaia} result; hence, the deviation is most likely a manifestation of errors in this timing solution.
    The timing parallax measurements for PSR J1955+2908 tended to be negative.
    The \textit{Gaia} counterpart of this pulsar was recognized as an unrelated foreground star in \citetads{2018ApJ...864...26J} and considered as the most likely nongenuine association in \citetads{2021MNRAS.501.1116A}.
    As a result, it is not surprising that the parallaxes derived from the timing and $\textit{Gaia}$ do not agree with each other.
    
    In the comparison between timing and VLBI parallaxes, more than 54\% of timing measurements were consistent with the VLBI measurements within their uncertainties.
    The parallax difference that is more confident than 99\% is found for only one measurement of PSR J1022+1001:
    0.72~$\pm$~0.19\,mas in \citetads{2016MNRAS.458.3341D} referred to DE421 and $1.39^{+0.04}_{-0.03}~{\rm mas}$ from VLBI.
    Again, this likely reflects errors in this timing measurement.
    For PSR J2010--1323 and PSR J1537+1155, the timing parallaxes tend to be smaller than those from VLBI, which may need further investigation but is beyond the scope of this work.
    
    There are two pulsars with parallax measurements from both \textit{Gaia} and VLBI astrometry.
    For PSR J1012+5307, the VLBI parallax measurement ($1.21^{+0.03}_{-0.08}\,{\rm mas}$) is only approximately 69\% of the \textit{Gaia} DR3 measurement (1.75\,$\pm$\,0.29\,mas).
    We noted that the \textit{Gaia} DR2 parallax of PSR J1012+5307 (1.33\,$\pm$\,0.41\,mas) matched well with the VLBI measurements \citepads{2020ApJ...896...85D}.
    It is surprising to see that the discrepancy between the \textit{Gaia} and VLBI parallax measurements increases when both the precision and accuracy of the \textit{Gaia} parallaxes improve.
    For PSR J0614+2229, the \textit{Gaia} DR3 parallax is negative, although the \textit{Gaia} proper motion is roughly consistent with the VLBI one.
    When checking the auxiliary parameter of the \textit{Gaia} DR3 table for this source, we found strong correlations between declination and parallax (correlation coefficients of $-0.52$) and between declination and declination proper motion (correlation coefficients of $-0.66$) .
    Since this pulsar is very close to the ecliptic plane, the observed parallactic motion is almost along the ecliptic plane, making it difficult to disentangle the parallactic motion from the proper motion (at least it seems that the \textit{Gaia} DR3 astrometric solution failed to do so).

\subsection{Comparison of proper motions}
    
    For the \textit{Gaia} pulsar sample, we found 11 pulsars with 40 proper motion measurements both in right ascension and declination derived from the timing positions.
    The comparison between \textit{Gaia} and timing proper motions is shown in Fig.~\ref{fig:timing-vs-gaia-pm}.
    We found the proper motion discrepancy significant at $>3\sigma$ for six pulsars in either right ascension or declination.
    Some of these proper motion discrepancies are already discussed in \citetads{2018ApJ...864...26J}, for example, PSR J1816+4510 and PSR 1302--6350.
    The proper motion differences between the timing and \textit{Gaia} measurements for PSR J1955+2908 again support that the \textit{Gaia} match of this pulsar is not genuine.
    PSR J1817$-$3618 only has one timing proper motion measurement \citepads{2019MNRAS.484.3691J}, which gives $\mu_{\alpha^*,{\rm T}}=-19\pm5\,{\rm mas\,yr^{-1}}$ and $\mu_{\delta,{\rm T}}=-16 \pm 17\,{\rm mas\,yr^{-1}}$, while the corresponding \textit{Gaia} values are much smaller: $\mu_{\alpha^*,{\rm G}}=-3.16 \pm 0.15\,{\rm mas\,yr^{-1}}$ and $\mu_{\delta,{\rm G}}=-6.09 \pm 0.11\,{\rm mas\,yr^{-1}}$.
    There is also one timing solution for PSR J0348+0432 \citepads{2013Sci...340..448A}, for which the significant proper motion difference only occurs in the declination component ($\mu_{\delta,{\rm T}}=3.44 \pm 1.35\,{\rm mas\,yr^{-1}}$ and $\mu_{\delta,{\rm G}}=-0.23 \pm 0.89\,{\rm mas\,yr^{-1}}$).
    The significant proper motion difference for PSR1024--0719 comes from the timing solution given in \citetads{1999MNRAS.307..925T}, while the recent timing solutions \citepads[e.g.,][]{2021MNRAS.507.2137R} give more consistent proper motion measurements with that of \textit{Gaia} DR3.

    For the VLBI pulsar sample, we found 55 of 62 pulsars with 109 timing measurements of proper motion.
    The timing proper motions were generally consistent with VLBI measurements within the quoted error, as evinced in Fig.~\ref{fig:timing-vs-vlbi-pm}.
    The timing solution for PSR J1022+1001 in \citetads{2004MNRAS.353.1311H} yields proper motion differences of $-346\,\mathrm{mas\,yr^{-1}}$ in right ascension and of $-906\,\mathrm{mas\,yr^{-1}}$ in declination, which vanishes in recent timing solutions such as the one published in \citetads{2021MNRAS.507.2137R}.
    Therefore, these large differences are likely due to errors in the corresponding time solution of \citetads{2004MNRAS.353.1311H}.
    PSR J0629+2415 only has two timing solutions as given in \citetads{2004MNRAS.353.1311H} and \citetads{2020PASJ...72...70L}, which both show proper motion differences in declination of $+405\,\mathrm{mas\,yr^{-1}}$ with respect to the VLBI proper motion.
    Noting that the quoted uncertainties of the timing proper motion in declination are as large as $300\,\mathrm{mas\,yr^{-1}}$, we suspect that the large proper motion differences are likely caused by the errors in the timing solutions.
    Three measurements (for PSR J2145--0750 and PSR J1537+1155) in the right ascension and five measurements (for PSR J0406+6138, PSR J1537+1155, PSR 1820--0427, and PSR J2145--0750) in the declination showed statistically significant differences.
    \citetads{2016ApJ...828....8D} reported the inconsistency in the proper motion of PSR J2145-0750 measured by the VLBI and timing, which was likely due to errors in the timing model as they concluded.
    
\section{Discussion}

\subsection{Comparison with previous results}

    \citetads{1982A&A...114..297S} claimed that the orientation of the DE200 frame was accurate within approximately 1\,mas with respect to the J2000 dynamical frame.
    \citetads{1994A&A...287..279F} determined the orientation offset between DE200 and extragalactic frames in 1988 to be approximately $-2\,\pm\,2$\,mas, $-12\,\pm\,3$\,mas, and $-6\,\pm\,3$\,mas on the $X$-, $Y$-, and $Z$-axes, respectively.
    Our results based on MSPs agree roughly with the results of \citetads{1994A&A...287..279F}.
    \citetads{2011CeMDA.111..363F} used a sample of four MSPs to determine the orientation offsets between three ephemeris frames (DE200, DE405, and DE421) and the ICRF with a precision of $\sim$5\,mas.
    Their results are roughly consistent with those of this work considering the uncertainties in both solutions.
    
    The alignment of the inner planet ephemeris system of DE405 to ICRF1 is performed with an accuracy of 1\,mas \citepads{standish1998jpl,2010ITN....36....1P}.
    \citetads{2017MNRAS.469..425W} observed a change of $2.16\pm0.33$\,mas in the ecliptic obliquity of the DE405 frame, that is, a rotation around the $X$-axis.
    We obtained a similar rotation around the $X$-axis in the comparison between the DE405 and VLBI frames.
    However, the rotation was in the opposite direction and statistically insignificant.
    We also noticed that this rotation largely diminished when combining the \textit{Gaia} and VLBI samples.
    Therefore, our results suggest that a misalignment of the DE405 frame is possible but likely no greater than $\sim$2\,mas.

    The frame of DE421 was aligned to ICRF1 with an accuracy of 0.25\,mas 
    thanks to the VLBI observations of spacecraft in orbit around Mars \citepads{2009IPNPR.178C...1F,2010ITN....36....1P}, which is supported by the comparison of DE421 versus VLBI in this work.
    Our results are also consistent with those presented in \citetads{2017MNRAS.469..425W} within the quoted uncertainties.
    
    The DE430 ephemeris is the first version of the DE series to align onto the ICRF2 \citepads{2015AJ....150...58F}, which is claimed to be precise at 0.2\,mas \citepads{2014IPNPR.196C...1F}.
    There were only two MSPs in our sample, making the estimate of orientation offsets of the DE430 frame dominated by the timing positional errors of non-MSPs and thus less reliable.
    The pulsar sample size for DE436 is too small to make a solid conclusion of the orientation offset.
    
    For the \textit{Gaia} pulsar samples, the results based on non-MSPs are significantly inconsistent with those based on MSPs.
    Since many \textit{Gaia} non-MSPs are only candidate associates waiting for further verification, the errors due to the misidentifications of pulsar companions could be severe, leading to unreliable results.
    This indicates that removing the candidate associations and relying only on highly secure associations (e.g., MSPs) would be a better approach.
    We also found that the inclusion of non-MSPs would worsen the alignment precision and lead to unreliable results.
    Therefore, we can reach a similar conclusion to \citetads{2017MNRAS.469..425W} that the non-MSPs would contribute little to improving the alignment precision between the ephemeris and the extragalactic frames.
    
    The formal uncertainties of the rotation parameters based on the \textit{Gaia} pulsars are always at least several times greater than those based on the VLBI pulsars.
    Therefore, at this moment, VLBI pulsars are preferred for alignment between the ephemeris and extragalactic frames.
    Several limiting factors on using \textit{Gaia} pulsars for frame alignment will be discussed in Sect.~\ref{subsect:limits}.

    The orientation agreement between the \textit{Gaia}-CRF and ICRF is found to be less than 0.1\,mas based on the quasar sample \citepads{2018AJ....156...13L,2018A&A...609A..19L,2020A&A...644A.159C,2020A&A...634A..28L} .
    There are only two common pulsars between the \textit{Gaia} DR3 and PSR$\pi$ data, so we cannot estimate the orientation offset between the \textit{Gaia}-CRF and ICRF by directly comparing the VLBI and \textit{Gaia} positions of the pulsar.
    An indirect assessment can be made by comparing the rotation parameters between the ephemeris frames and the \textit{Gaia}-CRF with those between the ephemeris frames and the VLBI frame (i.e., the values reported in Table~\ref{tab:rot-ang}).
    However, none of these results agrees with the results based on the quasar sample.
    Therefore, the pulsar may not be as suitable for comparison between \textit{Gaia}-CRF and the VLBI frame as the quasar, a fact already recognized by \citetads{2019ApJ...875..100D}.

\subsection{Current defects and limitations} \label{subsect:limits}
    
    There are several limitations on the current check of the orientation agreement of the ephemeris frames with the \textit{Gaia} and VLBI reference frames.
    The reduced chi-squared much larger than one shown in Figs.~\ref{fig:timing-vs-gaia-rot} and \ref{fig:timing-vs-vlbi-rot} suggests that the uncertainty of the computed positional offset must have been severely underestimated.
    This could be due to several factors: (i) underestimated uncertainties from one or more of the input positions; (ii) underestimated uncertainties for the \textit{Gaia} and VLBI proper motion used for the propagation between reference epochs; (iii) systematics in the \textit{Gaia} and VLBI proper motion; and (iv) possible contamination of stars in the \textit{Gaia} pulsar sample (especially for the non-MSPs). 
    The lack of sufficient common pulsars (for DE436) also makes the estimate of the orientation offset easily affected by individual pulsars.
    
\subsubsection{Limitations of using the published timing solutions} \label{sect:timing-input-err}
    
    We used the published timing solutions to form our timing catalogs, for which different ephemerides were usually used in computing the pulse times-of-arrival (TOAs).
    Due to this fact, we had to divide these timing solutions into several groups based on the ephemerides used, resulting in two shortcomings.
    
    The first is that for older ephemerides such as DE200, the timing solutions usually come from earlier publications, which would have shorter data spans and narrower bandwidths and adopt less sophisticated data reduction techniques, for example, treatment of dispersion measurement variations and modeling of red noise \citepads[e.g.,][]{2019MNRAS.489.3810P}.
    The systematics due to these effects might overwhelm the misalignment of the ephemeris frames and dominate the timing positional differences with respect to the \textit{Gaia} and VLBI positions, which is more pronounced for non-MSPs.
    However, it is difficult for us to estimate the implication of these effects on our determination of the orientation offsets between the ephemeris and extragalactic frames.
    
    The second is that the division significantly reduced the size of the input pulsar samples, leaving the estimation of the orientation offsets more vulnerable to errors in the positions of individual pulsars measured by timing, VLBI, or \textit{Gaia}.
    Considering these shortcomings, the results given in Table~\ref{tab:rot-ang} should be used with great caution, especially those based on non-MSPs.
    
    The intercomparisons of results between successive ephemerides cannot effectively reflect the systematics in the ephemeride systems due to the aforementioned shortcomings.
    A more sophisticated method is to reanalyse a given timing data set of more precisely timed pulsars using different ephemerides as done in \citetads{2017MNRAS.469..425W}, for example, the published TOAs from several pulsar timing arrays such as the International Pulsar Timing Array \citepads{2019MNRAS.490.4666P}.
    We plan to carry out such work in the near future.
    Another method of modeling the systematics in the ephemerides using the timing observations is to add in the TOA model the unknowns relating to the deficiencies of the ephemerides, such as the correction to the mass of outer planets \citepads{2020ApJ...893..112V}.
    This method does not need to choose a preferred ephemeris (reference ephemeris) in the comparison among various ephemerides, which is usually assumed to be ideal.
    However, it should also be noted that the results obtained via this method cannot be associated directly with misalignment of the ephemeris frames.
    
    To partly overcome the small sample size used in this work, we attempted to combine the \textit{Gaia} and VLBI samples to form an average extragalactic frame and then redetermined the orientation offsets of the ephemeris frames with respect to this frame.
    The results are presented in the lower panel of Table~\ref{tab:rot-ang}.
    Although the estimates of the rotation parameters were not always consistent with those based on the \textit{Gaia} or VLBI sample, we obtained smaller uncertainty for all ephmerides except DE200 when using all pulsars or MSPs.
    This implies that the combination of the \textit{Gaia} and VLBI pulsars could potentially improve the precision of the pulsar-based frame tie.
    However, including more non-MSPs would lead to a poorer estimate for frame rotation, which is most likely due to various underestimated errors in the input solutions of \textit{Gaia} pulsars discussed earlier in this section.
    
\subsubsection{Errors in the proper motion systems} \label{subsect:pm-error}
    
    The global spin of \textit{Gaia}-CRF3 is better than $\mathrm{10\,\mu as\,yr^{-1}}$ on each axis \citepads{2018A&A...616A..14G}, and it is ten times smaller for ICRF3 \citepads{2021arXiv211210079L}.
    Considering the difference of $\mathrm{\sim 22\,yr}$ between the median epoch of timing positions in the DE200 frame and the reference epoch for the \textit{Gaia} DR3 position, the orientation bias due to the spin of the \textit{Gaia}-CRF3 is approximately 0.2\,mas.
    The bias is reduced to 0.04\,mas in comparison with the DE436 frame, which is insignificant compared to the values given in Table~\ref{tab:rot-ang}.
    For the VLBI frames, this orientation bias is supposed to be ten times smaller than that of \textit{Gaia}-CRF3 and thus does not affect the estimation of the rotation parameter much.
    
    On the other hand, the errors in the $\textit{Gaia}$ and VLBI proper motion could be magnified in the position propagation, resulting in greater positional offsets, especially for the early timing measurements of pulsar positions in the DE200 frame.
    It is possible to consider using the timing proper motion instead.
    However, for only a few MSPs with long timing spans, the best precision of the timing proper motion measurements is five times smaller than that of the \textit{Gaia} or VLBI measurements.
    There are four such sources in the \textit{Gaia} sample (PSR J0348+0432, PSR J0437$-$4715, PSR J1012+5307, and PSR J1024$-$0719) and half of the MSPs in the VLBI sample (PSR J1012+5307, PSR J2010$-$1323, and PSR J2317+1439).
    Therefore, the \textit{Gaia} or VLBI proper motion measurements are more precise than or comparable to the timing measurement for the bulk of the pulsars.
    In addition, in the comparison of the celestial frames for which the systematics are concerned, it is preferred to use the \textit{Gaia} or VLBI proper motions that are estimated simultaneously with reference positions in the astrometric solutions for position propagation, even if the timing proper motions are more precise than those of \textit{Gaia} or VLBI in terms of the random error.

\subsection{Future prospects} \label{sect:potential-precision}

    \begin{figure}
        \centering
        \includegraphics[width=\columnwidth]{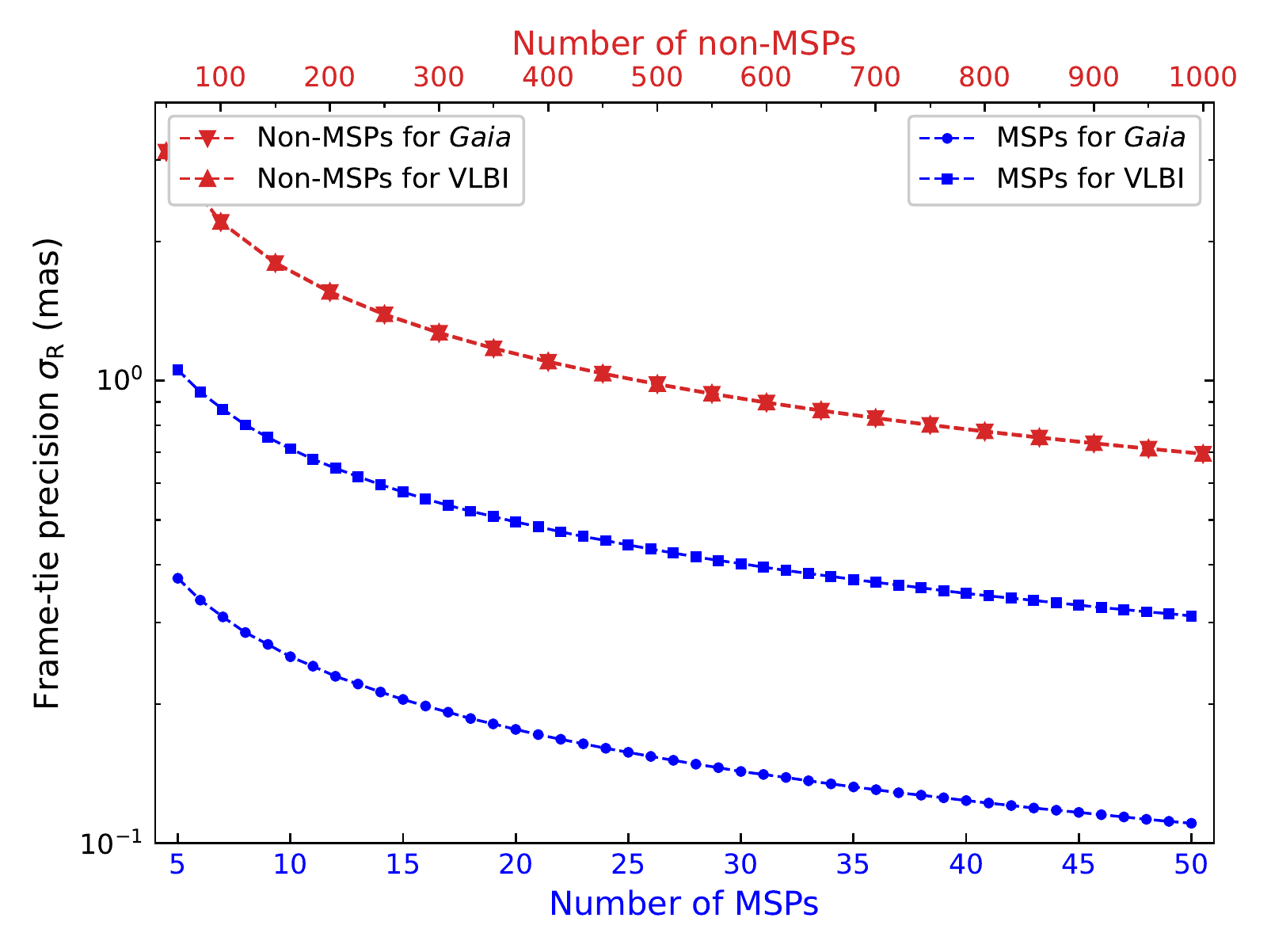}
        \caption{\label{fig:frame-tie-pre}%
        Simulated alignment precision of the ephemeris frame onto the \textit{Gaia} and VLBI reference frames as a function of the number of pulsars (blue filled circles for \textit{Gaia} MSPs, blue filled squares for VLBI MSPs, red inverted triangles for \textit{Gaia} non-MSPs, and red filled triangles for VLBI non-MSPs).
        The $X$-axis at the bottom corresponds to the number of MSPs while the $X$-axis at the top corresponds to the number of non-MSPs
                  }
    \end{figure}
    
    Since there are likely to be several tens of pulsars in the \textit{Gaia} and VLBI catalogs, a natural question would be at which level of precision the dynamical-kinematic celestial frame tie could be reached as the sample size increases.
    We performed a covariance analysis to address this question.
    We first assumed that the precision of the timing positions (either in right ascension or declination) for all these pulsars was distributed randomly, that is, following a normal distribution of $N(\sigma_{\rm P},\,\sigma_{\rm P}/10)$, where $\sigma_{\rm P}$ is the typical position precision.
    We assumed $\sigma_{\rm P}$ to be 0.3\,mas for \textit{Gaia} and 1.0\,mas for VLBI (Table~\ref{tab:psr-pos-err}).
    Considering that the achievable timing position precision depends strongly on the type of pulsar, it was assumed to be 0.2\,mas for MSPs \citepads[e.g.,][]{2021MNRAS.507.2137R} and 10\,mas for non-MSPs \citepads[e.g.,][]{2019MNRAS.489.3810P}.
    In addition, we assumed that the epochs of pulsar positions measured by \textit{Gaia}, VLBI, and timing are identical. 
    Then we used these simulated position uncertainties, together with the actual positions of a certain number of pulsars (MSPs or non-MSPs) randomly taken from the ATNF catalog to construct the normal matrix for determining the rotation parameters in Eq.~(\ref{eq:vsh01}).
    The covariance matrix of the rotation parameter estimates was the inversion of the normal matrix assuming the posterior reduced $\chi^2$ to be unity, from which we obtained the uncertainty in the rotation parameter estimation.
    This procedure was repeated 1000 times. 
    We took the mean values of the uncertainties in each axis and adopted their root-sum-squared $\sigma_{\rm R}$ as the final estimate of the precision of the frame tie.
    
    Figure~\ref{fig:frame-tie-pre} illustrates the potential alignment precision between the dynamical and kinematic frames via pulsars in our simulation.
    A sub-mas frame tie can be easily achieved with more than five MSPs, but more than 500 non-MSPs are needed, suggesting that it would be preferable to focus on the MSPs only for improving the frame tie.
    We found that the alignment precision between the timing and VLBI frames was limited to 0.3\,mas, which is not found in the case between the \textit{Gaia} and timing frames.
    This may support the claim in \citetads{2017MNRAS.469..425W} that the main contributor to the error budget for the frame tie between timing and VLBI frames comes from the VLBI.
    We note that we only used two pulsars in the MSPSR$\pi$ project; hence, our results are far from what can be expected from the full potential of the MSPSR$\pi$ project.
    The pulsar absolute VLBI position in the MSPSR$\pi$ project would be much improved with more precise positions of calibrators and the modeling of the core shift effect \citepads{2019ApJ...875..100D}, which should reduce the noise floor in the frame tie seen in Fig.~\ref{fig:frame-tie-pre}.
    
    On the other hand, this situation could also, in principle, be improved if the \textit{Gaia} positions are used instead.
    As suggested by the simulation, the alignment precision could reach 0.2\,mas with a sample of 15 MSPs with timing precision of 0.2\,mas, surpassing the current precision of the frame tie between DE440/DE441 and ICRF3 \citepads{2021AJ....161..105P}.
    However, these statements are too optimistic considering the fact that most precisely timed MSPs are too faint to be observed by \textit{Gaia}.
    A possible way to improve the frame-tie precision based on \textit{Gaia} pulsars might be to use deeper and more accurate optical observations of the companions in the binary MSP systems with referred to \textit{Gaia} sources, for instance, using the Vera C. Rubin Large Synoptic Survey Telescope \citepads[][]{2019ApJ...873..111I}.

\section{Summary}

    For the first time, we compared the timing reference frame with the \textit{Gaia} celestial reference frame, complementary to the comparison between the timing and VLBI reference frames.
    We identified 49 counterparts of pulsars in \textit{Gaia} DR3, among which 33 pulsars have published timing solutions.
    We also used 62 VLBI counterparts in the PSR$\pi$ and MPSR$\pi$ data archives, all with timing positions available.
    Based on the offsets of the timing position with respect to the \textit{Gaia} and VLBI positions, we estimated the rotation between the ephemeris frames and the \textit{Gaia} and VLBI frames.
    
    We found orientation offsets of $\sim$10\,mas in the DE200 frame relative to both the \textit{Gaia} and VLBI celestial frames.
    We noted that our results depend strongly on the subset used in the comparison: the results using the non-MSP sample generally differ from those based on the MSP sample, especially for \textit{Gaia} pulsars.
    This suggests that our results might be biased due to several limitations and error sources (as discussed in Sect.\ref{subsect:limits}) and thus should be used with caution.
    We also learned that the comparison of literature timing results to VLBI and/or \textit{Gaia} astrometry has significant limitations in the ability to measure frame rotation.
    A successful measurement of the frame rotation would require careful sample selection and dedicated reanalysis of timing data using modern approaches and with each solar system ephemeris of interest.
    
    Since many pulsars in the \textit{Gaia} DR3 do not have a timing solution and those with available timing positions were mostly observed more than a decade ago, the alignment between the timing and \textit{Gaia} celestial frames cannot be achieved better than 1\,mas.
    The median timing positional uncertainty for \textit{Gaia} MSPs is 2.3\,mas, which is much worse than what can be achieved for a typical MSP.
    Therefore, we anticipate that \textit{Gaia} pulsars, especially MSPs, can be observed in more timing experiments to improve the timing measurement of their positions, which should benefit both the astrophysical and astrometric applications of timing observations.
    
    For the VLBI pulsars, although we used the preliminary results of the PSR$\pi$ project and only included the VLBI solutions for two pulsars from the MSPSR$\pi$ project, the best-achieved alignment precision in a single axis is approaching the current alignment accuracy of DE440.
    As a result, it is quite promising that the outcome of the PSR$\pi$ and MSPSR$\pi$ projects will play an important role in bridging the ephemeris and extragalactic frames.

\begin{acknowledgements}
    We sincerely thank Prof. Adam Deller for his meticulous review, constructive comments, and useful suggestions, which greatly improve the work.
    N. Liu and Z. Zhu were supported by the National Natural Science Foundation of China (NSFC) under grant Nos~11833004 and 12103026.
    N. Liu was also supported by the Yuxiu Postdoctoral Institute at Nanjing University (``Yuxiu Young Scholars Program'') and the China Postdoctoral Science Foundation (Grant Number: 2021M691530).
    J. Antoniadis was supported by the Stavros Niarchos Foundation (SNF) and the Hellenic Foundation for Research and Innovation (H.F.R.I.) under the 2nd Call of ``Science and Society'' Action Always strive for excellence -- ``Theodoros Papazoglou'' (Project Number: 01431).
    N. Liu also thanks Dr. Emilie Parent for her patient instruction on the timing data and Prof. Qin Wang for her kind host of the postdoc program.
    This work has made use of data from the European Space Agency (ESA) mission
    {\it Gaia} (\url{https://www.cosmos.esa.int/gaia}), processed by the {\it Gaia}
    Data Processing and Analysis Consortium (DPAC,
    \url{https://www.cosmos.esa.int/web/gaia/dpac/consortium}). Funding for the DPAC
    has been provided by national institutions, in particular the institutions
    participating in the {\it Gaia} Multilateral Agreement.
    This research had also made use of the data from the PSR$\pi$ campaign.
    We thank all authors for publishing their timing solutions as used in this work.
    We used programming packages and tools such as IPython \citepads{2007CSE.....9c..21P}, 
    Numpy \citepads{2011CSE....13b..22V}, 
    Scipy \citepads{2020NatMe..17..261V},
    Astropy\footnote{\href{http://www.astropy.org}{http://www.astropy.org}} \citepads{2018AJ....156..123A}, 
    Astroquery \citepads{2019AJ....157...98G},
    Statsmodels \citepads{seabold2010statsmodels},
    the Python 2D plotting library Matplotlib \citepads{2007CSE.....9...90H}, 
    the SIMBAD database operated at CDS, Strasbourg, France, 
    and NASA's Astrophysics Data System.
\end{acknowledgements}

\bibliographystyle{aa} 
\bibliography{references} 

\begin{appendix} 

\onecolumn
\begingroup
\begin{landscape}

\section{Information for \textit{Gaia} pulsar sample}

Table~\ref{tab:gaia-psr-info} tabulates basic information for 49 pulsars found in the \textit{Gaia} DR3.

\begin{ThreePartTable}

\begin{TableNotes}\footnotesize
    \item[]\textbf{Notes.} The columns are pulsar names, \textit{Gaia} DR3 identifiers, the angular separation between the ATNF and \textit{Gaia} EDR3 positions, formal uncertainties of the ATNF and \textit{Gaia} DR3 positions, spin period, orbital period (the sign of ``\dots'' means an isolated pulsar), \textit{Gaia} DR3 $G$ magnitude, and techniques or instruments for deriving the positions quoted in ATNF.
\end{TableNotes}
  
\begin{longtable}{llccccrrrcc}
\caption{
    \label{tab:gaia-psr-info}
    Information for the 49 pulsars found in the \textit{Gaia} DR3. }\\
\toprule
    PSR     &\texttt{source\_id}     &$\theta$  &$\sigma_{\alpha^*,{\rm A}}$  &$\sigma_{\delta,{\rm A}}$  &$\sigma_{\alpha^*,{\rm G}}$  &$\sigma_{\delta,{\rm G}}$  &$P_s$  &$P_b$  &$G$  &Origin  \\
    & & $({}^{\prime\prime}$)    &(mas)  &(mas)  &(mas)  &(mas)  &(s)  &(d)  &(mag) &            \\ 
\midrule
\endfirsthead
\caption{continued.}\\
\toprule
    PSR     &\texttt{source\_id}     &$\theta$  &$\sigma_{\alpha^*,{\rm A}}$  &$\sigma_{\delta,{\rm A}}$  &$\sigma_{\alpha^*,{\rm G}}$  &$\sigma_{\delta,{\rm G}}$  &$P_s$  &$P_b$  &$G$  &Origin \\ 
    & & $({}^{\prime\prime}$)    &(mas)  &(mas)  &(mas)  &(mas)  &(s)  &(d)  &(mag) & \\ 
\midrule
\endhead

\midrule
\endfoot

\bottomrule
\insertTableNotes
\endlastfoot
    J0045--7319  & 4685849525145183232  &  0.510  & 301      &  200      &  0.038  &  0.038  & 0.926  &     51.17  &16.20  & Timing \\ 
    J0337+1715   &   44308738051547264  &  0.090  &  2   &  2        &  0.136  &  0.141  & 0.003  &      1.63  &18.05  & VLBA \\ 
    J0348+0432   & 3273288485744249344  &  0.040  &  0.1  &  0.2      &  0.696  &  0.646  & 0.039  &      0.10  &20.59  & Timing \\ 
    J0437--4715  & 4789864076732331648  &  1.136  &  0.006  &  0.007    &  0.459  &  0.516  & 0.006  &      5.74  &20.35  & Timing \\ 
    J0534+2200   & 3403818172572314624  &  0.367  &  70      & 60      &  0.073  &  0.061  & 0.033  &        \dots  &16.53  & Optical \\ 
    J0534--6703  & 4660152083015919872  &  0.517  & 585      &800      &  0.155  &  0.154  & 1.818  &        \dots  &18.86  & Timing \\ 
    J0540--6919  & 4657672890443808512  &  0.030  & 50      & 50      &  1.205  &  1.320  & 0.051  &        \dots  &20.77  & HST \\ 
    J0614+2229   & 3376990741688176384  &  0.406  &  1    &  1      &  0.544  &  0.535  & 0.335  &        \dots  &19.62  & VLBA \\ 
    J0857--4424  & 5331775184393659264  &  0.541  &  21      & 20      &  0.205  &  0.205  & 0.327  &        \dots  &18.39  & Timing \\ 
    J1012+5307   &  851610861391010944  &  0.287  &  0.01  &  0.14   &  0.204  &  0.204  & 0.005  &      0.60  &19.59  & Timing \\ 
    J1023+0038   & 3831382647922429952  &  0.167  &  0.5   &  0.5    &  0.058  &  0.055  & 0.002  &      0.20  &16.23  & VLBA \\ 
    J1024--0719  & 3775277872387310208  &  0.624  &  0.05  &  0.11   &  0.254  &  0.256  & 0.005  &        \dots  &19.15  & Timing \\ 
    J1036--8317  & 5192229742737133696  &  0.346  &  5      &  40     &  0.131  &  0.123  & 0.003  &      0.34  &18.57  & 4FGL-DR3 \\ 
    J1048+2339   & 3990037124929068032  &  0.276  &  1.0  &  2.0    &  0.307  &  0.380  & 0.005  &      0.25  &19.59  & Timing \\ 
    J1227--4853  & 6128369984328414336  &  0.248  &  10      &  3      &  0.097  &  0.061  & 0.002  &      0.29  &18.07  & Timing \\ 
    J1302--6350  & 5862299960127967488  &  0.106  &  0.08  &  0.08   &  0.009  &  0.010  & 0.048  &   1236.72  & 9.63  & VLBA \\ 
    J1305--6455  & 5858993350772345984  &  0.480  & 127      &100      &  0.027  &  0.028  & 0.572  &        \dots  &16.04  & Timing \\ 
    J1306--4035  & 6140785016794586752  &  0.274  & 228      &200      &  0.130  &  0.109  & 0.002  &      1.10  &18.09  & USNO-B1 \\ 
    J1311--3430  & 6179115508262195200  &  0.084  &  2    &  4      &  1.144  &  0.626  & 0.003  &      0.07  &20.44  & Timing \\ 
    J1417--4402  & 6096705840454620800  &  0.304  & 970      &900      &  0.040  &  0.033  & 0.003  &      5.37  &15.77  & 1FGL \\ 
    J1431--4715  & 6098156298150016768  &  0.223  &  2    &  4      &  0.082  &  0.107  & 0.002  &      0.45  &17.73  & Timing \\ 
    J1435--6100  & 5878387705005976832  &  0.412  &  4    &  7      &  0.125  &  0.178  & 0.009  &      1.35  &18.92  & Timing \\ 
    J1509--6015 & 5876497399692841088  &  0.199  & 744  &600      &  0.074  &  0.080  & 0.339  &        \dots  &17.76  & Timing \\ 
    J1542--5133 & 5886184887428050048  &  0.326  &1212  &30       &  0.210  &  0.187  & 1.784  &        \dots  &19.03  & Timing \\ 
    J1546--5302 & 5885808648276626304  &  0.558  & 902  &900      & 17.726  &  5.101  & 0.581  &        \dots  &21.11  & Timing \\ 
    J1622--0315 & 4358428942492430336  &  0.187  &  4    &  7    &  0.248  &  0.187  & 0.004  &      0.16  &19.21  & 3FGL \\ 
    J1624--4411 & 5992089027071540352  &  0.303  & 194  & 500      &  0.404  &  0.276  & 0.233   &        \dots  &19.88  & Timing \\ 
    J1624--4721 & 5941843098026132608  &  0.310  & 813  & 20      &  0.876  &  0.441  & 0.449  &        \dots  &20.39  & Timing \\ 
    J1653--0158 & 4379227476242700928  &  0.273  &  0.8   &  0.5   &  0.632  &  0.353  & 0.002  &      0.05  &20.45  & \textit{Gaia} DR2 \\ 
    J1723--2837 & 4059795674516044800  &  0.115  &  11  &  110    &  0.035  &  0.025  & 0.002  &      0.62  &15.54  & Timing \\ 
    J1810+1744   & 4526229058440076288  &  0.048  & 143  & 70      &  0.379  &  0.441  & 0.002  &      0.15  &20.00  & Optical \\ 
    J1816+4510   & 2115337192179377792  &  0.056  &  0.7  &  0.8   &  0.094  &  0.096  & 0.003  &      0.36  &18.20  & Timing \\ 
    J1817--3618  & 4038146565444090240  &  0.418  &  109  &  300      &  0.106  &  0.106  & 0.387  &        \dots  &17.62  & Timing \\ 
    J1839--0905  & 4155609699080401920  &  0.165  & 444  & 800      &  0.049  &  0.045  & 0.419  &        \dots  &16.51  & Timing \\ 
    J1851+1259   & 4504706118346043392  &  0.332  &  89  &  150    &  0.818  &  1.030  & 1.205  &        \dots  &20.50  & Timing \\ 
    J1852+0040   & 4266508881354196736  &  0.538  & 600  & 600      &  0.638  &  0.778  & 0.105  &        \dots  &20.21  & \textit{Chandra} \\ 
    J1903--0258  & 4261581076409458304  &  0.390  & 165  & 700      &  0.243  &  0.245  & 0.301  &        \dots  &18.93  & Timing \\ 
    J1928+1245   & 4316237348443952128  &  0.164  &  1  &  3   &  0.113  &  0.133  & 0.003  &      0.14  &18.23  & Timing \\ 
    J1946+2052   & 1825839908094612992  &  0.363  &  84  &  90      &  0.291  &  0.319  & 0.017  &      0.08  &19.86  & VLA \\ 
    J1955+2908   & 2028584968839606784  &  0.115  &  0.10  &  0.11  &  0.102  &  0.137  & 0.006  &    117.35  &18.70  & Timing \\ 
    J1957+2516   & 1834595731470345472  &  0.180  &  4  &  3    &  0.357  &  0.659  & 0.004  &      0.24  &20.28  & Timing \\ 
    J1958+2846  & 2030000280820200960  &  0.429  & 394  & 10      &  0.157  &  0.212  & 0.290  &        \dots  &19.34  & Timing \\ 
    J1959+2048   & 1823773960079216896  &  0.239  &  0.7  &  0.6   &  0.635  &  0.703  & 0.002  &      0.38  &20.17  & Timing \\ 
    J2027+4557   & 2071054503122390144  &  0.271  &  31  &  40      &  0.023  &  0.025  & 1.100  &        \dots  &15.71  & Timing \\ 
    J2032+4127   & 2067835682818358400  &  0.074  &  22  &  90      &  0.012  &  0.015  & 0.143  &  16835.00  &11.28  & Timing \\ 
    J2039--5617  & 6469722508861870080  &  0.242  &  1  &  1    &  0.125  &  0.102  & 0.003  &      0.23  &18.52  & \textit{Gaia} DR2 \\ 
    J2129--0429  & 2672030065446134656  &  1.389  &  15  &  80      &  0.060  &  0.055  & 0.008   &      0.64  &16.82  & Timing \\ 
    J2215+5135   & 2001168543319218048  &  0.062  &  4  &  13    &  0.176  &  0.200  & 0.003   &      0.17  &19.20  & \textit{Fermi} \\ 
    J2339--0533  & 2440660623886405504  &  0.212  &  149  & 30      &  0.159  &  0.140  & 0.003  &      0.19  &18.79  & Optical \\ 

\end{longtable}
\end{ThreePartTable}
\end{landscape}

\endgroup
\twocolumn

\section{Comparison of pulsar proper motion and parallax}

\begin{figure}[h!]
        \resizebox{0.9\hsize}{!}{\includegraphics{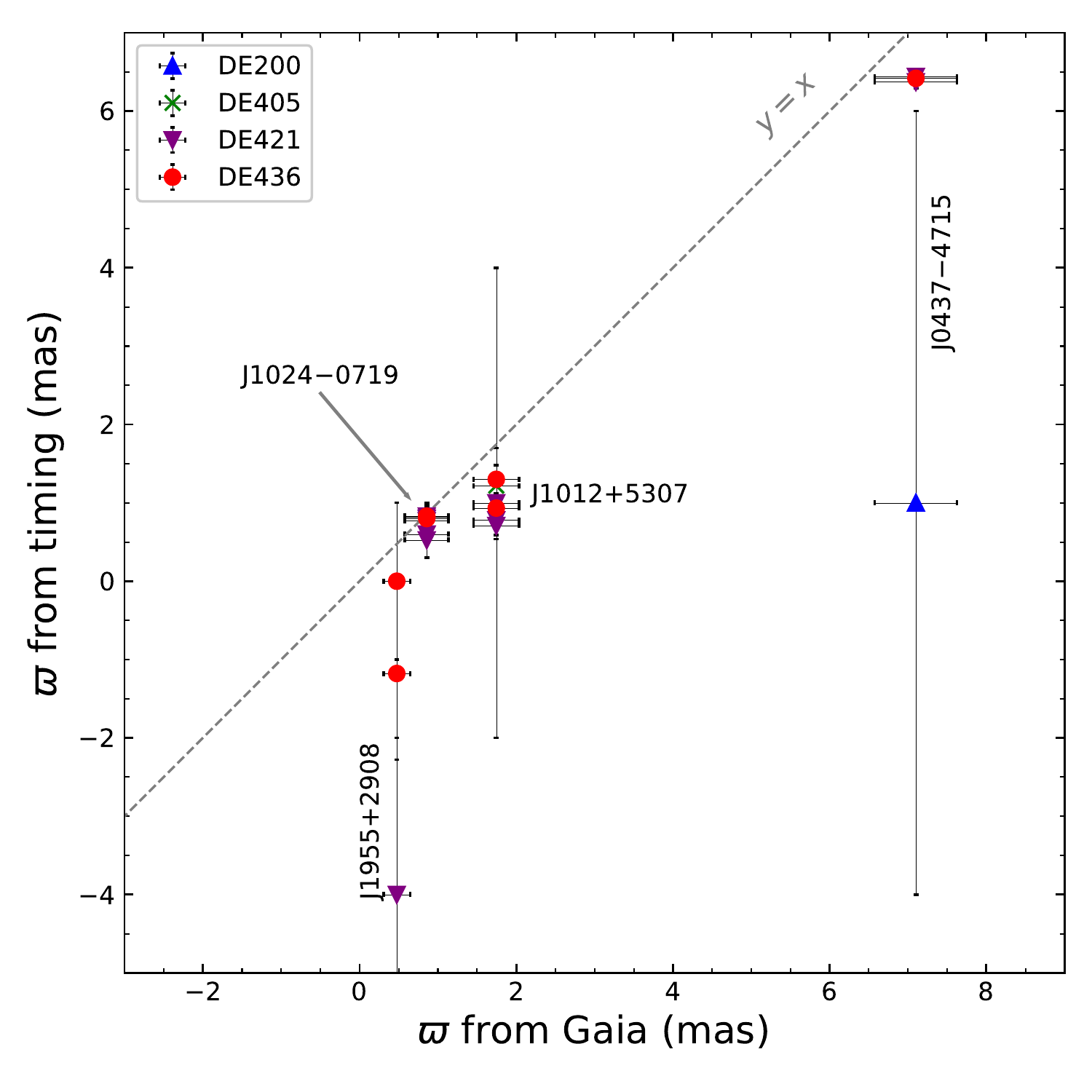}}
        \resizebox{0.9\hsize}{!}{\includegraphics{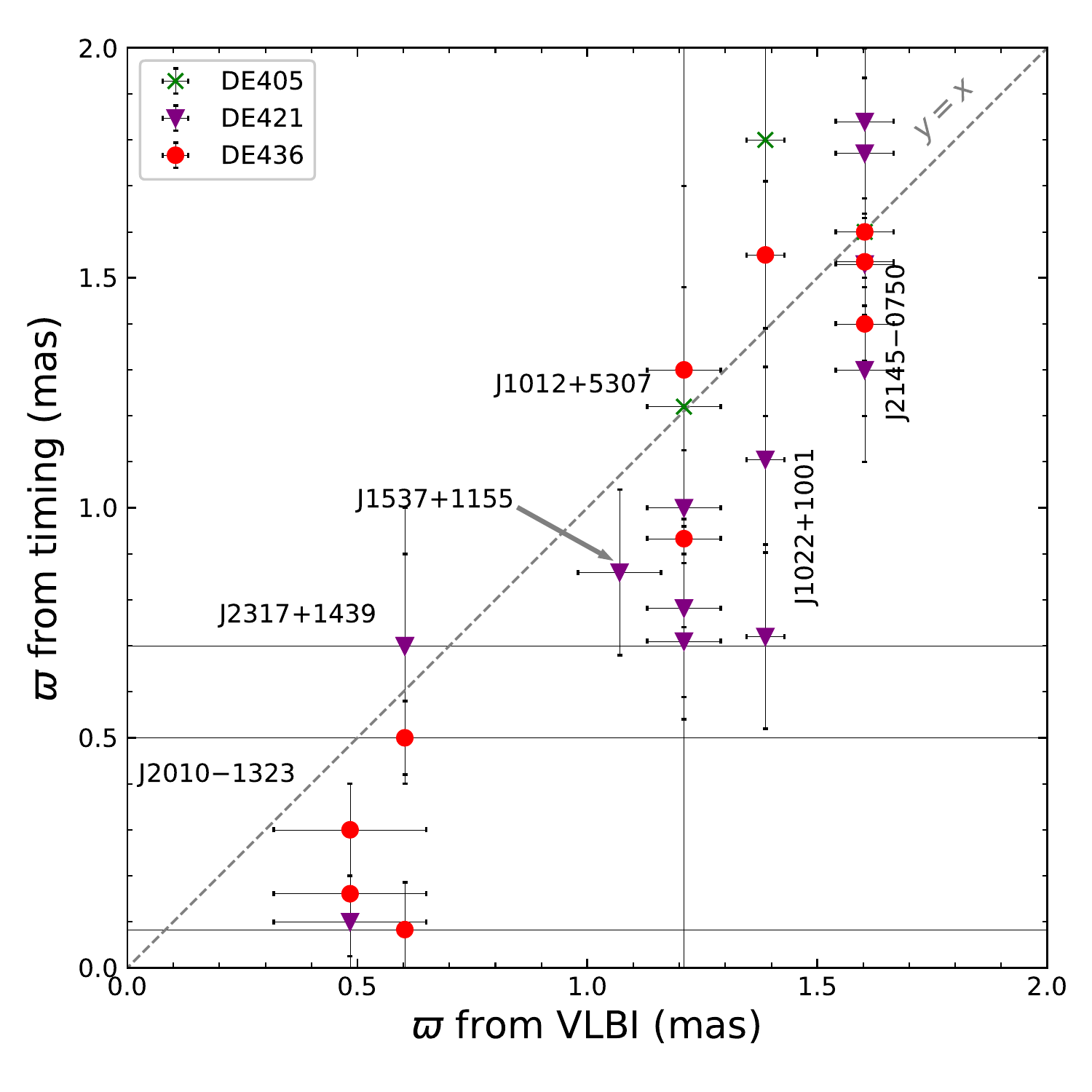}}
        \caption{
            Comparison of parallaxes from timing, VLBI, and \textit{Gaia} measurements.
            Top: timing versus \textit{Gaia};
            Bottom: timing versus VLBI.
            The error bars represent the formal uncertainties quoted from the published data, usually corresponding to a confidence level of 68\%.
            Different marks are used to distinguish pulsars whose timing solutions refer to different ephemerides.
            There are four pulsars with 20 timing parallax measurements for the \textit{Gaia} pulsar sample and five pulsars but with 25 timing parallax measurements for the VLBI pulsar sample.
            Different markers are used to distinguish the ephemerides used in the timing solution, that is, blue filled triangles for DE200, green crosses for DE405, purple filled inverted-triangles for DE421, and red filled circles for DE436.
                  }
        \label{fig:plx-comparison}%
    \end{figure}

\begin{figure}
        \resizebox{\hsize}{!}{\includegraphics{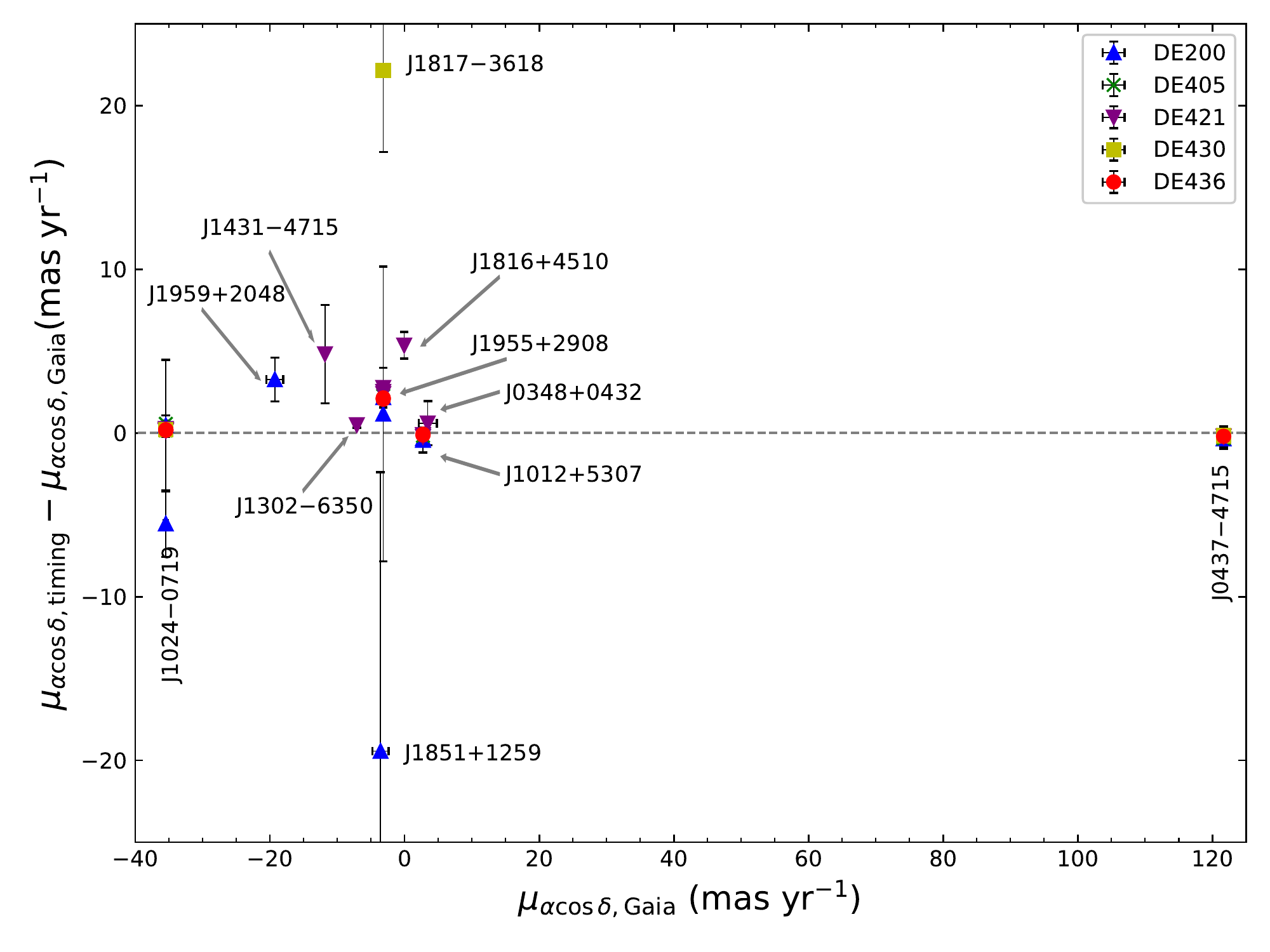}}
        \resizebox{\hsize}{!}{\includegraphics{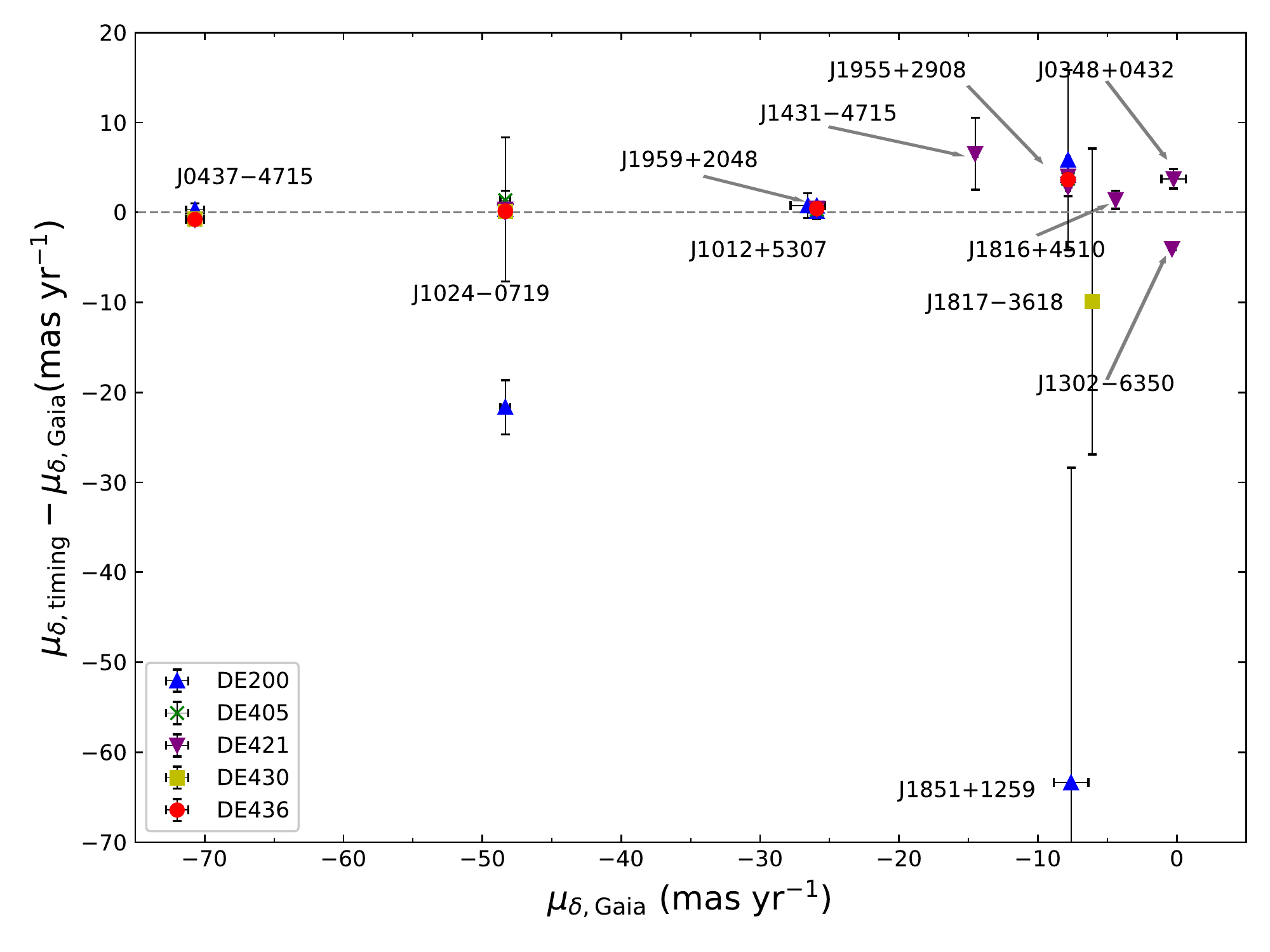}}
        \caption{
            Comparison of proper motions from timing and \textit{Gaia} measurements for 11 pulsars in common.
            Top: Right ascension; Bottom: Declination.
            The error bars represent the formal uncertainties quoted from the published data, usually corresponding to a confidence level of 68\%.
            Different marks are used to distinguish pulsars whose timing solutions were referred to different ephermerides, that is, blue filled triangles for DE200, green crosses for DE405, purple filled inverted-triangles for DE421, yellow filled squares for DE430, and red filled circles for DE436.
                  }
          \label{fig:timing-vs-gaia-pm}%
    \end{figure}

\begin{figure*}
        \centering
        \includegraphics[width=\columnwidth]{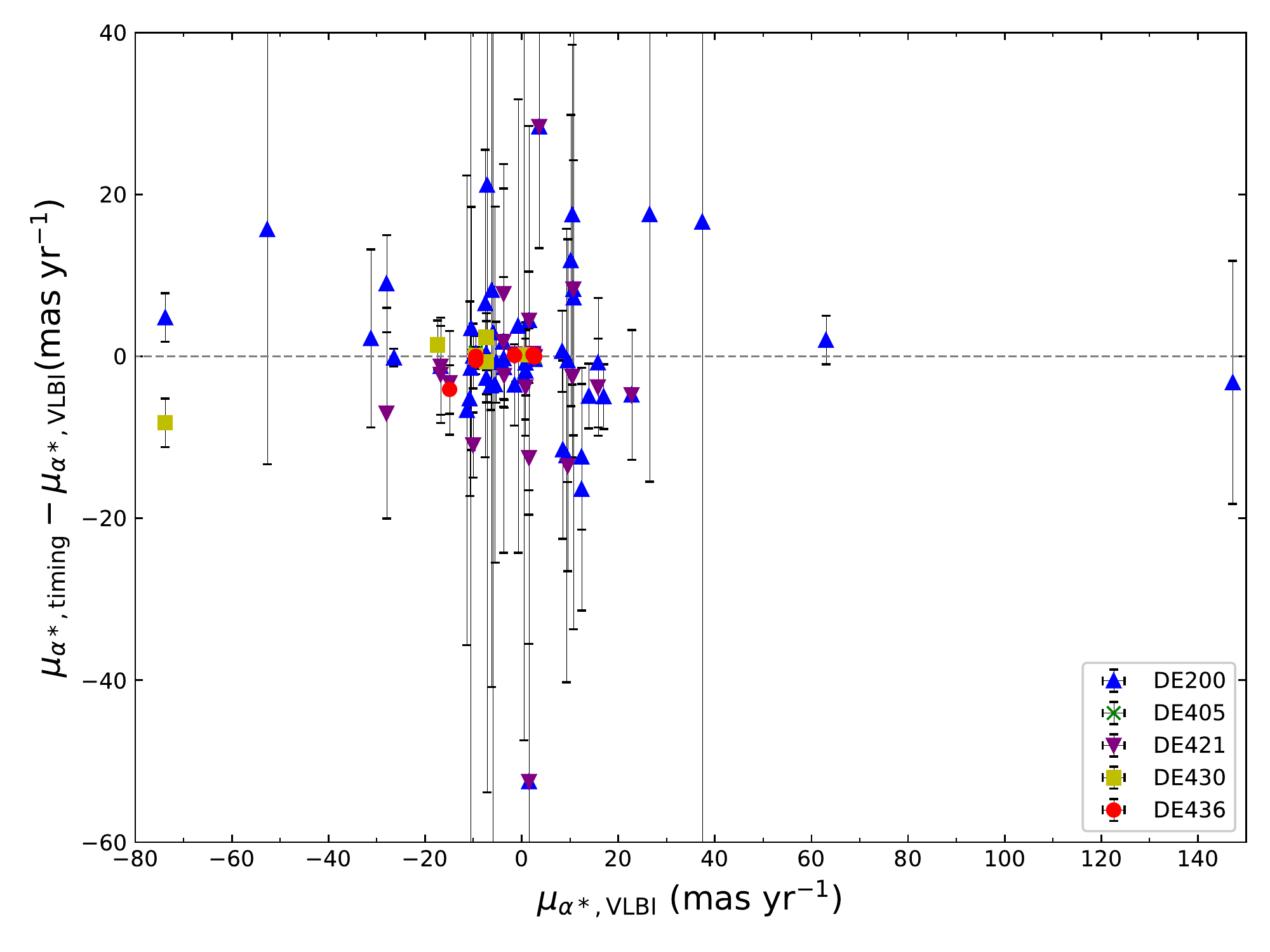}
        \includegraphics[width=\columnwidth]{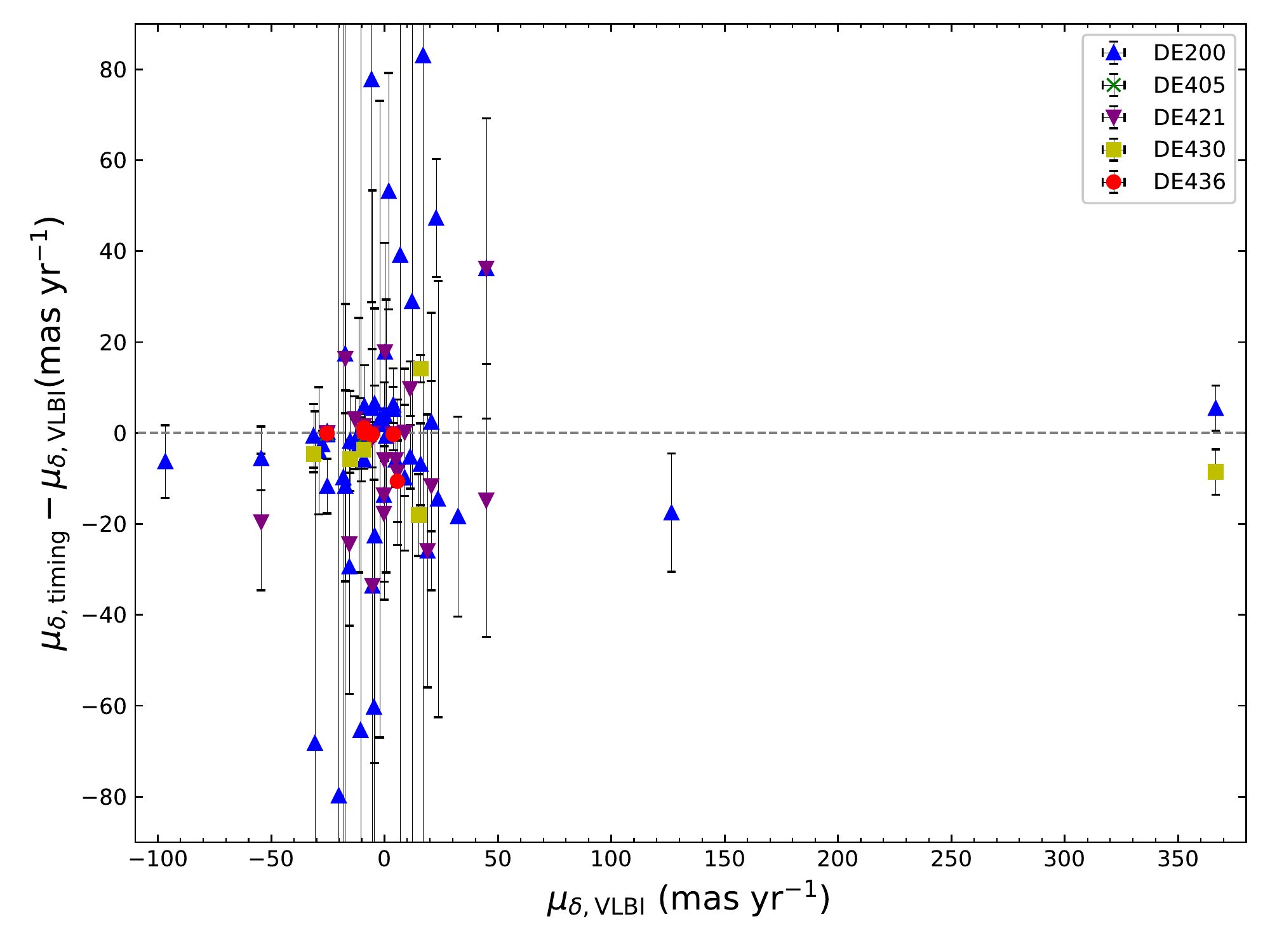}
        \includegraphics[width=\columnwidth]{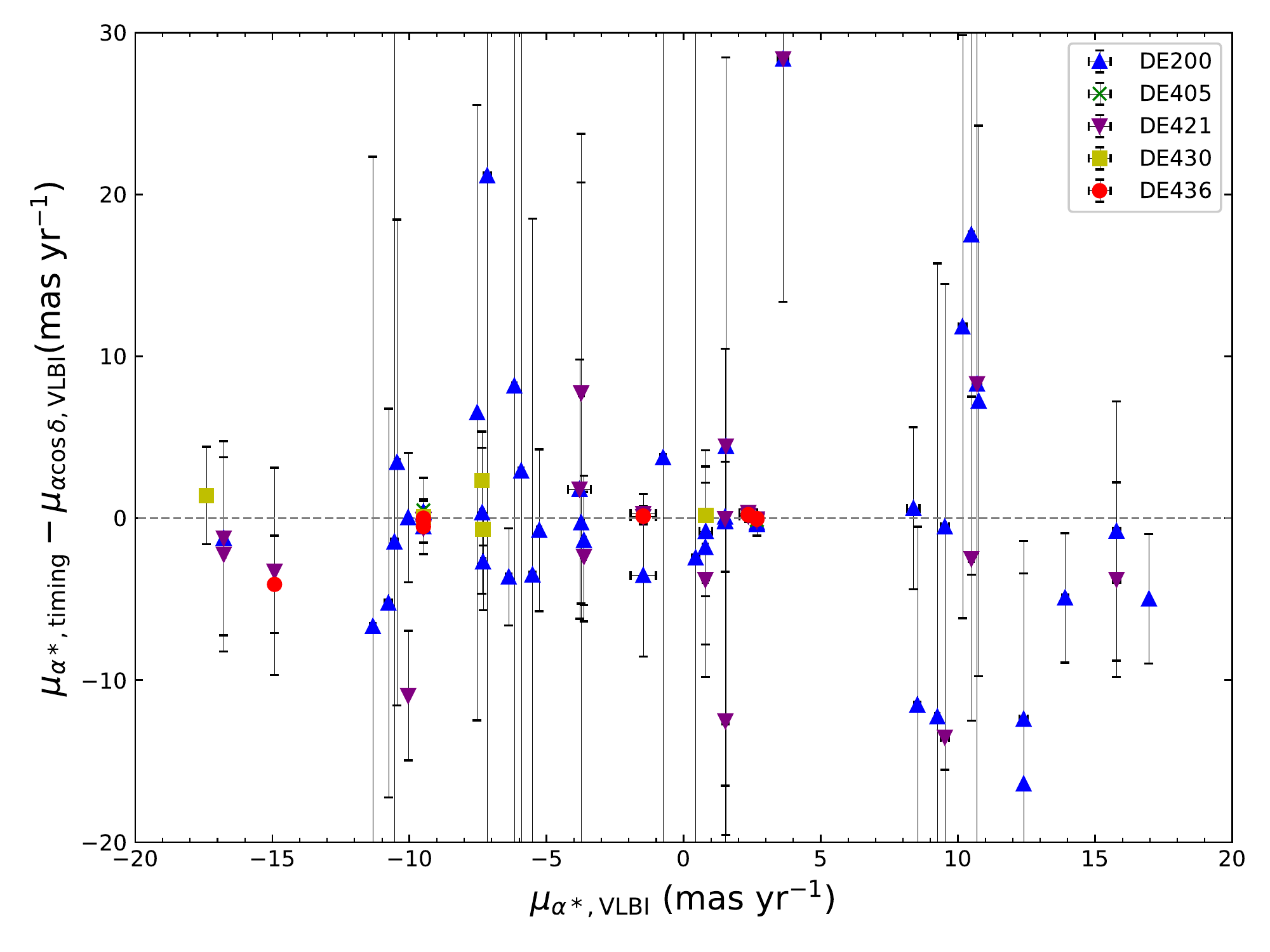}
        \includegraphics[width=\columnwidth]{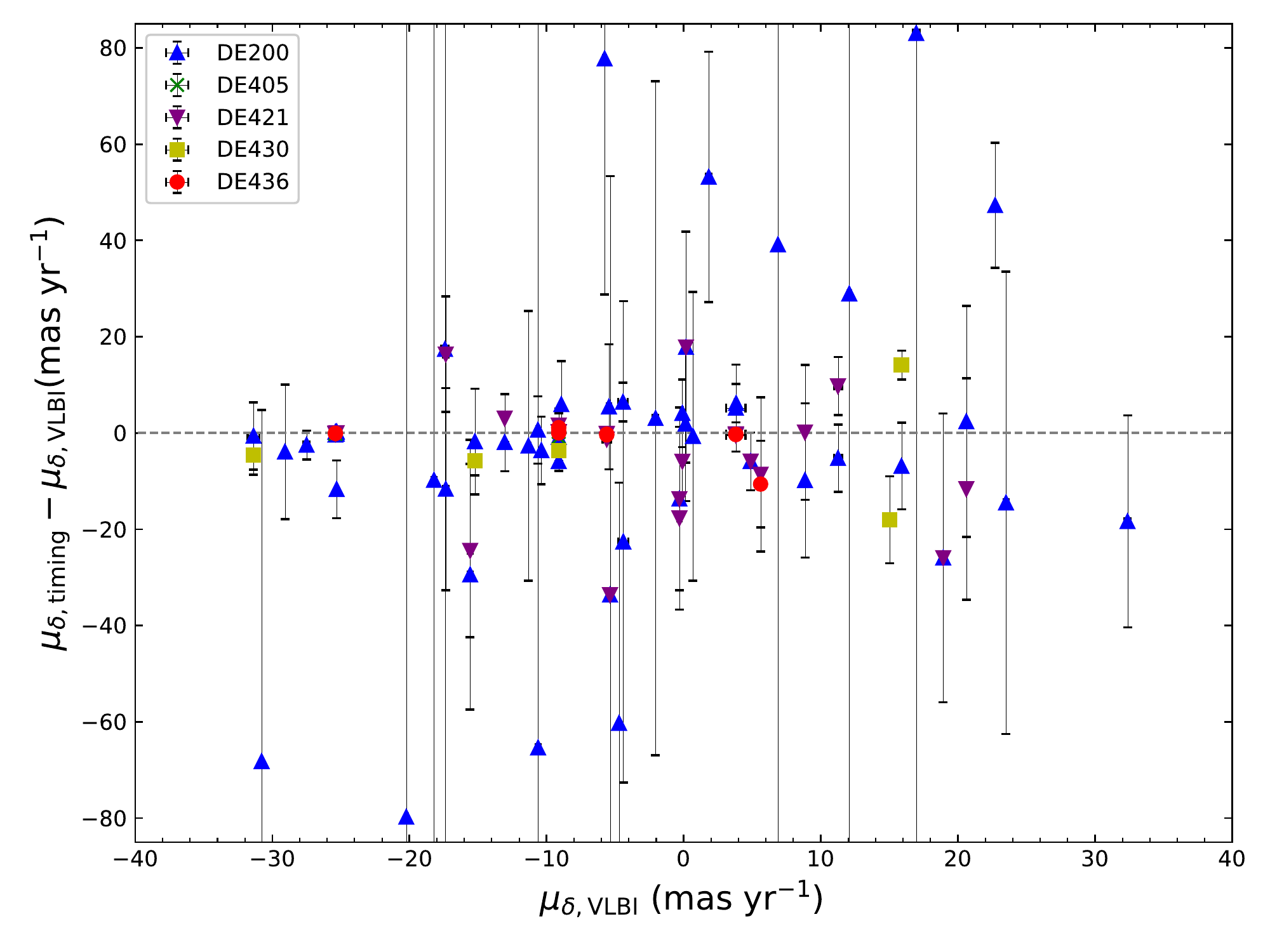}
        \caption{\label{fig:timing-vs-vlbi-pm}%
            Comparison of proper motions from timing and VLBI measurements.
            Top Left: Right ascension;
            Top Right: Declination;
            Bottom Left: Right ascension (zoom-in);
            Bottom Right: Declination (zoom-in);            
            The errorbars represent the formal uncertainties quoted from the published data, usually corresponding to a confidence level of 68\%.
            Different marks are used to distinguish pulsars whose timing solutions were referred to different ephermerides, that is, blue filled triangles for DE200, green crosses for DE405, purple filled inverted-triangles for DE421, yellow filled squares for DE430, and red filled circles for DE436.
                  }
    \end{figure*}

Figure~\ref{fig:plx-comparison} compares the parallax measurements from timing, VLBI, and \textit{Gaia}, while the comparisons of timing proper motions to those of \textit{Gaia} and VLBI are displayed in Fig.~\ref{fig:timing-vs-gaia-pm} and Fig.~\ref{fig:timing-vs-vlbi-pm}, respectively.

\end{appendix}

\end{document}